\documentclass [journal]{IEEEtran}

\usepackage{cite}

\ifCLASSINFOpdf
   \usepackage[pdftex]{graphicx}
\else
   \usepackage[dvips]{graphicx}
\fi

\usepackage[cmex10]{amsmath}

\ifCLASSOPTIONcompsoc
  \usepackage[caption=false,font=normalsize,labelfont=sf,textfont=sf]{subfig}
\else
  \usepackage[caption=false,font=footnotesize]{subfig}
\fi

\usepackage{multirow}
\usepackage{bm}
\usepackage{amssymb}
\usepackage{color}

\DeclareMathOperator*{\argmin}{arg\,min}

\begin{document}
%
\title{Soft  Delivery: Survey on A New Paradigm for Wireless and Mobile Multimedia Streaming}

\author {Takuya~Fujihashi,~\IEEEmembership{Member,~IEEE,}
        Toshiaki~Koike-Akino,~\IEEEmembership{Senior Member,~IEEE,}
        and Takashi~Watanabe,~\IEEEmembership{Member,~IEEE,}

\thanks{Takuya~Fujihashi and Takashi~Watanabe are with Graduate School of Information Science and Technology, Osaka University, Suita, Osaka, 565-0871 JAPAN e-mail: fujihashi.takuya@ist.osaka-u.ac.jp, watanabe@ist.osaka-u.ac.jp.}
\thanks{Toshiaki~Koike-Akino is with Mitsubishi Electric Research Laboratories (MERL), Cambridge, MA 02139 USA e-mail: koike@merl.com.}
\thanks{Manuscript received November 9, 2021.}}


\maketitle

\begin{abstract}
The increasing demand for video streaming services is the key driver of modern wireless and mobile communications.
For robust and high-quality delivery of video content over wireless and mobile networks, the main challenge is sending image and video signals to single and multiple users over unstable and diverse channel environments.
To this end, many studies have designed digital-based video delivery schemes, which mainly consist of a sequence of digital-based coding and transmission schemes. 
Although digital-based schemes perform well when the channel characteristics are known in advance, significant quality degradation, known as cliff and leveling effects, often occurs owing to the fluctuating channel characteristics.  
To prevent cliff and leveling effects irrespective of the channel characteristics of each user, a new paradigm for wireless and mobile video streaming has been proposed.
Soft delivery schemes skip the digital operations of quantization and entropy and channel coding while directly mapping the power-assigned frequency--domain coefficients onto the transmission symbols. This modification is based on the fact that the pixel distortion due to communication noise is proportional to the magnitude of the noise, resulting in graceful quality improvement, wherein quality is improved gradually, according to the wireless channel quality without any cliff and leveling effects.
Herein, we present a comprehensive summary of soft delivery schemes. First, we provide a brief introduction on  wireless and mobile video streaming.
Second, we discuss the issues associated with conventional digital-based video delivery schemes. Third, an overview of soft delivery is presented, and then the state of the art in soft delivery is summarized by considering energy compaction, power allocation, bandwidth utilization, and packet loss resilience. Finally, an excursion on the extensions needed for immersive content and future research directions are provided.
\end{abstract}

%
\IEEEpeerreviewmaketitle

\section{Introduction}
Video streaming over wireless and mobile networks is one of the major applications in wireless environments; according to Cisco Visual Networking Index studies, approximately four-fifths (82\%) of the world’s mobile data traffic will be video content by 2022~\cite{bib:cisco2017}.
The explosive growth of data traffic, especially video traffic, poses a huge challenge to wireless and mobile networks. In recent years, immersive content, such as virtual reality~(VR), augmented reality~(AR), and mixed reality~(MR), have shown very good potential to be the next important applications for networks. Together with the development of fifth generation~(5G) technology and smart wearable devices, which enable technology for all extended reality~(XR) applications, the growth of such immersive applications is rapidly increasing.

In general, wireless video streaming systems transmit images and video signals to a single or multiple users with different channel characteristics. 
For high-quality video streaming applications, the main challenge is the difficulty in fully utilizing each user’s channel capacity and providing each user with the best video quality possible under his/her channel conditions; this will provide users with an improved quality of experience~(QoE).
To address this challenge, conventional streaming systems, which consist of video coding and transmission technologies, have been proposed based on digital-based solutions. In terms of the video coding part, H.265/High-Efficiency Video Coding (HEVC)~\cite{bib:pcs_hevc}, which has been standardized by the Joint Collaborative Team on Video Coding~(JCT-VC), can be used to encode VR/360-degree videos. As the successor of HEVC, the future video coding standard, named, H.266/Versatile Video Coding (VVC), has been developed by the Joint Video Experts Team (JVET); VVC takes camera-view video, high dynamic range  video, and VR/360-degree video into account. In addition, video- and geometry-based point cloud coding~\cite{bib:LiDAR_journal,bib:Danillo2020} have been standardized by the Motion Picture Experts Group~(MPEG) for volumetric video encoding and decoding.
In the video transmission part, the source bits are channel-coded with time interleaving to exhibit robustness against a certain level of channel errors. The channel-coded bits are then mapped into the transmit data symbols corresponding to  arbitrary modulation schemes, such as binary phase shift keying ~(BPSK), quadrature phase shift keying~(QPSK), or quadrature amplitude modulation~(QAM).
To  choose an appropriate source and channel coding rate according to the user's channel condition, the channel statistics are generally required to be known at the time of source and channel coding. Once both the source and channel coding processes are completed, the conventional systems work optimally only for a specific channel condition. 

If the observed channel quality, i.e., the channel signal-to-noise ratio (SNR), falls below a threshold, the decoding process tends to break down completely. 
This phenomenon is called the cliff effect~\cite{bib:cliff2011}.
In contrast, if the observed channel quality increases beyond the threshold, it does not improve the performance unless an adaptive rate control of the source and channel coding is performed in real-time according to the rapid fading channels. 
This phenomenon is known as the leveling effect. 

Thus, accurate channel estimation and real-time rate control of the source and channel coding are desired for conventional streaming systems.
However, the channel conditions of wireless and mobile networks may vary drastically and unpredictably, resulting in imperfect channel estimation and rate control owing to this variation. 
Conventional streaming systems tend to utilize the channel capacity conservatively to prevent cliff and leveling effects, taking into account the fact that rate control may be inaccurate.

A new paradigm of wireless video delivery, namely, soft delivery~ \cite{bib:softcast1,bib:softcast2,bib:softcast3}, has been proposed to fully utilize the instantaneous channel capacity without cliff and leveling effects. 
It is essentially a scheme with ``lossless compression and lossy transmission.'' 
The compression stage is solely a transform to decorrelate the image and video signals into frequency domain coefficients, leaving out the conventional quantization and entropy coding. 
The transmission stage skips digital-based channel coding. Instead, it scales each transform coefficient individually and modulates it directly to a dense constellation for transmission. Here, the scaling operation serves
the purposes of both power allocation and unequal signal protection
against channel noises and fading effects to maximize the reconstruction quality.
At the receiver end, the image and video signals are reconstructed by demodulating the
received signals and inverting the scaling and transform operations. 
The soft delivery scheme was shown to not only provide a graceful performance transition in a wide channel SNR  range but also achieve competitive performance compared with the conventional digital-based delivery schemes.

\begin{figure*}[t]
  \begin{center}
   \includegraphics[scale=0.5]{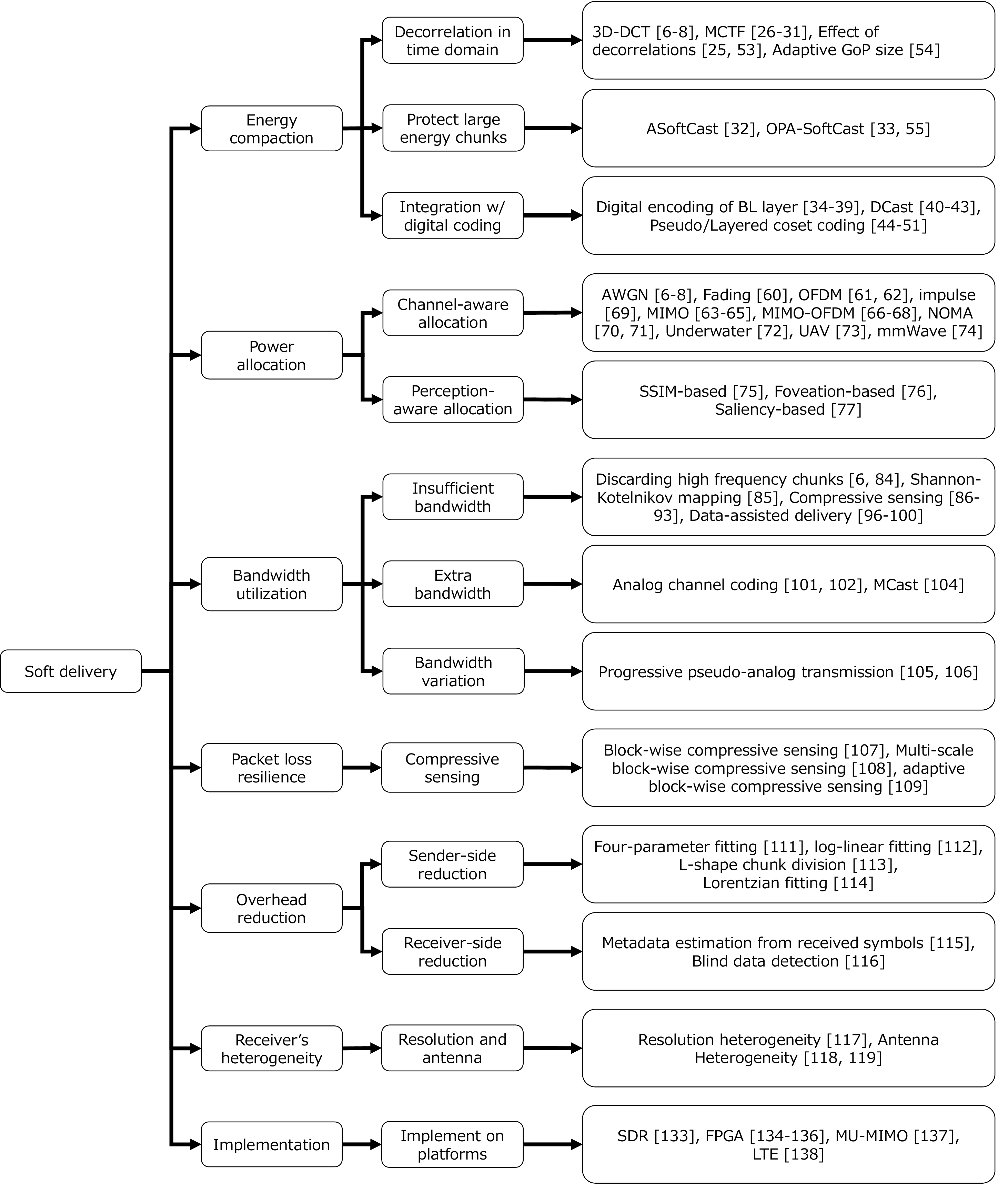}
   \caption {Taxonomy of the studies on soft delivery schemes.}
   \label{fig:list}
  \end{center}
\end{figure*}

\subsection{Contributions of the Study}
This work entails a comprehensive survey of soft delivery schemes, including an overview on existing techniques, extension for immersive content, and future research directions. 
Some existing studies have focused on soft delivery schemes, which are shown in Fig.~\ref{fig:list}, with a brief description of the related topics and key techniques.
To the best of our knowledge, this survey is the first to introduce methodologies and approaches for soft delivery to transmit high-quality image and video signals via unstable and diverse wireless and mobile channel environments. 
The main contributions of this study are summarized as follows:
\begin{itemize}
\item An overview of the conventional digital-based and soft delivery schemes, as well as the benefits of the latter, is presented. 
\item The existing soft delivery techniques, such as energy compaction, power allocation, bandwidth utilization, and packet loss resilience, are surveyed. In this context, the abstraction and key contributions of these techniques are reviewed and summarized. 
\item The extensions needed for immersive video streaming using the soft delivery are classified, including free-viewpoint video, 360-degree video, and point cloud. 
\item We finally review the future research directions related to soft delivery.
\end{itemize}

\subsection{Survey Structure}

The remainder of this paper is organized as follows:
\begin{itemize}
\item Section~\ref{sec:digital} describes
an overview of conventional digital-based delivery schemes and their issues, such as the cliff, leveling, and staircase effects. 
\item Section~\ref{sec:softcast} presents the basic principles of the pioneering work on soft delivery to solve the aforementioned effects in wireless and mobile video streaming applications.
\item Section~\ref{sec:aspect} presents a review of the existing techniques on soft delivery. We classify these techniques into energy compaction, power allocation, bandwidth utilization, packet loss resilience, overhead reduction, receiver heterogeneity, and discuss their implementation, as well as their contributions.
\item Section~\ref{sec:immersive} lists the existing techniques vis-à-vis soft delivery for immersive contents, i.e., future multimedia applications. 
\item Section~\ref{sec:future} suggests the future directions of the soft delivery approach based
on the overall trends observed from the survey results. 
\item Section~\ref{sec:conclusion} concludes the paper.
\end{itemize}

\begin{figure*}[t]
  \centering
   \subfloat[Digital-based Scheme]{\includegraphics[height=0.21\linewidth]{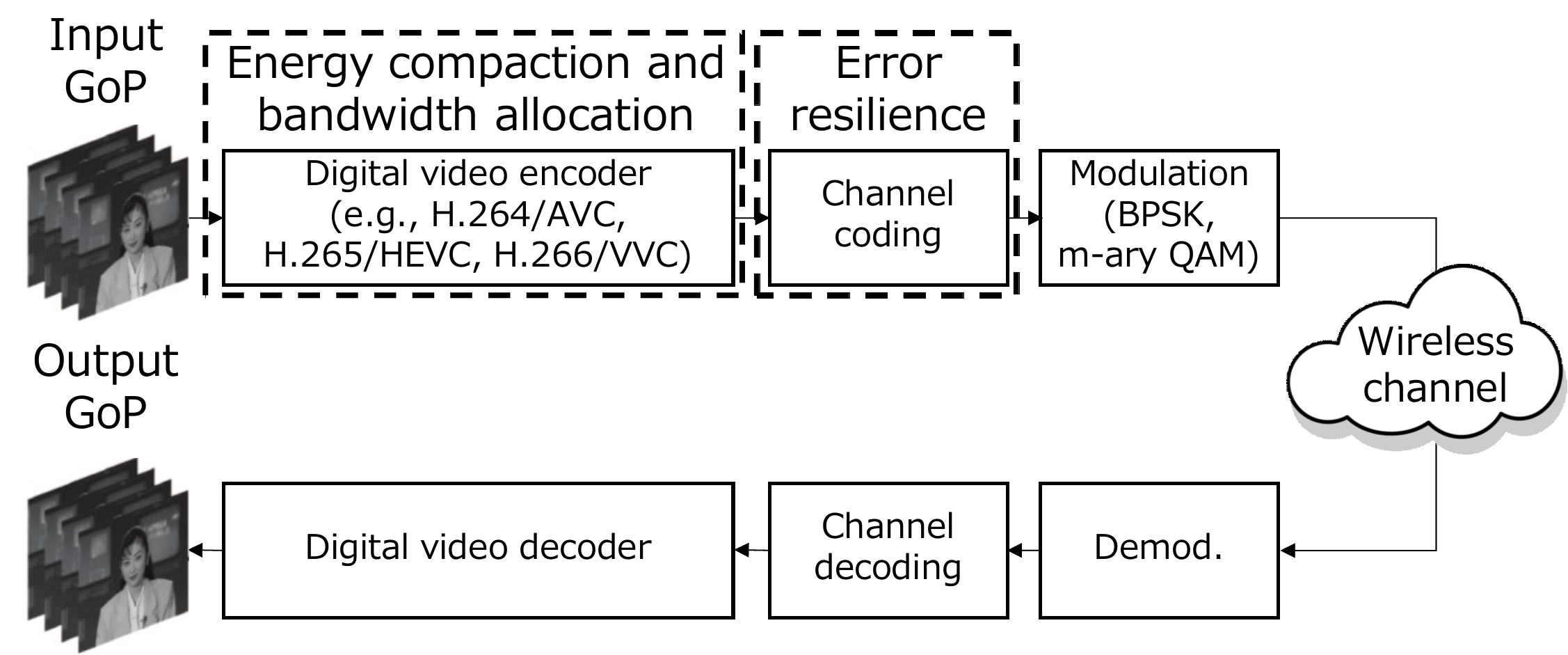}} 
   \hfill
   \subfloat[SoftCast]{\includegraphics[height=0.21\linewidth]{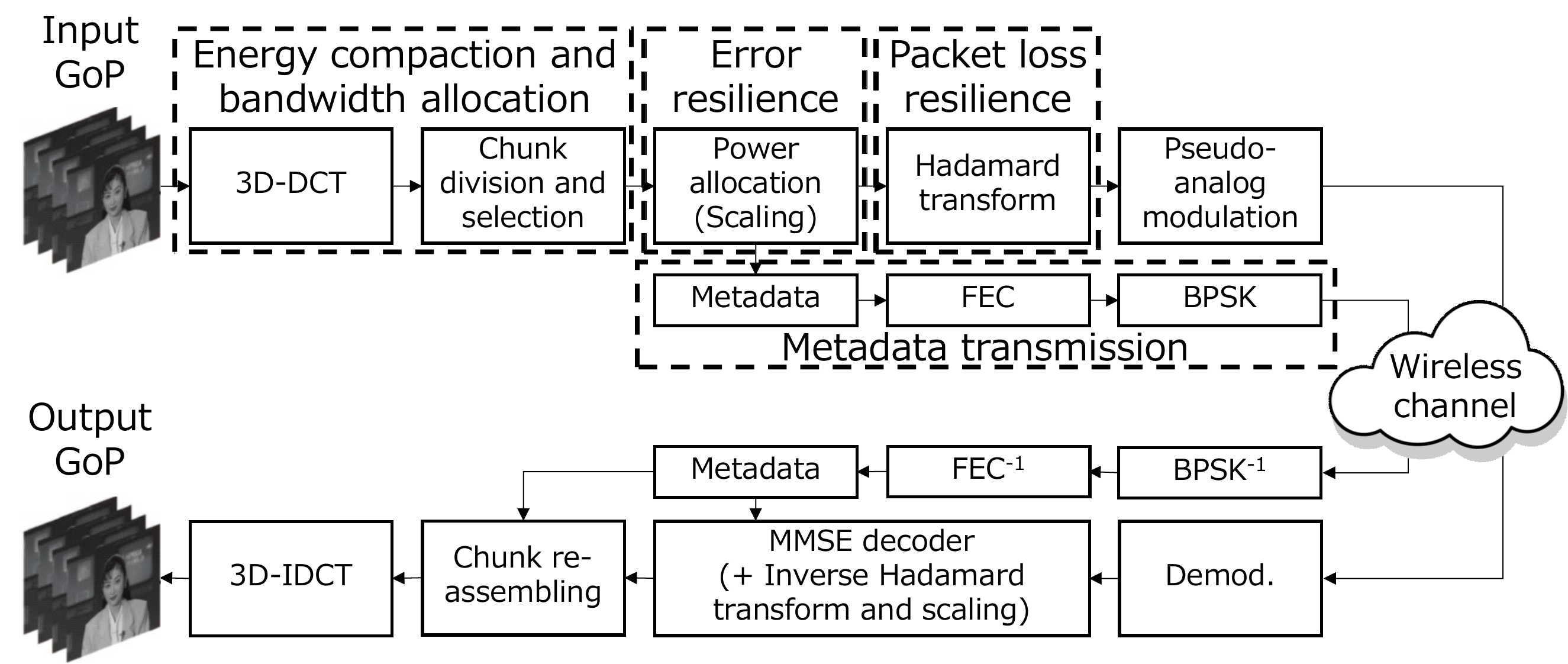}}
 \caption[]{Framework of the conventional digital-based delivery and SoftCast schemes.}
 \label{fig:digital_analog}
\end{figure*}

\begin{figure*}[t]
  \centering
   \subfloat[Cliff and leveling effects in digital-based schemes]{\includegraphics[height=0.22\linewidth]{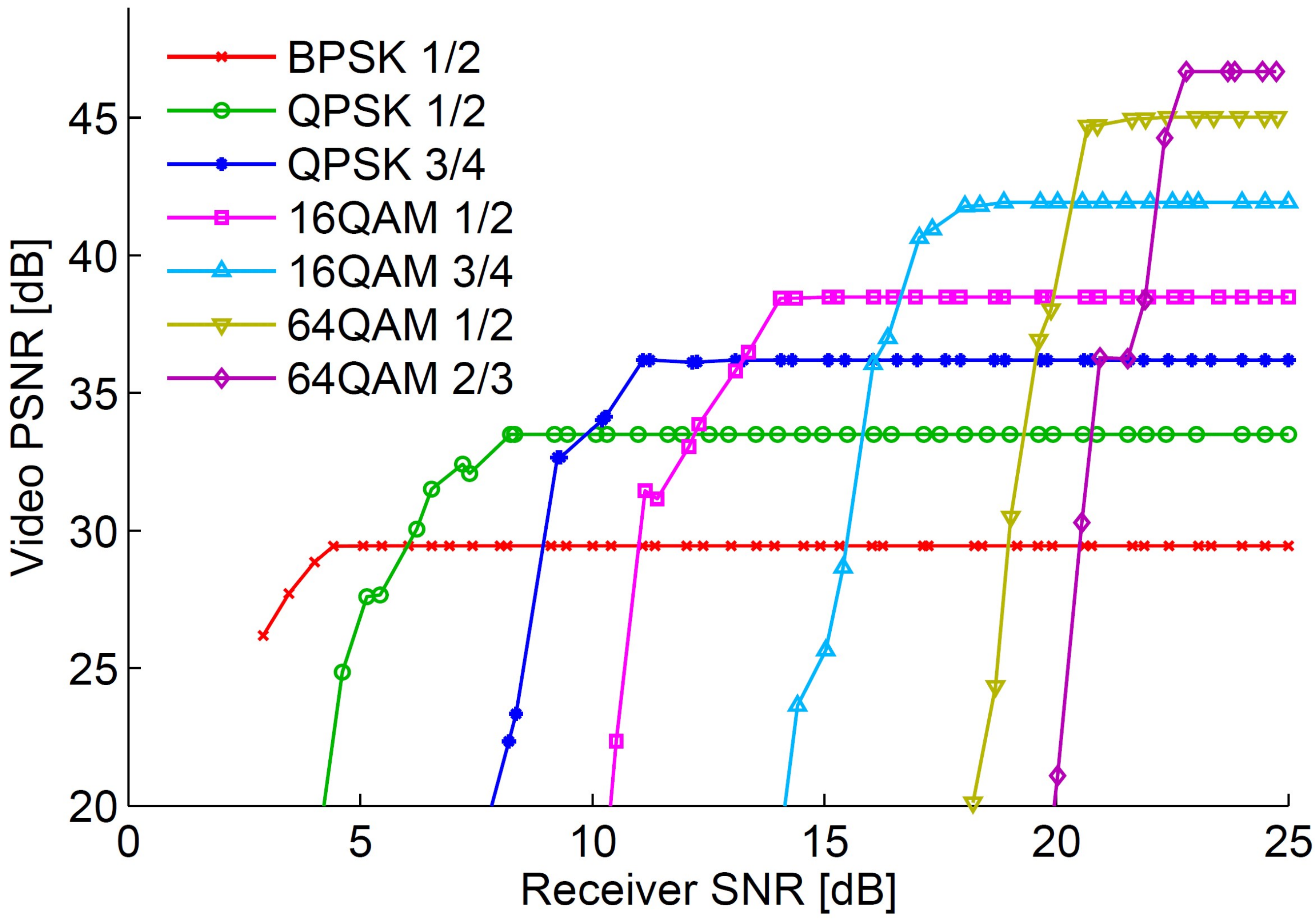}} 
   \hfill
   \subfloat[Staircase effect in layered coding with hierarchical modulation schemes]{\includegraphics[height=0.22\linewidth]{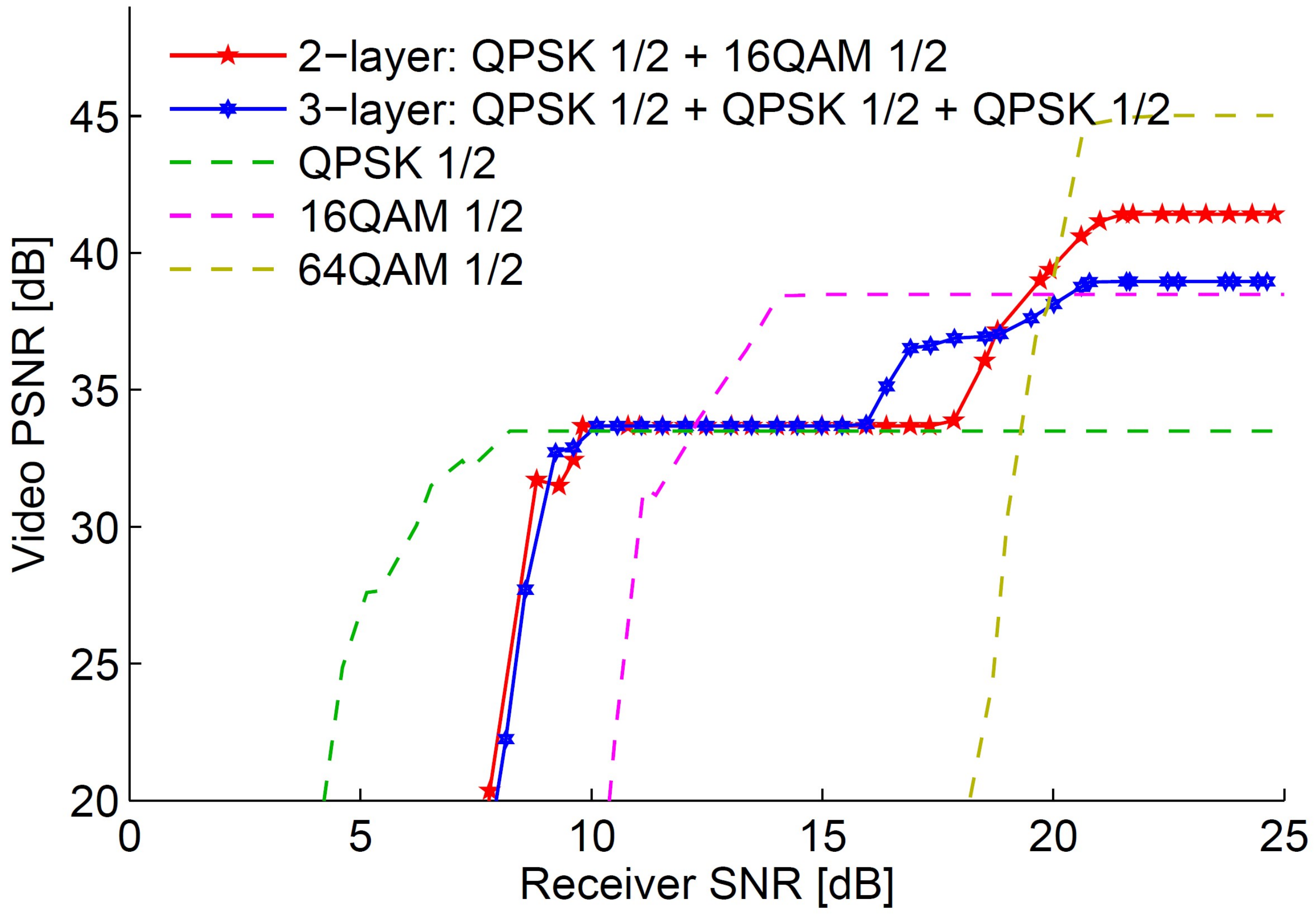}}
   \hfill
   \subfloat[Soft delivery scheme]{\includegraphics[height=0.22\linewidth]{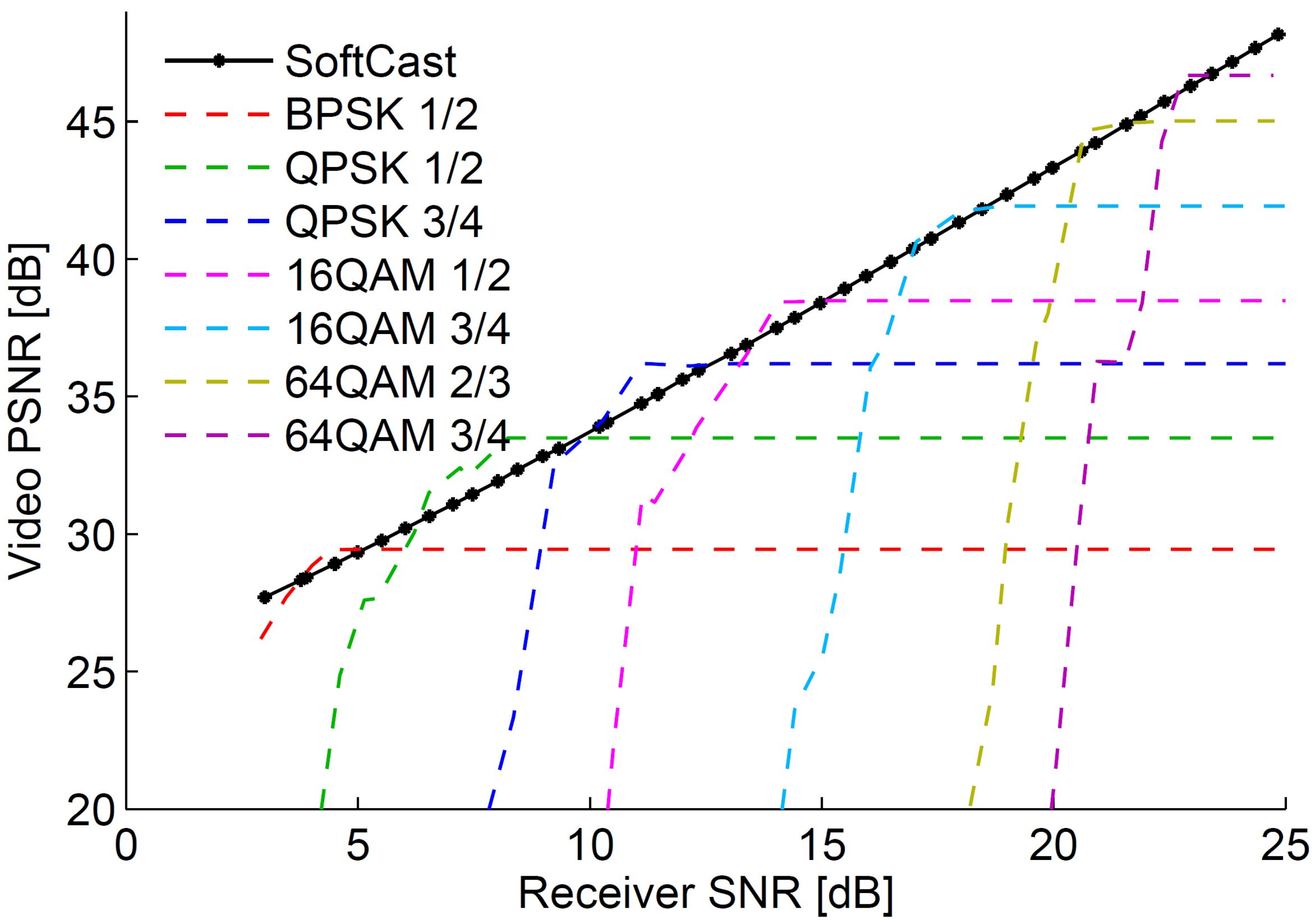}}
 \caption[]{Video quality of the conventional digital-based schemes and SoftCast scheme via wireless networks~\cite{bib:softcast1}: (a) Cliff and leveling effects in H.264/AVC over 802.11 under different MCSs. (b) Staircase effect in two-layered video coding and three-layered video coding shown in red and blue, respectively. For reference, the dashed lines are the three equivalent single-layer H.264/AVC videos. (c) Performance of SoftCast versus single-layer H.264/AVC.}
 \label{fig:cliff}
\end{figure*}

\section{Conventional Digital-based Delivery}
\label{sec:digital}
\subsection{Overview}
One of the major issues in wireless video delivery
is sending high-quality videos within the considerably limited capacity of wireless links. For this purpose, standardized digital video compression is carried out for video frames in conventional video delivery schemes~\cite{bib:wireless_streaming1, bib:wireless_streaming2, bib:wireless_streaming3}, as shown in Fig.~\ref{fig:digital_analog}~(a) to remove redundancy among video frames. 
In particular, H.264/Advanced Video Coding (AVC)~\cite{bib:MPEG_Standard}, H.265/HEVC, and H.266/VCC standards are typical video coding standards for generating a compressed bitstream from video frames.
In such standard video encoders, the video frames are classified into I-, P-, and B-frames. The I-frame is a reference frame encoded independently from other video frames. 
In each I-frame, pixel values are divided into blocks and transformed into frequency domain coefficients using discrete cosine transform (DCT) or discrete sine transform (DST), and then non-uniformly quantizing the coefficients according to a quantization parameter~(QP). A large QP indicates a larger quantization step, leading to a smaller bit rate. Finally, the quantized coefficients are compressed by an entropy coder, which removes statistical redundancy in the coefficients. Variable length coding (VLC) is widely deployed for entropy coding because of its efficiency and simplicity.
The P-frame obtains residuals between the previous reference and the current video frames using block-based motion estimation and then encodes the residuals using the aforementioned procedures. The B-frame also obtains and encodes residuals between the previous and past reference video frames and the current video frame using motion estimation. The P-frame and B-frame have lower traffic than the I-frame.

After passing through the digital video compression, the bitstream is then passed to the wireless transmission part in sequence. 
Channel coding is first used for the bitstream to protect against channel errors. For example, binary convolutional codes and low-density parity-check are widely used as forward error correction in Wi-Fi systems. The coded bitstream is then mapped onto in-phase and quadrature (I and Q) components using digital modulation formats such, as QPSK and m-ary QAM. In both wireless and mobile networks, a combination of modulation formats and different channel coding rates, for example, 1/2 and 3/4, is defined in the modulation and coding scheme~(MCS). According to the measured wireless channel SNRs, the sender adapts its MCS value to maximize the link data rate. 
At the receiver end, bit errors may occur in channel-coded bits owing to effective noise and/or fading effects. The receiver then tries to reconstruct video frames from the received bits using inverse procedures, i.e., demodulation, channel decoding, and video decoding.

\begin{table*}[t]
\caption{Critical issues regarding video quality in wireless and mobile video streaming}
    \centering
    \begin{tabular}{c c c c c}
    \hline
        Phenomenon & \begin{tabular}{c} Effect on \\ video quality \end{tabular} & \begin{tabular}{c} Direction of channel \\ SNR fluctuation \end{tabular} & Cause & Solution in soft delivery \\ \hline
        Cliff effect & Sudden degradation & Degrade & \begin{tabular}{c} All-or-nothing behavior in \\ entropy and channel codings \end{tabular} & \begin{tabular}{c} Skip entropy and channel codings \\ to prevent all-or-nothing behavior \end{tabular} \\ \hline
        Leveling effect & \begin{tabular}{c} Constant irrespective of \\ channel quality \end{tabular} & Improve & \begin{tabular}{c} Unrecoverable quantization \\ error in video coding \end{tabular} & \begin{tabular}{c} Skip quantization and adopt \\ pseudo-analog modulation for \\ recoverable errors at the \\ receiver end \end{tabular} \\ \hline
        Staircase effect & \begin{tabular}{c} Step function of \\ channel quality \end{tabular} & Both & \begin{tabular}{c} All-or-nothing behavior and \\ quantization error in \\  layered coding \end{tabular}  & \begin{tabular}{c} Skip hierarchical operations and \\ adopt pseudo-analog modulation for \\ linear video quality 
        \end{tabular}   
        \\\hline
    \end{tabular}
    
    \label{tab:phenomenon}
\end{table*}

\subsection{Critical Issues on Quality}
If the measured wireless channel quality is stable during video transmission, conventional digital-based schemes can provide high-quality video frames for users. However, the channel quality of each user fluctuates over time owing to a combination of user mobility, multipath propagation, and obstacle shadow. Table~\ref{tab:phenomenon} lists three critical issues regarding the video quality of the digital-based schemes because of the channel quality fluctuation: cliff, leveling, and staircase effects.

\subsubsection{Cliff Effect}
Digitally encoded bits are known to be susceptible to errors during wireless transmission. Because entropy coding schemes have an all-or-nothing behavior, even a single bit error can cause the loss of entire data~\cite{bib:VLC}. As mentioned earlier, channel coding schemes are adopted to correct burst and random bit errors. 
However, they generally exhibit an all-or-nothing behavior for error correction.
When the instantaneous channel quality, i.e., the SNR, falls below a certain threshold, possible errors that occur in the bitstream during wireless communications will disable video decoding. 

A collapsed signal reconstruction causes a cliff effect.
The cliff effect is a phenomenon whereby the quality of the received information abruptly drops as soon as the channel quality falls below the threshold, as shown in Fig.~\ref{fig:cliff}~(a). For example, the video quality of the BPSK modulation format with 1/2-rate channel coding drops below the wireless channel SNR of 4~dB. 
In modern network environments (e.g., content delivery, mobile, and wireless networks), the cliff effect becomes a major impediment when video frames are transmitted over diverse channel conditions to heterogeneous users.
In this case, users whose channel quality is below the critical point receive unwatchable video frames. 

Some solutions have addressed the cliff effect associated exclusively with channel coding, such as hybrid automatic repeat request and rateless coding schemes~\cite{bib:rateless,bib:rateless2,bib:iCast,bib:simcast,bib:flexcast}.
They adapt the number of transmissions to changing channel conditions for error prevention.
However, these schemes are not well suited for streaming multiple users under diverse channel conditions. In addition, they do not reduce the quantization error at the video encoder end; thus, the leveling effect still occurs in video quality. 

\subsubsection{Leveling Effect}
Once the channel quality surpasses the threshold, the video quality remains constant as shown in Fig.~\ref{fig:cliff}~(a). As mentioned earlier, the cliff effect is caused when the receiver SNR is below 4~dB in the BPSK modulation format with 1/2-rate channel coding, whereas the channel gain does not reflect on the video quality above the wireless channel SNR of 4~dB. 
Digital-based schemes determine the parameters of the video coding and wireless transmission parts based on the channel
estimation. If the instantaneous channel quality is better than the estimated one, 
no additional gain can be obtained because the distortion of the video coding cannot be reconstructed for each user.  

\subsubsection{Staircase Effect}
To mitigate the cliff and leveling effects, some layered coding schemes, referred to as schemes with scalable video coding~(SVC)~\cite{bib:svc1} with a combination of hierarchical modulation~(HM)~\cite{bib:hm1}, have been proposed for wireless and mobile video streaming~\cite{bib:hm2, bib:hm3}. These layered coding schemes encode video frames into one base layer~(BL) and several enhancement layers~(ELs). 
The BL is used to ensure that all the users in the target channel SNR range can receive the baseline quality of video frames, whereas  the ELs are used to enhance the video quality of the users in high-channel SNRs. Each SVC layer is then mapped onto the corresponding HM layer. Notably, HM provides unequal error protection to the transmitted video frames according to their relative importance. However, SVC with HM cannot completely remove the cliff effect; it only divides one big cliff into multiple stairs according to the number of layers, as shown in Fig.~\ref{fig:cliff}~(b). In addition, because the assigned transmission power to each layer is lower than that of the single-layer coding schemes, the cliff shifts to higher wireless channel SNRs.

\section{SoftCast: A Pioneer Work on Soft Delivery}
\label{sec:softcast}
\subsection{Overview}
To prevent the cliff, leveling, and staircase effects in wireless video delivery, a pioneer soft delivery work, namely, SoftCast was proposed in~\cite{bib:softcast1,bib:softcast2,bib:softcast3}.
The block diagram of SoftCast is illustrated in Fig.~\ref{fig:digital_analog}~(b).
SoftCast’s design is based on a simple principle that ensures that the transmitted signal samples are linearly related to the original pixel values. This principle naturally enables a sender to satisfy multiple receivers with diverse channel qualities, as well as a single receiver, where different packets experience different channel qualities.

The sender first takes a group of pictures~(GoP) and uses a full-frame 3D-DCT as the decorrelation transform.  The DCT
frames are then divided into $N$ small rectangular blocks of transformed coefficients called chunks. 
The coefficients in each chunk are then scaled to match the transmission power constraints. 
Specifically, the scaling coefficients are chosen to minimize the reconstruction mean square error~(MSE).
A Walsh--Hadamard transform (WHT) is then applied to the scaled chunks for power normalization across the chunks to provide packet loss resilience.
This process transforms the chunks into slices. Each slice is a
linear combination of all scaled chunks.
Finally, the coefficients in the slices are directly mapped to the I and Q components in a pseudo-analog manner for transmission.
Here, channel coding operations are skipped for the coefficients.

\begin{figure*}[t]
  \centering
   \subfloat[16-QAM]{\includegraphics[height=0.22\linewidth]{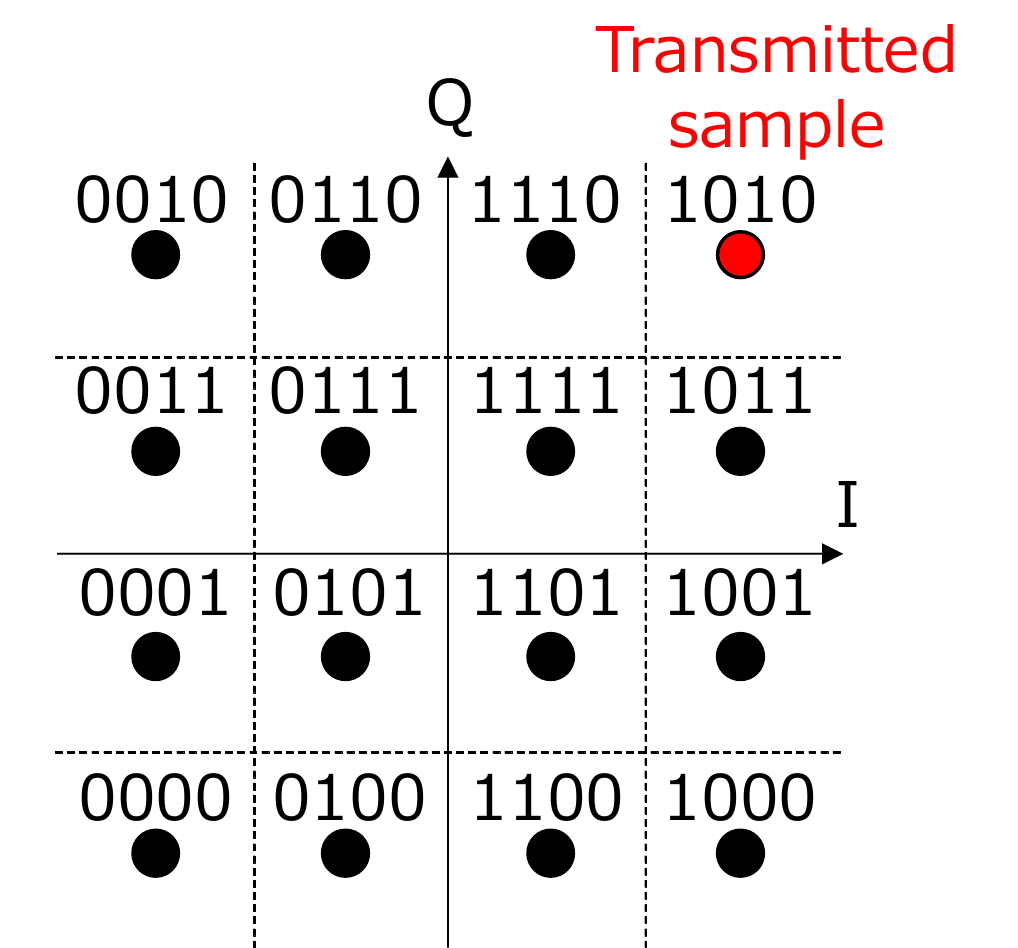}} 
   \hfill
   \subfloat[16-QAM in high channel SNRs]{\includegraphics[height=0.22\linewidth]{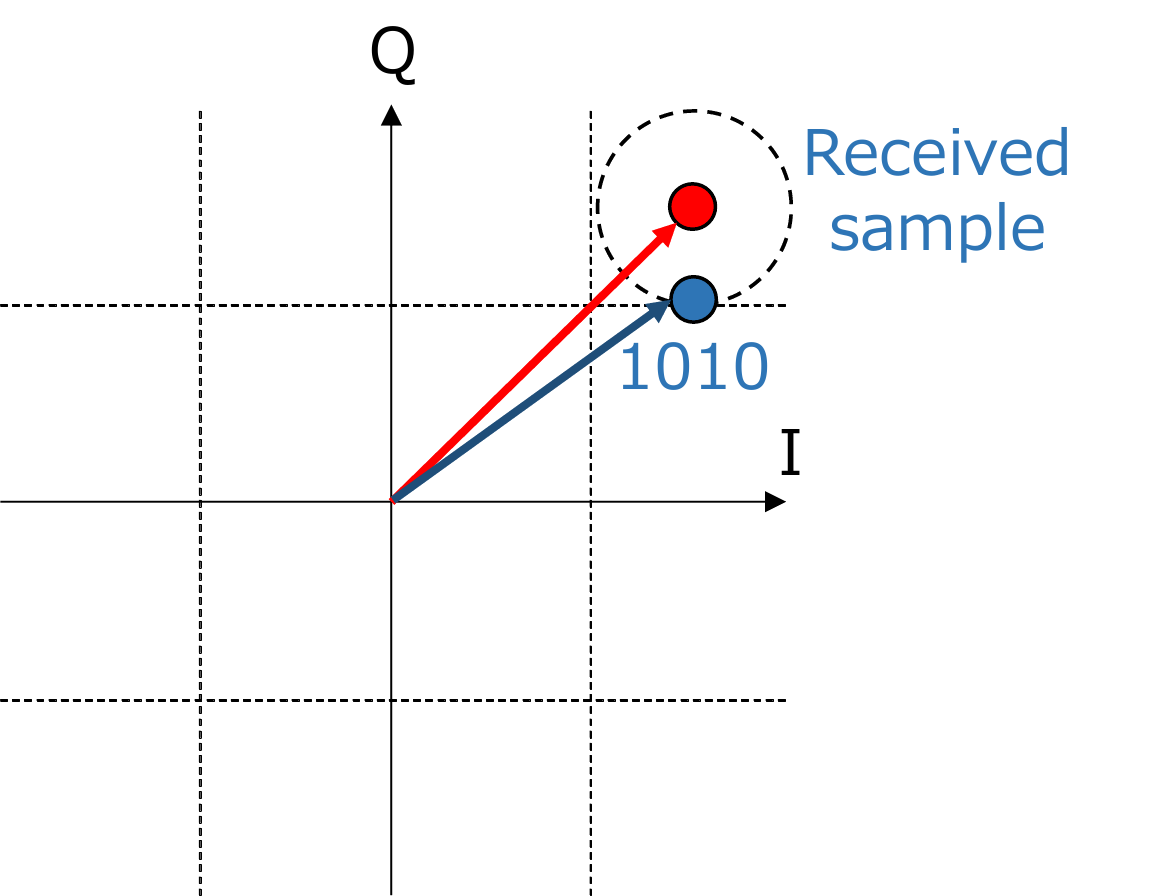}}
   \hfill
   \subfloat[16-QAM in low channel SNRs]{\includegraphics[height=0.22\linewidth]{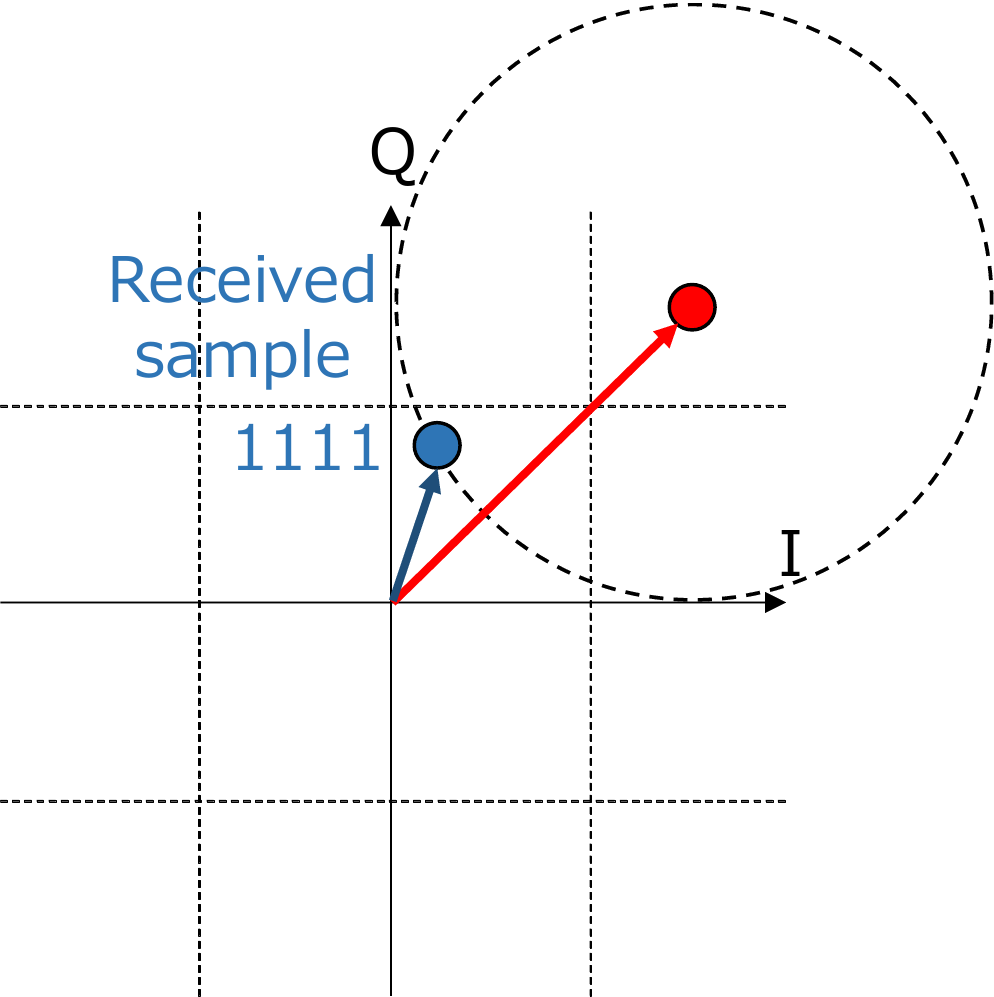}}
   \\
   \subfloat[SoftCast]{\includegraphics[height=0.22\linewidth]{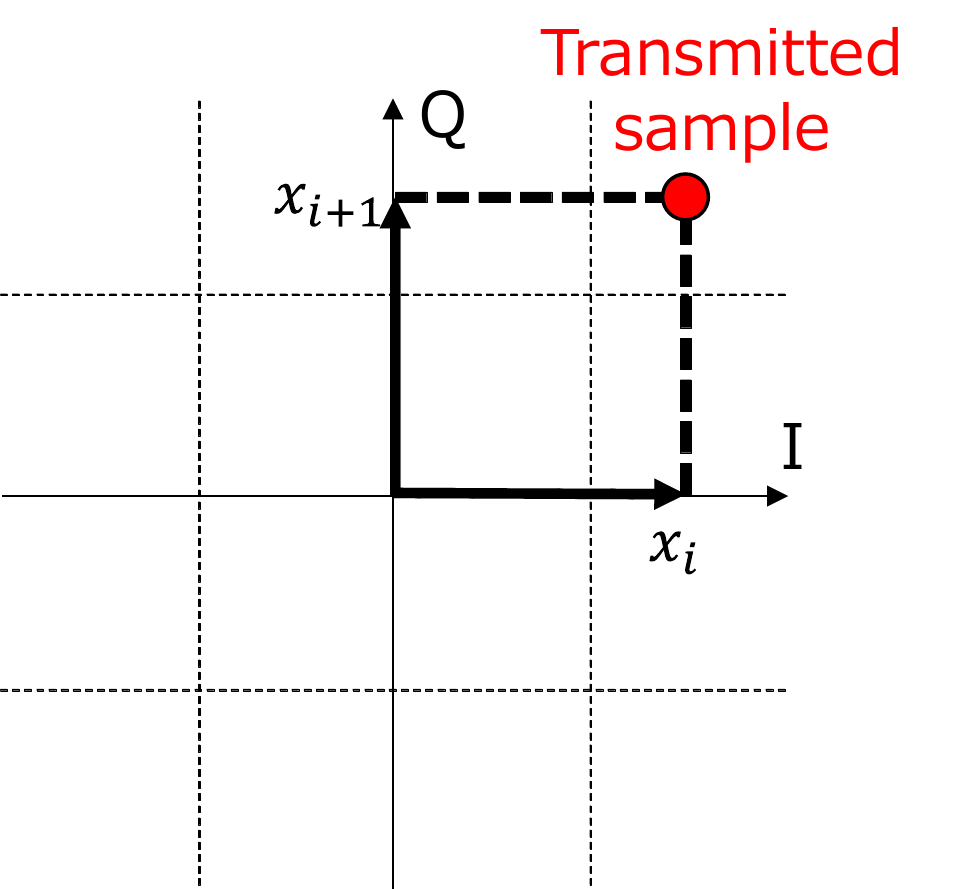}} 
   \hfill
   \subfloat[SoftCast in high channel SNRs]{\includegraphics[height=0.22\linewidth]{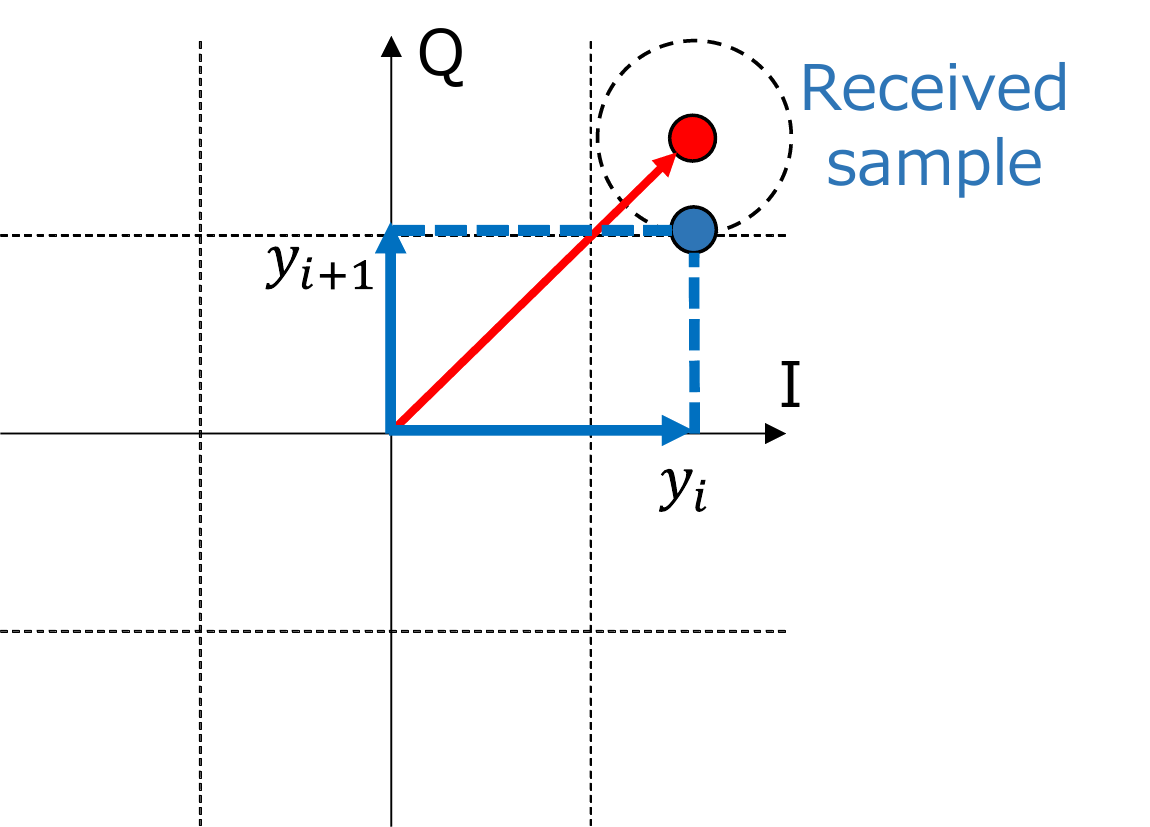}}
   \hfill
   \subfloat[16-QAM in low channel SNRs]{\includegraphics[height=0.22\linewidth]{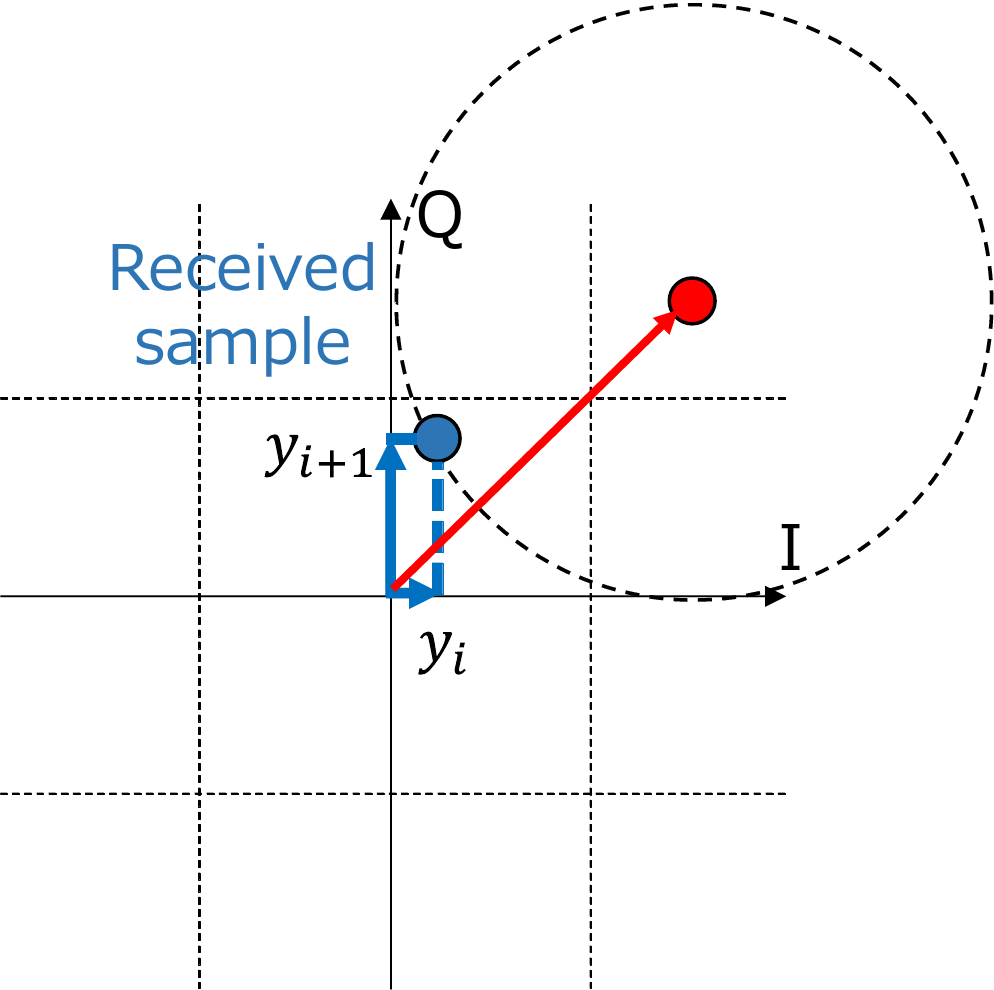}}
 \caption[]{Mapping coded video to I/Q components of the transmitted signal. (a) Traditional 16-QAM maps a bit sequence to the complex number corresponding to the point labeled with that sequence. (d) By contrast, SoftCast’s PHY treats pairs of coded values as the real and imaginary parts of a complex number.}
 \label{fig:phy}
\end{figure*}

Fig.~\ref{fig:phy}~(a) through (f) shows the conventional digital-based modulation, i.e., 16-QAM, and pseudo-analog modulation proposed in SoftCast. 
Conventional modulation modulates channel-coded bits to produce real-value digital samples that are transmitted to the channel.
For example, 16-QAM modulation takes sequences of four bits and maps each sequence to a complex I/Q number, as shown in Fig.~\ref{fig:phy}~(a). After modulation, the wireless physical layer~(PHY) of the sender transmits the mapped complex numbers to the receiver. In this study, we consider a transmitted signal sample of 1010. 
Because of the broadcast nature of the wireless medium, multiple receivers hear the transmitted samples but with different noise levels. 
For example, in Figs.~\ref{fig:phy}~(b) and (c), a receiver with a high channel SNR can distinguish
which of the 16 small squares the original signal sample belongs to, and hence, correctly decode the transmitted sample. 
A receiver with a low channel SNR can distinguish only the quadrant of the transmitted sample, and hence, can decode only the two bits of the transmitted sample. In this case, these bit errors may cause a collapsed signal reconstruction during digital video decoding. 

In contrast to the existing modulation design, SoftCast outputs the real values of the DCT coefficients that are already coded for error protection. 
The pseudo-analog modulation directly maps pairs of the scaled DCT coefficients to the I and Q of the digital signal samples, as shown in Fig.~\ref{fig:phy}~(d).
As mentioned earlier, multiple receivers hear the transmitted samples under different channel SNRs. 
Although the transmitted samples are distorted according to their SNR, the receiver regards the received samples as scaled DCT coefficients. 
This process avoids all cliff, leveling, and staircase effects because the sender does not need to estimate the channel condition, and the noise level in the received samples faithfully reflects the instantaneous channel condition~\cite{bib:mos}. 
Consequently, pseudo-analog modulation ensures that the received video quality is proportional
to the instantaneous channel quality, as shown in Fig.~\ref{fig:cliff}~(c).

In parallel, SoftCast sends an amount of data, referred to as metadata, for signal reconstruction.
These metadata consist of the mean and variance of each
transmitted chunk as well as a bitmap. 
The mean of each chunk is used to obtain the chunk approximate zero-mean distributions by subtracting the mean of all pixels in each chunk~\cite{bib:preprocessing}.
The variance of each chunk is used to find the per-chunk scaling factors such that the reconstruction error is minimized.
The bitmap indicates the positions of the discarded chunks into the GoP.
When the available channel bandwidth for SoftCast is less than the required bandwidth, SoftCast discards chunks with less energy. 
Specifically, when the available and required bandwidths for SoftCast are $M$ chunks and $N (> M)$ chunks, respectively, SoftCast discards less energy $M-N$ chunks to meet the bandwidth requirement.  
On the receiver side, these discarded chunks are replaced by null values. 
The discarded chunks are registered as a bitmap and then compressed using run-length encoding.
Metadata are strongly protected and transmitted in a robust way (e.g., BPSK modulation format with a low-rate channel code) to ensure correct delivery and decoding.

At the receiver side, a minimum MSE~(MMSE) decoder is used to estimate the content of the chunks due to channel noise. 
The MMSE provides a high-quality estimate of the DCT coefficients by leveraging the knowledge of the statistics of the DCT coefficients, i.e., chunk variance, as well as the statistics of the channel noise. 
Using the metadata, the denoised chunks are properly reassembled and undergo an inverse 3D-DCT, thereby providing the corresponding GoP.

\begin{figure*}[t]
  \begin{center}
   \includegraphics[scale=0.45]{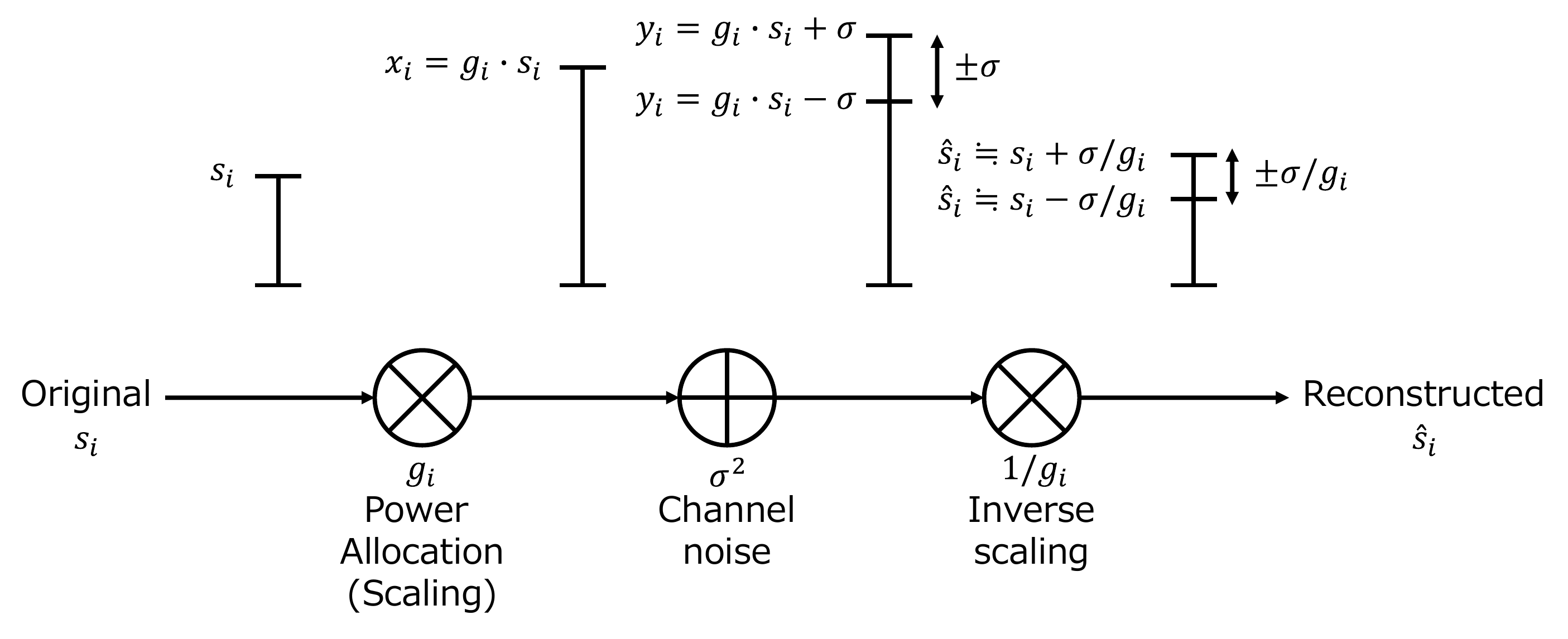}
   \caption{Scaling and inverse scaling operations in soft delivery schemes.}
   \label{fig:scaling}
  \end{center}
\end{figure*}

\subsection{Details of Scaling and Inverse Scaling Operations}
\label{sec:power}
In contrast to digital-based delivery schemes, soft delivery schemes use power allocation and MMSE filters, i.e., scaling and inverse scaling operations, for analog-modulated symbols to realize error protection over wireless channels. 
Figure~\ref{fig:scaling} illustrates the procedures of obtaining the reconstructed DCT coefficients at the receiver end.
SoftCast realizes chunk-wise power allocation and filter operations according to the statistics of the chunks and channel conditions.
Let $x_{i}$ denote the $i$th analog-modulated symbol. 
Each analog-modulated symbol is scaled by $g_{i}$ for noise reduction:
\begin{equation}
x_{i} = g_{i} \cdot s_{i}.
\end{equation}
Here, $s_{i}$ is the $i$th DCT coefficient, and $g_{i}$ is the scale factor for the coefficient power allocation.
The sender performs optimal power control for $g_i$ to achieve the highest video quality. Specifically, the best $g_{i}$ is obtained by minimizing the MSE under the power constraint with the total power budget $P$ as follows:
\begin{equation}
\min \quad \mathsf{MSE} = \mathbb{E} \left [ \left (x_{i} - \hat{x}_{i}\right)^2\right] = \sum_{i}^{N} \frac{\sigma^2 {\lambda}_{i}}{g_{i}^2{\lambda}_{i} + \sigma^2},
\end{equation}
\begin{equation}
\mathrm{s.t.} \quad \frac{1}{N}\sum_{i}^{N}  g_{i}^2{\lambda}_{i} = P,
\end{equation}
where $\mathbb{E}[\cdot]$ denotes the expectation, $\hat{x}_{i}$ is an estimate of the transmitted symbol, ${\lambda}_{i}$ is the power of the $i$th DCT coefficient, $N$ is the number of DCT coefficients, and $\sigma^2$ is the receiver noise variance.
The near-optimal solution is expressed as:
\begin{equation}
\label{water}
g_{i} = {\lambda}_{i}^{-1/4} \sqrt{\frac{P}{\sum_j{\lambda}_{j}}}.
\end{equation}

After transmission over the wireless channel, each symbol at the receiver end can be modeled as follows:
\begin{equation}
y_{i} = x_{i} + n_i,
\end{equation}
where $y_{i}$ is the $i$th received symbol and $n_{i}$ is an effective noise with a variance of $\sigma^2$.
The receiver extracts DCT coefficients from the I and Q components and reconstructs the coefficients using the MMSE filter \cite{bib:softcast1} as follows:
\begin{equation}
\hat{s}_{i} = \frac{g_{i} {\lambda}_{i}^2}{g_{i}^2 {\lambda}_{i}^2 + \sigma^2} \cdot y_{i}.
 \label{eq:mmse}
\end{equation}
The receiver then obtains the corresponding video sequence using the inverse 3D-DCT for the filter output $\hat{s}_i$. 

\section{Technical Solutions for Soft Delivery}
\label{sec:aspect}
Because SoftCast skips nonlinear digital-based encoding and decoding operations corresponding to motion estimation, quantization, and entropy coding, it realizes a linear quality improvement associated with channel quality improvement. In particular, SoftCast has shown outstanding performance compared with the conventional digital-based delivery schemes when receivers are highly diverse and/or the channel condition of each receiver varies drastically.
Conversely, SoftCast's design is simplistic, so there remains much scope for improvement in adopting soft delivery in practical scenarios, including stable channel conditions, band-limited, and/or error-prone environments.
For this purpose, many studies have been conducted to improve the performance of soft delivery. 
The existing works on soft delivery schemes can be classified into seven types, as shown in Fig.~\ref{fig:list}: energy compaction, optimal scaling, bandwidth utilization, resilience to packet loss, overhead reduction, multi-user heterogeneity, and hardware implementation.

\begin{table*}[t]
\caption{Brief introduction to typical energy compaction techniques for soft delivery schemes}
    \centering
    \begin{tabular}{c c c c c}
    \hline
        Papers & Techniques & Features & Pros & Cons  \\ \hline
        \cite{bib:decorrelation1} & \begin{tabular}{c} 2D-DCT/\\2D-DWT \end{tabular} & \begin{tabular}{c} Take DCT/DWT operation for \\ each video frame \end{tabular} & Reduce spatial redundancy & No temporal filter \\ \hline
        \cite{bib:softcast1,bib:softcast3} & 3D-DCT  & \begin{tabular}{c} Take DCT operation \\ for each GoP \end{tabular} & \begin{tabular}{c} Reduce both spatial and \\ temporal redundancy \end{tabular} & Weak temporal filter \\ \hline
        \cite{bib:3ddwt-soft,bib:3ddwt-soft2,bib:3ddwt-soft3,bib:wavecast,bib:cactus,bib:adaptive_hybrid} & MCTF & \begin{tabular}{c} Take wavelet transform for \\ temporal filtering \end{tabular} & Further reduce temporal redundancy & \begin{tabular}{c} Computational cost for \\ temporal filtering \end{tabular} \\ \hline
        \cite{bib:2ddwt-soft,bib:AGDCast} & \begin{tabular}{c} Component \\ protection \end{tabular} & \begin{tabular}{c} Send lower frequency \\ coefficients as metadata \end{tabular} & \begin{tabular}{c} Distribute transmission power to \\ higher frequency coefficients \end{tabular} & \begin{tabular}{c} Communication overhead \\ Significant degradation \\ due to metadata error \end{tabular} \\ \hline
        \cite{bib:layeredsoftcast,bib:layeredcast1,bib:structure,bib:visual_hybrid, bib:G-Cast, bib:CG-Cast} & \begin{tabular}{c} Layered \\ operation \end{tabular} & \begin{tabular}{c} Divide video frames into BL and \\ ELs and send them in digital and \\ pseudo-analog ways, respectively \end{tabular} & \begin{tabular}{c} Provide baseline quality via the BL \\ while enhancing the quality via the ELs \end{tabular} & \begin{tabular}{c} ELs will be meaningless \\ if bit errors occur in BL \end{tabular} \\ \hline
        \cite{bib:dcast,bib:dcast2,bib:dcast3,bib:dcast4,bib:coset1,bib:magnitude_shift,bib:hyperspectrum,bib:layered_coset, bib:layered_coset2,bib:relay1, bib:relay2, bib:relay3} & \begin{tabular}{c} Coset \\ coding \end{tabular} & \begin{tabular}{c} Partition coefficients into several cosets \\ and transmit the coset residual codes \end{tabular} & \begin{tabular}{c} Bring lower entropy  \\ according to a coset step \end{tabular} & \begin{tabular}{c} Accuracy of coset step and \\ side information is crucial \\ for reconstruction \end{tabular}
        \\\hline
    \end{tabular}
    
    \label{tab:compaction}
\end{table*}

\subsection{Energy Compaction of Source Signals}
In soft delivery schemes via linear mapping~(from source signals to channel signals), the reconstruction quality greatly depends on the performance of the energy compaction technique for the source signals. 
Specifically, the existing study in~~\cite{bib:hybrid_theory} clarified that the performance of the soft delivery schemes degrades as the ratio of maximum energy to minimum energy of the source component
increases.
To yield better quality under both stable and unstable channel conditions, existing studies have adopted different energy compaction techniques listed in Table~\ref{tab:compaction} for the source signals.

Typical solutions are to adopt wavelet-based signal decorrelation methods. Specifically, some studies~\cite{bib:3ddwt-soft,bib:3ddwt-soft2,bib:3ddwt-soft3,bib:wavecast,bib:cactus,bib:adaptive_hybrid} have adopted a motion-compensated temporal filter (MCTF), which is a temporal wavelet transform method, to remove inter-frame redundancy by realizing motion compensation in soft delivery.
The MCTF recursively decomposes video frames into low- and high-frequency frames according to a predefined level.
For example, WaveCast~\cite{bib:wavecast} adopted a 3D-discrete wavelet transform~(DWT), i.e., the integration of 2D-DWT and MCTF, to remove temporal and spatial redundancy.
Although SoftCast exploits a full-frame 3D-DCT to remove the intra- and inter-frame redundancy for energy compaction, WaveCast can further improve the reconstruction quality by fully exploiting the inter-frame redundancy using motion compensation.
A detailed discussion on the effects of other decorrelation methods is presented in~\cite{bib:decorrelation1,bib:decorrelation2}.
\cite{bib:gop} also utilized inter-frame redundancy by designing an adaptive GOP size mechanism. It adaptively controlled the GoP size based on shot changes and the spatio-temporal characteristics of the video frames and took a full-frame 3D-DCT for energy compaction across the video frames in one GoP.

Another typical solution is to send large energy coefficients as metadata, and thus prevent the transmission of such coefficients using pseudo-analog modulation. 
\cite{bib:2ddwt-soft} designed Advanced SoftCast (ASoftCast) to send low-frequency coefficients as the metadata. ASoftCast decomposed the original images into frequency components using 2D-DWT; the frequency component was then divided into
two parts: the lowest frequency sub-band and other sub-bands. The wavelet coefficients in the lowest-frequency sub-band are processed by run-length coding; they are then channel-coded and digitally modulated for additional metadata transmissions. 
The optimized power allocation for the SoftCast scheme in~\cite{bib:AGDCast} selected and sent high-energy coefficients as the metadata to reduce the energy of the analog-modulated symbols. These results can assign a high transmission power to low-energy coefficients to improve the received quality. Here, determining the high-energy coefficients for each GoP is computationally complex owing to the use of an exhaustive search. 
To reduce the computational complexity, \cite{bib:sfc1} adopted a zigzag scan to select the side information.
Other studies in \cite{bib:layeredsoftcast,bib:layeredcast1,bib:structure,bib:visual_hybrid, bib:G-Cast, bib:CG-Cast} divided the video into BL and ELs, which were coded and sent in digital and pseudo-analog ways, respectively.  
For example, the base layer in gradient-based image SoftCast~(G-Cast)~\cite{bib:G-Cast} sent the DC and low-frequency coefficients of the image, while the enhancement layer extracted and sent an image gradient, which represents the edge portion of the image, using a gradient transform. The receiver then created a final estimation of the image via a gradient-based reconstruction~(GBR) procedure, utilizing both the image gradient at the enhancement layer and the low-frequency coefficients provided by the base layer.

Other solutions adopted a nonlinear encoder and decoder for source signals to decrease the ratio of the maximum to the minimum energy of the analog-modulated symbols.
The typical solution is to introduce coset coding~\cite{bib:coset_theory1,bib:coset_theory2}, which is a typical technique in distributed source coding, for soft delivery. Coset coding partitions the set of possible source values into several cosets and transmits the coset residual codes to the receiver. With the received coset codes and the predictor, the receiver can recover the source value by choosing the one in the coset closest to the predictor. 
DCast~\cite{bib:dcast,bib:dcast2,bib:dcast3,bib:dcast4} first introduced coset coding for the soft delivery of inter frames. 
The coset coding in DCast divides each frequency domain coefficient $s_i$ by a coset step $q$ and obtains the coset residual code $l_i$ as follows:
\begin{equation}
    l_i = s_i - \left \lfloor \frac{s_i}{q} + \frac{1}{2} \right \rfloor q,
\end{equation}
where $\left \lfloor \frac{s_i}{q} + \frac{1}{2} \right \rfloor$ represents the coset index.
At this time, the sender only needs to transmit the coset residual code for energy compaction. 
At the user side, with the received coset residual code $\hat{l}_i$ and the side information $\bar{s}_i$ (i.e., the predicted DCT coefficient obtained from the reference video frame), the receiver reconstructs the DCT coefficients by coset decoding. Given the coset residual code $l_i$, there are multiple possible reconstructions of $s_i$ that form a coset $C$:
\begin{equation}
C = \{ \hat{l}_i, \hat{l}_i \pm q, \hat{l}_i \pm 2q, \hat{l}_i \pm 3q, \ldots\}.
\end{equation}
DCast is then selected in coset $C$ that is nearest to the side information $\bar{s}_i$ as the reconstruction of the DCT coefficient:
\begin{equation}
\hat{s}_i = \argmin_{c \in C} |c - \bar{s}_i|.
\end{equation}
In this case, the value of each coset step $q$ is crucial for the coding performance of DCast. The value of $q$ is calculated by estimating the noise at the receiver end shown in~\cite{bib:dcast3,bib:dcast4}. However, the reconstruction quality of DCast also depends on the side information quality. If the side information $\bar{s}_i$ is rough, the receiver may make wrong decisions
with a smaller $q$. \cite{bib:SIRcast} introduced a side information refinement (SIR) algorithm~\cite{bib:sir} to refine the side information for the quality enhancement of DCast. 

The concept of coset coding has been widely applied in other studies on soft delivery for the same purpose. 
For example, \cite{bib:coset1,bib:magnitude_shift,bib:layered_coset,bib:layered_coset2,bib:hyperspectrum} utilized pseudo-coset coding for lower frequency components and sent the coset index using the digital framework. Here, the residuals in the lowest-frequency components and other frequency components are sent using pseudo-analog modulation. The main difference between coset coding and pseudo-coset coding is the sending of the coset index as additional metadata.
The layered coset coding and adaptive coset coding were applied to the soft delivery scheme in~\cite{bib:layered_coset} and \cite{bib:layered_coset2}, respectively.
LayerCast in~\cite{bib:layered_coset} introduced layered coset coding to simultaneously accommodate heterogeneous users with diverse SNRs and bandwidths. The layered coset coding used large to small coset steps to obtain coarse to fine layers from each chunk. The coarse layer, i.e., BL, is sufficient to reconstruct a low-quality DCT chunk for narrowband users, whereas each fine layer, i.e., EL, provides refinement information of the DCT chunk for wideband users.
\cite{bib:relay1, bib:relay2, bib:relay3} utilized the coset coding for cooperative soft delivery systems, i.e., a three-node relay network.
A sender broadcasts the DCT coefficients obtained from the video frames using pseudo-analog modulation to the relay node and the destination node. If the channel quality between the sender and the destination node is higher than a threshold, the destination node reconstructs the video frames from the soft-delivered DCT coefficients. If the channel condition is lower than the threshold, the relay node sends the coset residual code to the destination node, and then the destination node reconstructs the video frames using the received coset residual code and the side information obtained from the softly delivered DCT coefficients from the sender. 

\begin{table}[t]
\caption{Overview of power allocation techniques for soft delivery schemes}
    \centering
    \begin{tabular}{c c c}
    \hline
    Papers & Channel Consideration & \begin{tabular}{c} Quality Metric for \\ Optimization \end{tabular}  \\ \hline \cite{bib:softcast1} & AWGN & MSE \\ \hline
        \cite{bib:fading1}  & Fading & MSE \\ \hline
        \cite{bib:soft-fading,bib:carrier} & OFDM & MSE \\ \hline
        \cite{bib:mimo1,bib:waterfilling, bib:per-carrier} & MIMO & MSE \\ \hline
        \cite{bib:parcast,bib:ParCast+,bib:ecast} & MIMO-OFDM & MSE \\ \hline
        \cite{bib:impulse} & Impulse noise & MSE \\ \hline
        \cite{bib:noma2,bib:noma1} & NOMA & MSE \\ \hline
        \cite{bib:underwater} & Underwater acoustic networks & MSE \\ \hline
        \cite{bib:UAV} & UAV-enabled networks & MSE \\ \hline
        \cite{bib:mmwave} & mmWave lens MIMO & MSE \\ \hline
        \cite{bib:ssim-softcast} & AWGN & SSIM \\ \hline
        \cite{bib:foveacast} & AWGN and MIMO & FWD \\ \hline 
        \cite{bib:saliency} & AWGN & EQMSE \\ \hline
        \cite{bib:scast} & AWGN & \begin{tabular}{c} Foreground and \\ background distortions \end{tabular} \\\hline
    \end{tabular}
    
    \label{tab:power}
\end{table}

\subsection{Channel-Aware and Perception-Aware Power Allocation}
As mentioned in Section~\ref{sec:power}, the power allocation in SoftCast minimizes the MSE between the original and reconstructed video signals over additive white Gaussian noise~(AWGN) channels. There are several drawbacks toward adopting SoftCast in practical scenarios: 1)  practical wireless channels have more complex characteristics, e.g., fading caused by multipath and impulse noise, than the AWGN channels, and 2) MSE is not an effective index for describing the perceptual fidelity of images/videos.

For the first drawback, the existing studies re-designed the power allocation for practical wireless channels, including fading~\cite{bib:fading1} and frequency-selective fading, i.e., orthogonal frequency-division multiplexing (OFDM)
~\cite{bib:soft-fading,bib:carrier}, impulse noise~\cite{bib:impulse}, multiple-input and multiple-output ~(MIMO)~\cite{bib:mimo1,bib:waterfilling, bib:per-carrier}, and MIMO-OFDM channels~\cite{bib:parcast,bib:ParCast+,bib:ecast}.  
In~\cite{bib:fading1}, the authors designed an optimal power allocation for fading channels. In fading channels, a fading effect, i.e., multiplicative noise, will degrade the reconstruction quality. Although SoftCast assumes that multiplicative noise can be canceled with exact channel estimation at the receiver end, no algorithm can guarantee an error-free channel estimation. In addition to the power allocation design, the authors analyzed the effect of the channel estimation error on the reconstruction quality at the receiver end.

\begin{figure}[t]
  \begin{minipage}{0.5\textwidth}
  \centering
   \subfloat[2D $8 \times 8$ block DCT coefficient energy]{\includegraphics[width=0.9\linewidth]{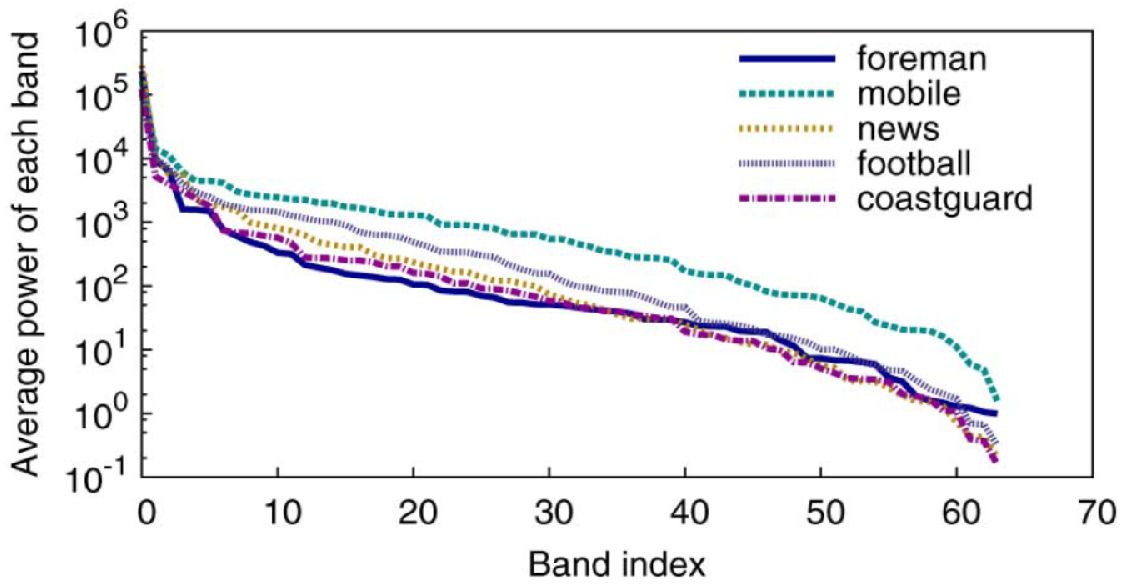}} \\ 
   \subfloat[$3 \times 3$ MIMO-OFDM subchannel gains]{\includegraphics[width=0.9\linewidth]{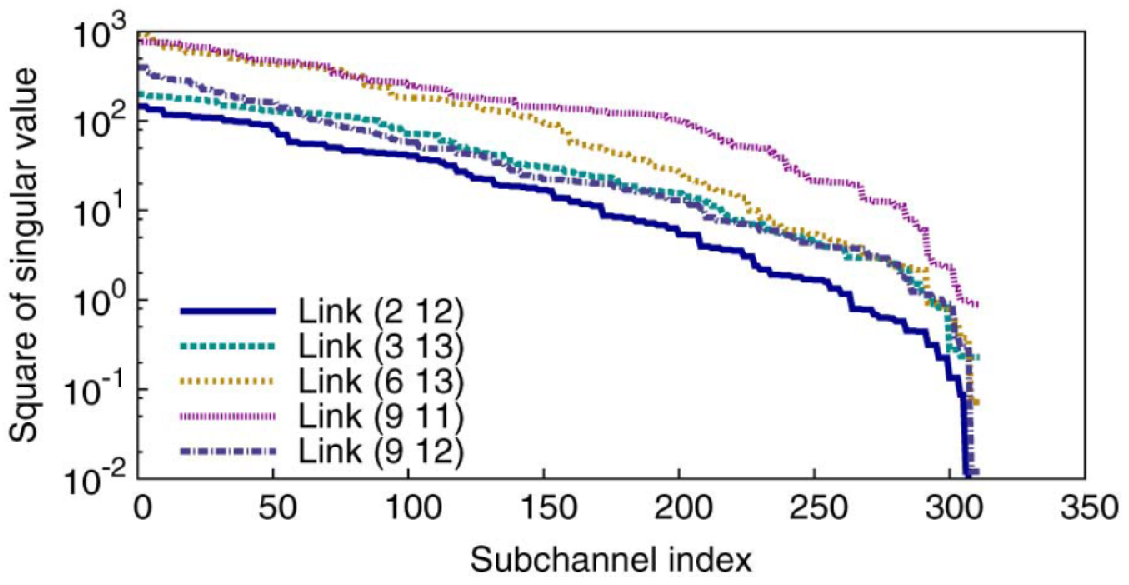}}
\end{minipage} \hfill
\caption{Characteristics of the source and channel components~\cite{bib:ParCast+}.}
  \label{fig:similarity}
\end{figure}

For frequency-selective fading channels, such as OFDM and MIMO-OFDM channels, the key issue is how to match the analog-modulated symbols to the independent subcarriers/subchannels for high-quality image/video reconstruction.
In~\cite{bib:parcast,bib:ParCast+}, they observed similarities between the source and channel characteristics and exploited the similarities for subcarrier/subchannel matching. 
Specifically, Figs.~\ref{fig:similarity}~(a) and (b) show a similar spread between the highest- and lowest-energy source and channel components, respectively. It seems natural to match both sides so that high-energy DCT components are transmitted on high-gain subchannels to prevent them from acting against each other. 
ParCast~\cite{bib:parcast} and the extended version of ParCast$+$~\cite{bib:ParCast+} assigned the more important DCT coefficients to higher gain channel components and allocated power weights for each DCT coefficient with joint consideration of the source and channel for video unicast systems.
ECast in~\cite{bib:ecast} extended the source and channel matching and power allocation for video multicast systems. For multicast systems, it is necessary to deal with the large overhead of channel feedback from multiple receivers. In ECast, multiple users simultaneously send tone signals for the channel feedback, and the sender receives the superposition of multiple tone signals. 
Although the sender cannot distinguish each of the channel gains, the weighted harmonic means of channel gains can be obtained from the superposed tone signals; thus, ECast utilizes the channel gain for the source and channel matching and power allocation.

Other studies solved power allocation problems in modern wireless systems, including non-orthogonal multiple access~(NOMA)~\cite{bib:noma2,bib:noma1}, underwater acoustic OFDM~\cite{bib:underwater}, unmanned aerial vehicle~(UAV)-enabled~\cite{bib:UAV}, and mmWave lens MIMO systems~\cite{bib:mmwave}. 
For example, in NOMA systems, source signals are coded into BL and ELs and then transmitted simultaneously through superposition coding~(SC). With successive interference cancellation~(SIC), near users with strong channel gains can decode both BL and EL signals, whereas far users with weak channel gains may only decode BL signals. In the existing studies, both BL and ELs are analog-coded in~\cite{bib:noma2}, whereas BL and ELs are digital- and analog-coded, respectively, in~\cite{bib:noma1}. They solved the power allocation across the BL and ELs to minimize the distortion for all receivers with heterogeneous channel conditions. 
In underwater acoustic OFDM~\cite{bib:underwater} and mmWave lens MIMO systems~\cite{bib:mmwave}, the error behavior differed substantially across channel components, and the channel characteristics showed a similar tendency, as depicted in Fig.~\ref{fig:similarity}~(b). They solved the source and channel matching and power allocation problems, which are also discussed in frequency-selective fading channels , to minimize the distortion at the receiver end.

For the second drawback, some studies~\cite{bib:ssim-softcast,bib:foveacast,bib:saliency,bib:scast} also redesigned the power allocation with perceptual considerations, including structural similarity~(SSIM)~\cite{bib:ssim-softcast}, foveation~\cite{bib:foveacast}, and saliency~\cite{bib:saliency}. 
In these studies, determining the perception-aware weights for each source component is challenging. Specifically, in SoftCast, the scaling factor for each coefficient is obtained from its power information to minimize the MSE.
\begin{equation}
    g_i \varpropto \lambda_i^{-1/4}.
\end{equation}
These studies considered the perception-aware weight for the $i$th coefficient $w_i$ in the scaling factor to minimize the perceptual distortion as follows:
\begin{equation}
    g_i \varpropto w_i^{1/4} \lambda_i^{-1/4}.
\end{equation}
For this purpose, \cite{bib:ssim-softcast} demonstrated the relationship between the MSE in the DCT coefficients and the SSIM distortion to obtain the weight for the $i$th DCT coefficients of all the chunks $w_i$. 
They found that the weight for the high-frequency coefficients was larger than that for the low-frequency
coefficients, which was consistent with the characteristics of the human visual system~(HVS).
FoveaCast in~\cite{bib:foveacast} introduced the foveation-based HVS~\cite{bib:foveation} and the corresponding HVS-based visual perceptual quality metric, called foveated weighted distortion~(FWD), for the optimization objective.
For a given foveation point $(f_x, f_y)$ in the pixel and frequency domains, the error sensitivity for each pixel/frequency coefficient at location $(x, y)$ can be defined in the foveation-based HVS.
FoveaCast regarded the error sensitivity in the DWT domains as the weight $w_i$ and performed foveation-aware power allocation.
In~\cite{bib:saliency}, visual saliency maps were introduced for the perception-aware power allocation. 
Saliency maps represent the attended regions in an image when a user watches the image owing to the visual attention mechanism of the human brain.
In this case, the weight for the $i$th pixel $w_i$ is based on the normalized visual saliency defined from any arbitrary visual saliency model, such as the Itti--Koch--Niebur model~\cite{bib:salience_model}. 
Based on the weight, it allocates considerable transmission power to salient regions to minimize the eye-tracking weighted MSE~(EQMSE).

\subsection{Bandwidth Utilization}
The source bandwidth of soft delivery schemes depends on the number of transmitted analog-modulated symbols every second, i.e., baud rate. 
In the aforementioned designs, the source bandwidth is mainly considered sufficient to send all the transmitted non-zero analog-modulated symbols over the wireless medium. However, when the channel bandwidth is lower than the source bandwidth, some analog-modulated symbols are discarded at the receiver side. Here, the loss of the important coefficients, i.e., the low-frequency coefficients, may have a significant impact on the reconstruction quality. Specifically, the expected distortions in soft delivery schemes for single and multiple contents owing to the bandwidth constraint under the transmission power constraint are discussed in~~\cite{bib:tradeoff} and \cite{bib:MUCast,bib:MUCast2}, respectively. 
To meet the bandwidth constraint, the typical method is to selectively discard the chunks in higher frequency components to fill the bandwidth~\cite{bib:softcast1,bib:unequalblock}. When the sender discards some chunks, the receiver regards all the coefficients in the discarded chunks as zeros. Because it needs to send the locations of the discarded chunks to the receiver, SoftCast sends the location information as a bitmap. 
Although SoftCast assumes equal-size chunks across low- to high-frequency components, \cite{bib:unequalblock} adopted smaller chunk sizes in high-frequency components to realize a fine-grained control to meet the bandwidth limitation. 
Another study in~\cite{bib:SKmapping} used bandwidth-reducing Shannon--Kotelnikov~(SK) mappings to increase the number of chunks transmitted over bandwidth-constrained channels. The SK mappings are typical $N$:1 bandwidth-reducing or 1:$M$ bandwidth-expanding
non-linear mappings. In this study, 2:1 SK mappings are used to encode several pairs of chunks with less energy to send more chunks with medium energy within the channel bandwidth. 

Other studies~\cite{bib:cs2,bib:satelite-cs,bib:visual-cs, bib:cs1,bib:cs4,bib:sparsecast,bib:cs6,bib:adaptcast} introduced compressive sensing~(CS) techniques~\cite{bib:CS_theory_orig, bib:CS_theory} for soft delivery over bandwidth-constrained wireless channels. 
Notably, CS is a sampling paradigm that allows the simultaneous measurement and compression of signals that are sparse or compressible in some domains. In general, recovering source signals from compressed signals is impossible because the system is underdetermined. However, if the source signals are sufficiently sparse in some domains, the CS theory indicates that the source signals can be reconstructed from the compressed signals by solving the $\ell_1$ minimization problem.
The advantage of CS-based soft delivery is the recovery of chunks in high-frequency coefficients using CS-based signal reconstruction algorithms, such as approximate message passing~(AMP) and iterative thresholding, even though the chunks are discarded at the sender’s end. 
For high-quality reconstruction, adaptive rate control and reconstruction algorithms are mainly adopted for CS-based soft delivery.
For instance, \cite{bib:visual-cs} adaptively controlled the compression rate based on visual attention, i.e., both the texture complexity and visual saliency, to satisfy the bandwidth constraint while maintaining better perceptual quality. \cite{bib:cs6} adaptively selected reliable columns from the measurement matrix and compressed source signals using the selected columns. 
In view of the reconstruction algorithm, \cite{bib:cs1} designed an adaptive transform for noisy measurement signals to obtain sparser transform coefficients for clean reconstruction. 
\cite{bib:cs4} and \cite{bib:sparsecast} designed grouping methods for measurement signals to utilize the similarity between video frames for the reconstruction.

\begin{figure}[t]
  \begin{center}
   \includegraphics[width=\hsize]{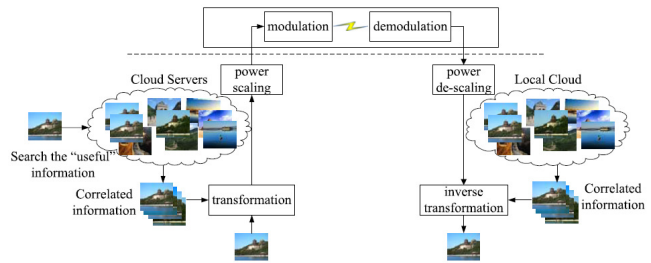}
   \caption{Data-assisted soft delivery schemes~\cite{bib:kmvcast3}. They utilize the stored images and videos on the cloud to limit the number of transmission symbols.}
   \label{fig:data-assisted}
  \end{center}
\end{figure}

Other studies utilized stored images/videos on the cloud to reduce the bandwidth requirement in soft delivery. 
Specifically, data-assisted communication of mobile image (DAC-Mobi)~\cite{bib:dacmobi}, data-assisted cloud radio access network (DaC-RAN)~\cite{bib:dacran}, and knowledge-enhanced mobile video broadcasting (KMV-Cast) schemes~\cite{bib:kmvcast3, bib:kmvcast2, bib:kmv-cast1}, which are referred to as data-assisted soft delivery schemes, have been proposed for high-quality image/video transmission. 
Fig.~\ref{fig:data-assisted} presents an overview of the data-assisted soft delivery schemes. 
The main contributions of the data-assisted soft delivery schemes are 1) a sender sends a limited number of analog-modulated symbols and 2) the receiver reconstructs images/videos using correlated images, i.e., side information, obtained from a cloud. 

In DaC-Mobi~\cite{bib:dacmobi}, successive coset encoders were introduced to divide the DCT coefficients into three layers of bit planes: most significant bits (MSBs) in low-frequency coefficients, MSBs in other frequency coefficients and middle bits, and least significant bits (LSBs).
Here, MSBs in low-frequency coefficients and LSBs were transmitted to the receiver in digital and pseudo-analog manners, respectively, whereas MSBs in other frequency coefficients and middle bits were discarded. 
Based on the received MSB in the low-frequency coefficients, the receiver reconstructs a down-sampled image to retrieve correlated images in the cloud. The retrieved correlated images were used as side information to resolve ambiguity due to discarded bits and reconstruct the entire image.
DaC-RAN~\cite{bib:dacran} and the extended version of KMV-Cast~\cite{bib:kmvcast3, bib:kmvcast2, bib:kmv-cast1} adopted Bayesian reconstruction algorithms that utilize correlated images/videos.
in the cloud as prior information to reduce the required bandwidth for soft delivery. 
The main difference between the DaC-RAN and KMV-Cast schemes is that the former assumes that the same images/videos exist in the cloud, whereas the latter does not require that the same images/videos exist at the receiver end by designing prior knowledge broadcasting in a digital manner.

The aforementioned studies considered the channel bandwidth to be lower than the source bandwidth. If the channel bandwidth is greater than the source bandwidth, the soft delivery schemes become less efficient. In this case, the soft delivery schemes utilize the extra bandwidth by retransmission. 
\cite{bib:analog_code} and \cite{bib:analog-coded} designed an analog channel coding to use an extra channel bandwidth for quality enhancement. For example, \cite{bib:analog-coded} proposed a chaotic function-based analog encoding~\cite{bib:chaotic} for soft delivery. Although the existing chaotic function-based analog coding is designed for uniformly distributed sources, the analog coding for Gaussian distributed sources significantly amplifies source signals and thus consumes unnecessary transmission power. They designed a chaotic map function for Gaussian distributed source signals to prevent power increments compared to the input power.
Mcast in~\cite{bib:channel} also utilized extra bandwidth for quality improvement. As mentioned earlier, the sender can send the source data multiple times if an extra bandwidth is available. In this case, the utilization of extra time slots for quality improvement is a key issue. To overcome this issue, MCast optimized the assignment of the chunks of the DCT coefficients to available channels in multiple time slots to fully exploit the time and frequency diversities.

In contrast to the aforementioned studies, \cite{bib:progressive,bib:analog_progressive} dealt with bandwidth variations. When the available bandwidth is less than the expected bandwidth at the sender’s end, some important chunks will not have the opportunity to be transmitted before the playback deadline. They grouped several chunks into a tile and sent the tile with a large variance and high priority to dispatch important coefficients before the playback deadline.

\begin{figure}[t]
  \begin{center}
   \includegraphics[width=\hsize]{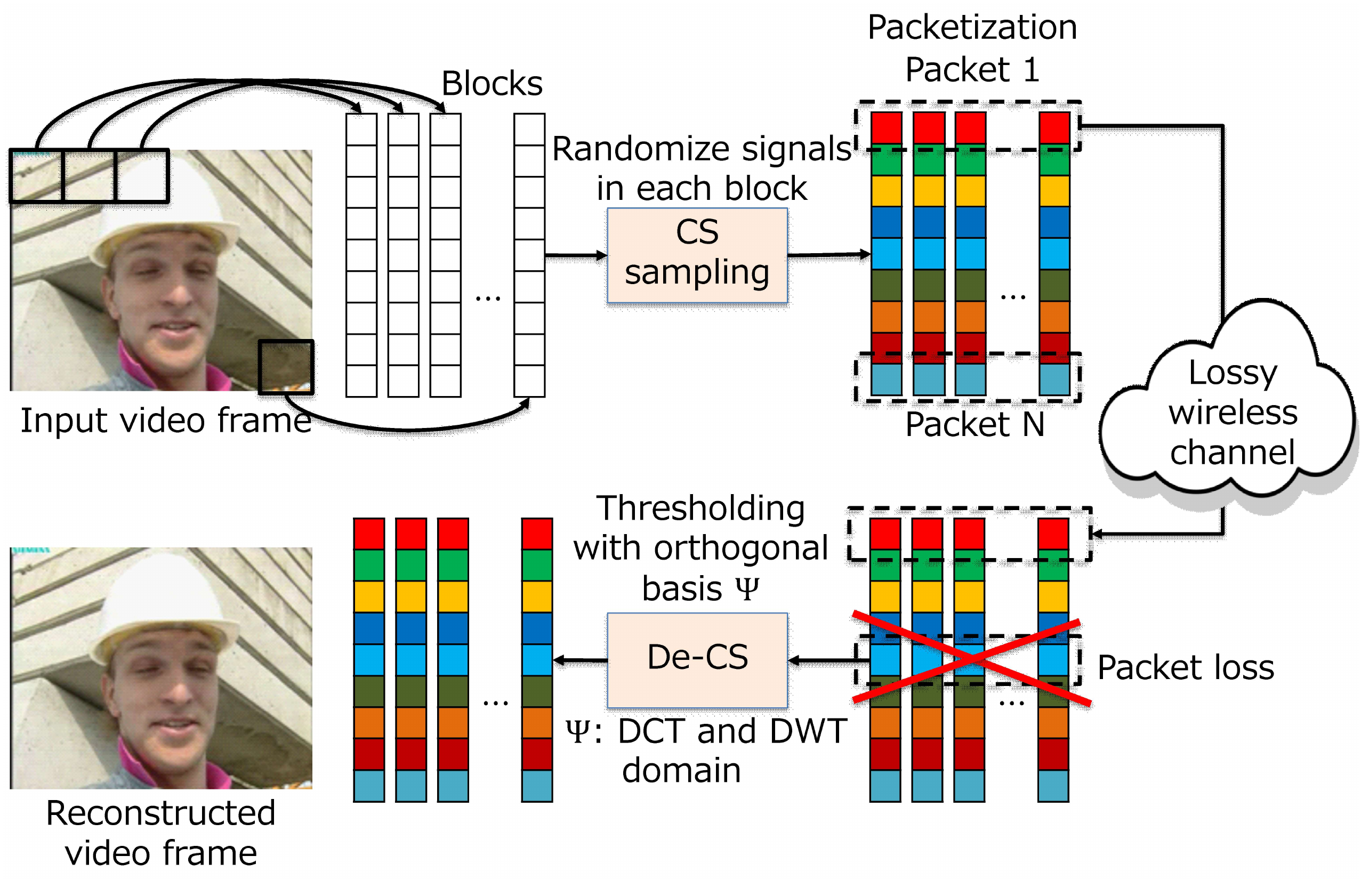}
   \caption{Brief description of DCS-Cast~\cite{bib:CS_analog1}. DCS-Cast introduces block-wise CS sampling and reconstruction for packet loss resilience.}
   \label{fig:bcs}
  \end{center}
\end{figure}

\subsection{Packet Loss Resilience}
Even though the channel bandwidth is sufficient to send all the non-zero analog-modulated symbols, some analog-modulated symbols are discarded at the receiver side owing to loss-prone wireless channels. Specifically, the packet loss owing to strong fading and interference may have a significant impact on the reconstruction quality if important chunks and coefficients are lost. 
SoftCast used the WHT to redistribute the energy of the source signals across whole packets for resilience against packet loss. 
However, each packet still contains a large amount of energy, and thus, degradation owing to packet losses remains considerable.

\begin{figure*}[t]
  \centering
   \subfloat[Sender-side Overhead Reduction]{\includegraphics[height=0.20\linewidth]{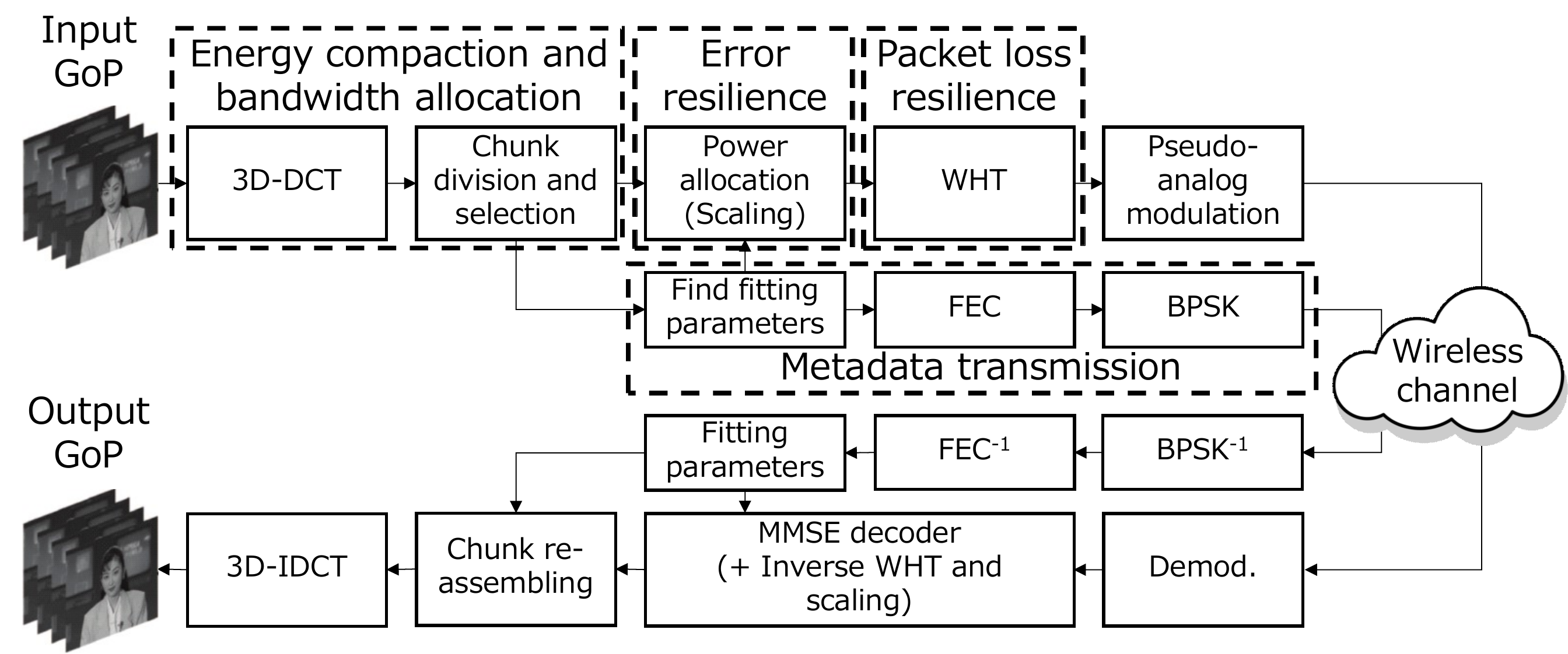}} 
   \hfill
   \subfloat[Receiver-side Overhead Reduction]{\includegraphics[height=0.20\linewidth]{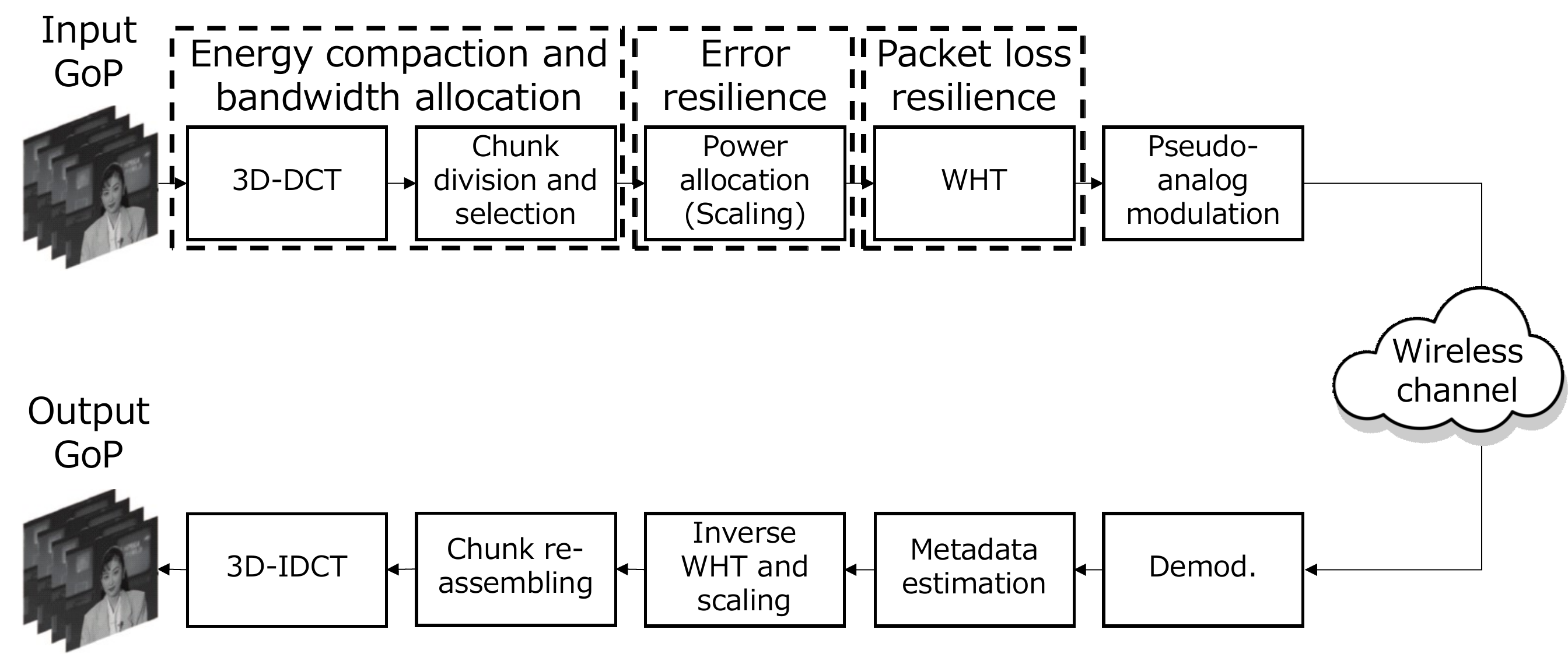}}
 \caption[]{Block diagram of the sender-side and receiver-side overhead reduction methods. (a) Sender utilizes fitting functions to obtain power information with fewer parameters. (b) Receiver estimates the power information only from the received symbols.}
   \label{fig:overhead}
\end{figure*}

To maintain better reconstruction quality in error-prone wireless channels, some related studies~\cite{bib:CS_analog1,bib:cs3,bib:cs5} have introduced CS techniques, i.e., block-wise CS~\cite{bib:BCS_SPL}, for packet loss resilience. The CS technique is suitable for wireless transmission with random packet loss owing to its random measurement. Random measurement considers all packets as of equal importance. In contrast to typical CS techniques, block-wise CS can reduce the storage and computational costs of the reconstruction. A pioneering work on packet loss resilience is the distributed compressed sensing-based multicast scheme (DCS-cast)~\cite{bib:CS_analog1}. Figure~\ref{fig:bcs} shows an overview of the sender and receiver operations in the DCS-Cast.
In the DCS-Cast, each image is first divided into blocks and the coefficients in each block are randomized using the same measurement matrix across the blocks. One coefficient in every block is packetized to normalize the importance across packets. Even though some packets may be lost over loss-prone wireless channels, the receiver obtains noisy pixel values using the same measurement matrix at the sender and reconstructs the lost pixel values using the CS reconstruction algorithm in the DCT/DWT domains. Because the lost pixel values can be recovered from the reconstruction algorithm, DCS-Cast maintains high image/video quality in the loss-prone channels. 
To further improve the reconstruction quality, multi-scale~\cite{bib:cs3} and adaptive~\cite{bib:cs5} block-wise CS algorithms have been adopted for soft delivery.  
The multi-scale block-wise CS algorithm~\cite{bib:cs3} decomposes each video frame into a multi-level 2D-DWT and then optimizes the sampling rate for each DWT level according to its importance. 
However, the adaptive block-wise CS algorithm~\cite{bib:cs5} divides several video frames into one reference frame and subsequent non-reference frames and adaptively determines whether direct or predictive sampling should be used for each block in a non-reference frame. Direct sampling randomizes the signals in the block, whereas predictive sampling calculates the residuals between the blocks in the reference and non-reference frames and randomizes residuals to utilize the inter-frame similarity for the reconstruction.

\subsection{Overhead Reduction}
In soft delivery schemes without chunk division, a sender needs to let the receiver know the power information of all the DCT coefficients to demodulate the signals.
For the receiver to carry out the MMSE filtering in Eq.~(\ref{eq:mmse}), the sender needs to transmit ${\lambda}_{i}$ of all coefficients without errors as metadata, which may constitute a large overhead. For example, when the sender transmits eight video frames with a resolution of $352\times288$, the sender needs to transmit metadata for all DCT coefficients, i.e., $352\times 288\times 8 = 811{,}008$ variables in total, to the receiver.
This overhead may induce performance degradation owing to the rate and power losses in the transmission of analog-modulated symbols.

To reduce the overhead, SoftCast divides the DCT coefficients into chunks and carries out chunk-wise power allocation using an MMSE filter.
However, overheads are still high, and chunk division causes performance degradation due to improper power allocation.

To achieve better quality under a low overhead requirement, the related studies can be classified into two types, as shown in Figs~\ref{fig:overhead}~(a) and (b): 1) sender-side overhead reduction and 2) receiver-side overhead reduction. Studies on the sender-side overhead reduction~\cite{bib:overhead1,bib:overhead2,bib:l-shape,bib:fuji_GMRF} designed fitting functions to obtain the power information with fewer parameters. In this case, the sender and receiver share the same fitting function in advance and send the parameters as metadata for overhead reduction. Specifically, \cite{bib:overhead1} designed a fitting function with four parameters for each chunk, and \cite{bib:overhead2} designed a log-linear function with two parameters for each chunk. Another study in~\cite{bib:l-shape} found that equal-size chunk division was not suitable for chunk-wise fitting, and thus, an adaptive chunk division, i.e., L-shaped chunk division, was designed for an accurate fitting. In addition, \cite{bib:fuji_GMRF} exploited a Lorentzian fitting function with seven parameters based on a Gaussian Markov random field for each GoP.

Studies on receiver-side overhead reduction~\cite{bib:analogcast,bib:blind} estimate the power information only from the received signals without any additional computational cost at the sender side. 
\cite{bib:analogcast} is a pioneer work to estimate the power information from the received signals, and blind data detection (BDD)~\cite{bib:blind} was proposed to decode the received analog-modulated symbols without the power information at the receiver. Specifically, BDD uses a zero-forcing estimator and the sign of the received signals to approximate the source signals. 

We note that both types of overhead reduction cause quality degradation owing to estimation errors. In~\cite{bib:overhead2}, the effect of modeling accuracy on the reconstruction quality in soft delivery was analyzed.

\subsection{Antenna and Resolution Heterogeneity}
In SoftCast and later soft delivery schemes, the channel heterogeneity of the receivers can be solved by pseudo-analog modulation with linear coding and decoding operations. However, the heterogeneity of other aspects still impairs each user's experience. 
To mitigate the impairment caused by the heterogeneity across multiple receivers, some studies extended soft delivery to deal with the heterogeneity of resolution~\cite{bib:hetero_resolution} and receiver antennas~\cite{bib:hetero1,bib:airscale}.
In~\cite{bib:hetero_resolution}, the authors dealt with the resolution heterogeneity of receivers in video broadcast systems. Specifically, they designed a novel spatial decomposition method based on linear projections to provide differentiated resolution demands. After decomposition, the input videos can be divided into BL and multiple ELs. A base layer guaranteed the base resolution of the video content, while the ELs progressively enlarged the video resolution.

For the antenna heterogeneity in MIMO systems, the CS algorithm was adopted in~\cite{bib:hetero1}, whereas AirScale in~\cite{bib:airscale} designed a combination of multiple similar description~(MSD) coding and multiplexed space--time block coding (M-STBC).
In~\cite{bib:hetero1}, the transformed coefficients were randomized using a measurement matrix before transmission. The randomized symbols are transmitted from multiple antennas; the receiver with an insufficient number of antennas may receive a limited number of symbols via wireless channels. Even though the receiver does not receive some symbols owing to the number of antennas, the CS-based signal reconstruction algorithm recovers the lost symbols to decrease the degradation owing to the antenna heterogeneity.
In AirScale~\cite{bib:airscale}, the MSD coding produces highly similar descriptions from video frames in one GoP to provide an additional feature that any linear combination of the descriptions can be used to reconstruct the source signal. M-STBC is then adopted for the descriptions to achieve either multiplexing gain or diversity gain for multiple receivers with a diverse number of antennas. When there is an insufficient number of receiver antennas, M-STBC puts similar symbols to corresponding space--time positions to enhance the reconstruction quality using the linear combinations of the received descriptions. When there is a sufficient number of receiver antennas, different symbols are used to achieve multiplexing gain.

 \begin{figure*}[t]
\centering
  \begin{minipage}{\textwidth}
  \centering
   \subfloat[Free viewpoint video]{\includegraphics[height=3.cm]{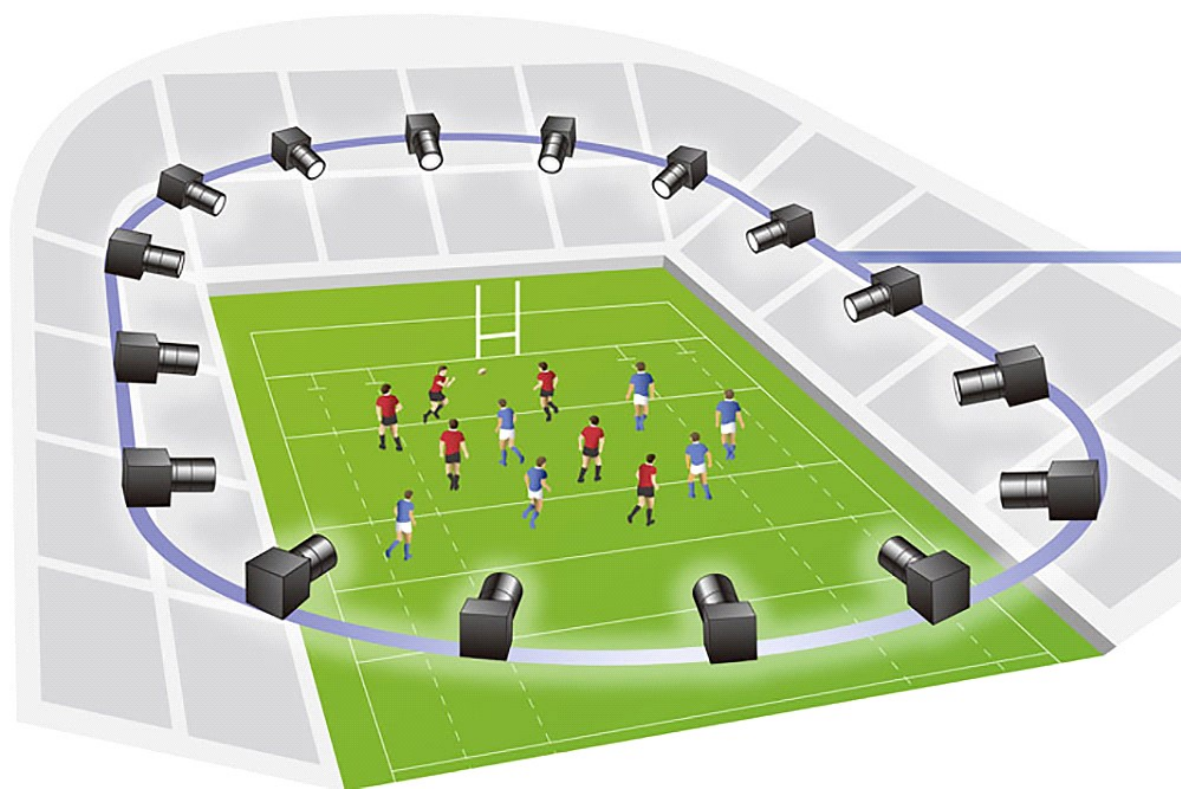}} 
   \hfill
   \subfloat[360-degree video]{\includegraphics[height=3.cm]{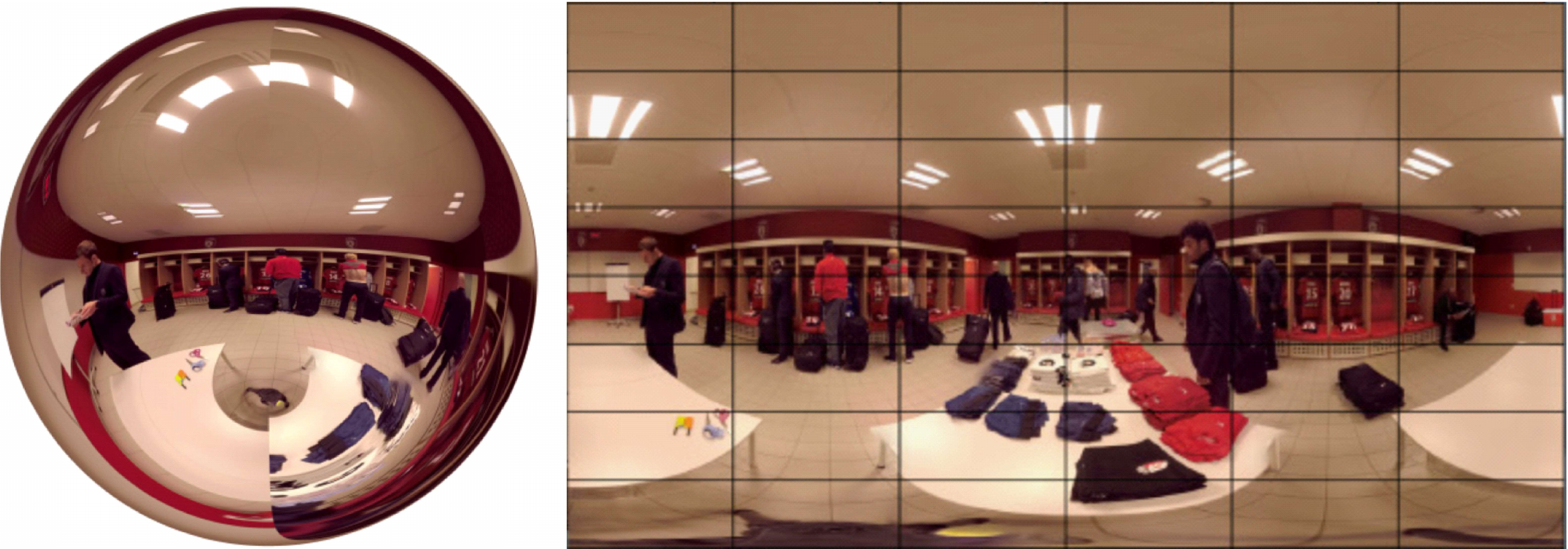}}
   \hfill
   \subfloat[Point cloud]{\includegraphics[height=3.cm]{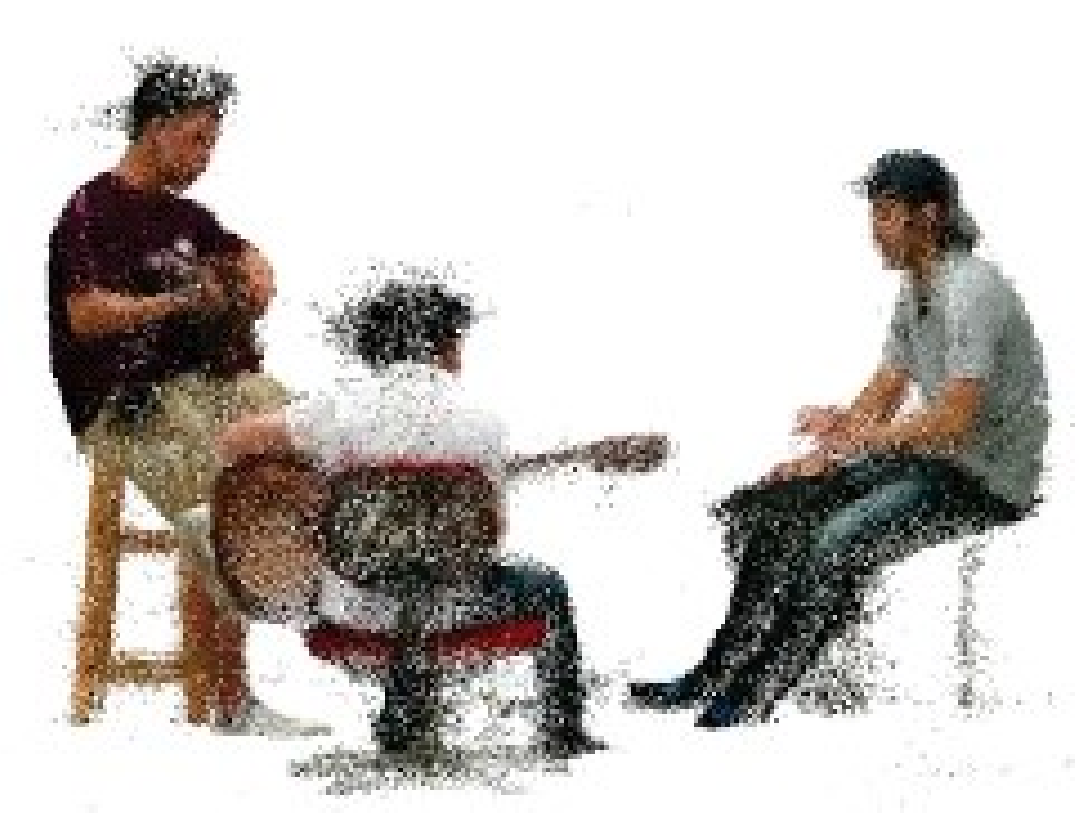}}
\end{minipage} \hfill
\caption{Typical immersive contents. (a) free-viewpoint video. (b) 360-degree video. (c) point cloud.}
  \label{fig:immersive}
\end{figure*}

\begin{table*}[t]
\caption{Typical immersive content and its features}
    \centering
    \begin{tabular}{c c c c c}
    \hline
        Papers & Content & Acquisition & Display & Key issues  \\ \hline
        \cite{bib:free,bib:FreeCast,bib:freecast2,bib:Luo2019,bib:depth,bib:Nu_MVV1,bib:Nu_MVV2} & Free viewpoint video & \begin{tabular}{c} Large number of \\ closely spaced RGB \\ and IR camera arrays \end{tabular} & \begin{tabular}{c} Synthesize virtual cameras \\ using rendering and freely \\ switch the viewing camera \end{tabular}  & \begin{tabular}{c} Resource allocation for \\ each RGB and IR camera \\ to maximize the viewing quality \end{tabular}  \\ \hline
        \cite{bib:360-fuji,bib:omnicast,bib:360cast,bib:360cast+} & 360-degree video  & 360-degree camera & \begin{tabular}{c} Playback viewport through \\ VR headset \end{tabular} & \begin{tabular}{c} Predict future viewport and \\ allocate resource to the viewport \\ for quality maximization \end{tabular} \\ \hline
        \cite{bib:holocast,bib:HoloCastGivens} & Point cloud & Laser scanner & \begin{tabular}{c} Playback 3D points through \\ AR and MR headsets and \\ holographic display \end{tabular}  & \begin{tabular}{c} Compress and send numerous \\ and irregular structure of \\ 3D points  \end{tabular} 
        \\\hline
    \end{tabular}
    
    \label{tab:immersive}
\end{table*}

\subsection{Implementation}
The aforementioned studies mainly discussed performance improvements in theoretical analyses and simulations. Some studies implemented a soft delivery scheme on software-defined radio~\cite{bib:Tang2020_1} platform and field-programmable gate array (FPGA)~\cite{bib:implementation1,bib:implementation2,bib:implementation5} to empirically demonstrate the benefits of soft delivery in practical wireless channels. 
In~\cite{bib:Tang2020_1}, the authors used Ettus Research X310 and GNU Radio for implementation and evaluated the visual quality of soft delivery in both simulations and experiments. Conversely, in~\cite{bib:implementation1,bib:implementation2,bib:implementation5}, they exploited the Xilinx Virtex7 FPGA for implementation and tested the reconstruction quality as a function of wireless channel SNRs. 
Other studies~\cite{bib:implementation3,bib:implementation4} implemented soft delivery on the prototypes of multi-user MIMO(MU-MIMO) and long-term evolution (LTE) systems, respectively. 
For example, in~\cite{bib:implementation3}, SoftCast is implemented on BUSH, which is a large-scale MU-MIMO prototype that performs scalable beam user selection with hybrid beamforming for phased-array antennas in legacy WLANs. They performed experiments to evaluate the video quality in terms of PSNR and SSIM over a lossy MU-MIMO channel.

\section{Extension for Immersive Contents}
\label{sec:immersive}
SoftCast and other soft delivery schemes mentioned in the previous sections were designed for conventional images and video signals. In modern wireless and mobile communication scenarios, the streaming of immersive content will be a key application for reconstructing 3D perceptual scenes that provide full parallax and depth information for human eyes. 
The immersive contents can be applied to various applications, such as three to six degrees-of-freedom (6-DoF) entertainment, remote device operation, medical imaging, vehicular perception, VR/AR/MR, and simulated training.
Figures~\ref{fig:immersive}~(a) through (c) show the typical immersive contents of free-viewpoint video~\cite{bib:ftv1, bib:ftv3, bib:ftv4}, 360-degree video, and point cloud~\cite{bib:mpeg}. 
Even in immersive contents, the video frames are compressed in a digital manner, and the compressed bitstream is then channel-coded and modulated in sequence. 
This means that cliff and leveling effects still occur in the streaming of the immersive contents owing to the variation in the channel conditions.
To prevent cliff, leveling, and staircase effects, some studies have extended soft delivery schemes toward immersive content for future wireless multimedia services.

\subsection{Free Viewpoint Video}
Free-viewpoint videos enable us  to observe a 3D scene from freely switchable angles/viewpoints. Fig.~\ref{fig:immersive}~(a) shows an example of a free-viewpoint video wherein numerous closely spaced RGB and infrared (IR) camera arrays are deployed to capture the texture and depth frames of a 3D scene, such as a football game. 
Even though the number of deployed cameras in the field is limited owing to physical constraints, the receiver can synthesize intermediate virtual viewpoints using rendering techniques, e.g., depth image-based
rendering~\cite{bib:dibr,bib:dibr1} to obtain numerous switchable viewpoints.
To synthesize intermediate virtual viewpoints using the rendering technique, the sender encodes and transmits the texture and depth frames of two or more adjacent viewpoints, the format of which is known as multi-view plus depth~(MVD)~\cite{bib:mvd}. 

For conventional MVD video streaming over wireless links, digital video compression for MVD video frames, e.g., MVC+D~\cite{bib:mvc_d} or 3D-AVC~\cite{bib:3davc}, fully utilizes the redundancy between the cameras and texture-depth for compression.
In this case, the streaming schemes need to solve view synthesis problems in addition to cliff and leveling effects to yield better video quality even in the synthesized virtual viewpoints.  
Specifically, the video quality of the virtual viewpoint is determined by the distortion of each texture and depth frame. In digital-based MVD schemes, the distortion depends on the bit and power assignments for each texture and depth frame. It is often cumbersome to achieve the best quality at a target virtual viewpoint using parameter optimization owing to the combinatorial problem with nonlinear quantization.

Some studies~\cite{bib:free,bib:FreeCast,bib:freecast2,bib:Luo2019,bib:depth} designed a soft delivery scheme for a free-viewpoint video. 
Specifically, FreeCast in~\cite{bib:free,bib:FreeCast,bib:freecast2} is the first scheme for a free-viewpoint video. Because MVD video frames have redundancy of cameras and texture-depth, FreeCast jointly transforms texture and depth frames using 5D-DCT to exploit inter-view and texture-depth correlations for energy compaction. In addition, FreeCast can simplify the optimization problems of view synthesis by reformulating it into a simple power assignment problem. This is because bit allocation, i.e., quantization, is not required in FreeCast. They found that the power assignment problem for the texture and depth frames can be solved using a quadratic function to yield the best quality at the desired virtual viewpoint. Another study ~\cite{bib:Luo2019} focuses on the view synthesis problems under the 3D-DCT operations for each camera's texture and depth frames and designs the power assignment method for solving the problem. The main difference between the power assignment methods of FreeCast and \cite{bib:Luo2019} is that FreeCast pre-assigns transmission power before the decorrelation, whereas \cite{bib:Luo2019} controls the transmission power after the decorrelation. 
Other studies in~\cite{bib:Nu_MVV1,bib:Nu_MVV2} adopt soft delivery to collect video frames from multiple wireless and mobile cameras via wireless channels, even though the aforementioned studies assume that the video frames at the server are error-free.

\subsection{360-Degree Video}
Notably, 360-degree video contents build a synthetic virtual environment to mimic the real world with which the users interact. Each user can watch 360-degree videos through a traditional computer-supported VR headset or an all-in-one headset (e.g., Oculus Go). When the user requests the 360-degree video, the sender sends the 360-degree video frames, and the user may play a part of the 360-degree video frames, which is referred to as the viewport, through the user’s headset. 
Here, 360-degree videos are mainly captured by an omnidirectional camera or a combination of multiple cameras and saved in a spherical format. Before transmissions, the sphere frames are mapped onto the 2D plane using a certain projection method, e.g., equirectangular and cube map projections. 

In 360-degree video streaming, the major issue is to yield better video quality in the user’s viewport by effectively reducing perceptual redundancy within 360-degree video frames. Because each user only watches the viewport via the headset at each time instance, a large video traffic is created if the sender sends the full resolution of the 2D-projected video frames with an identical quantization parameter. One of the simplest methods to reduce perceptual redundancy is viewport-only streaming~\cite{bib:viewport_only}.
In video playback, the user may move a viewing viewport according to the user’s head/eye movement. Based on the movement, the user requests a new viewport to the sender, and the sender sends back the corresponding viewport. Because the sender transmits one viewport at each time instant, viewport-only streaming can mitigate the video traffic. However, the user needs to receive a new viewport from the sender in every viewport switching, which causes a long switching delay. A long switching delay, i.e., approximately 10 ms, may cause \textit{simulator sickness}~\cite{bib:sickness}. Owing to a long delay in the standard Internet, it is difficult for viewport-only streaming schemes to satisfy the switching delay requirements. To prevent simulator sickness, conventional schemes~\cite{bib:viewport} divide 360-degree video frames into multiple tiles and independently encode them with different quantization parameters to yield better viewport quality within the bandwidth constraint. 

The studies~\cite{bib:360-fuji,bib:omnicast,bib:360cast,bib:360cast+} on soft delivery schemes focus on the quality optimization of the user's viewport in addition to cliff and leveling effect prevention. 
\cite{bib:360-fuji} is the first scheme for viewport-aware soft 360-degree video delivery. According to the viewing viewport, the sender first adopts pixel-wise power allocation to reduce the perceptual redundancy in 360-degree video frames and then carries out the combination of one-dimensional DCT~(1D-DCT) and spherical wavelet transform~(SWT) as a decorrelation to utilize the redundancy in the sphere and time domains. 
OmniCast~\cite{bib:omnicast} further considers the feature of 360-degree videos into quality optimization. Specifically, they analyze the relationship of the distortion between the spherical and projected 2D domains as the spherical distortion for each projection method, and design power allocation to realize the optimal quality in the 2D-projected 360-degree videos. 
360Cast~\cite{bib:360cast} and the extended version of 360Cast$+$~\cite{bib:360cast+} adopt viewport prediction based on linear regression and foveation-aware power allocation within the predicted viewport to further reduce the perceptual redundancy. 

\subsection{Point Cloud}
Volumetric content delivery provides highly immersive experiences for users through XR devices.
The point cloud~\cite{bib:mpeg} is arguably the most popular volumetric data structure for representing 3D scenes and objects on holographic displays~\cite{bib:holodisplay, bib:holodisplay2}. 
A point cloud typically consists of a set of 3D points, and each point is defined by 3D coordinates, i.e., (X, Y, Z), and color attributes, i.e., (R, G, B). 
In contrast to conventional 2D images and videos, 3D point cloud data are neither well aligned nor uniformly distributed in space.

The major challenge in volumetric delivery over wireless channels is how to efficiently compress and send numerous and irregular structures of the 3D point cloud within a limited bandwidth.
Some compression methods have been proposed for point clouds to deliver 3D data. Specifically, Draco~\cite{bib:draco} employs $k$d tree-based compression~\cite{bib:kd} and a point cloud library (PCL) using octree-based compression~\cite{bib:PCL, bib:PCL2, bib:octree}.
To further reduce the amount of data traffic in point cloud delivery, two transform techniques have been proposed for energy compaction of the non-ordered and non-uniformly distributed signals: Fourier-based transform, e.g., graph Fourier transform~(GFT) and wavelet-based transform, e.g., region-adaptive Haar transform ~\cite{bib:DeQueiroz2016}.
For example, recent studies used GFT for the color components~\cite{bib:DigitalGFT} and 3D coordinates~\cite{bib:graph_PCC2} of graph signals for signal decorrelation. They used quantization and entropy coding for the compression of decorrelated signals.

HoloCast~\cite{bib:holocast} is a pioneering work on soft 3D point-cloud delivery for unstable wireless channels. 
Specifically, they regard 3D points as vertices in a graph with edges between nearby vertices to deal with the irregular structure of the 3D points motivated by~\cite{bib:graph_PCC2,bib:graph_PCC3}. 
HoloCast uses GFT for such graph signals to exploit the underlying correlations among adjacent graph signals and directly transmits linear-transformed graph signals as a pseudo-analog modulation over the channel. However, it has been found that graph-based coding schemes need to send the graph-based transform basis matrix used in GFT as additional metadata for signal decoding. For example, the sender needs to send $N^2$ real elements of the graph-based transform basis matrix as the metadata when the number of 3D points is $N$. 
In~\cite{bib:HoloCastGivens}, Givens  rotation~\cite{bib:givens,bib:givens2} was used for GFT basis matrix compression. Givens rotation is used to selectively introduce zeros into a matrix to create an identity matrix from the basis matrix using angle parameters. The angle parameters are quantized prior to the metadata transmission for overhead reduction.

\section{Future Directions}
\label{sec:future}
As mentioned in the previous sections, soft delivery schemes have been studied to overcome the issues of conventional image and video streaming in modern wireless and mobile networks since 2010. 
The main concept of soft delivery schemes is to replace conventional nonlinear operations with only linear operations, thus preventing the cliff and leveling effects, which are caused by such nonlinear operations.

Moreover, the reconstruction quality of soft delivery schemes highly depends on the performance of the linear encoding and decoding operations.
Recent studies integrate nonlinear encoding and decoding operations in soft delivery to take advantage of further quality improvements. 
Specifically, the studies in~\cite{bib:hybrid_first_conf,bib:hybrid_first} integrate low-rate digital-based encoding and decoding into soft delivery. 
Although high-rate digital-based operations are sensitive to channel quality fluctuations and thus have a cliff effect at the receiver end, operations with a relatively low rate can compact the signal energy and prevent bit errors even with channel fluctuations. In this case, the reconstruction quality was significantly low. For quality enhancement, the hybrid digital-analog delivery is utilized to send the residual signals; then, the receiver adds the received residuals to the digitally coded images/videos. In this case, the reconstruction quality can be gradually improved based on the wireless channel quality.

Other recent studies utilize deep neural network~(DNN) architectures for nonlinear encoding and decoding operations. In particular, deep convolutional neural networks~(DCNNs) have been successfully applied to image-based tasks~\cite{bib:DCNN2,bib:DCNN3}. The DCNN first learns the weights of the mapping function using noisy and original images based on a massive number of typical image datasets. Because the mapping function can represent linear/nonlinear noise, DCNN-based nonlinear encoding and decoding operations offer better performance compared with conventional nonlinear operations. 

\begin{figure*}[t]
  \begin{center}
   \includegraphics[scale=0.45]{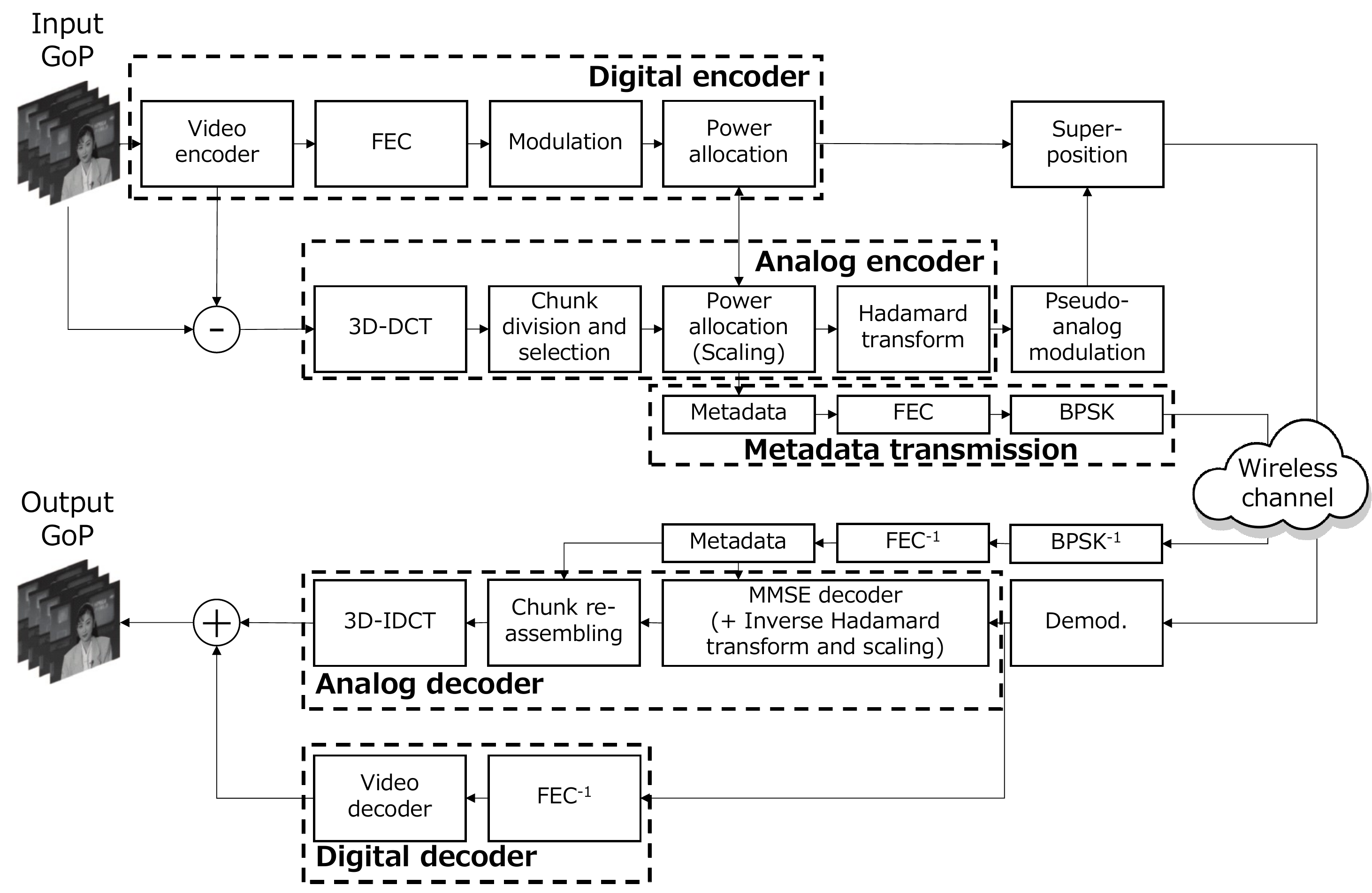}
   \caption{Typical framework of hybrid digital--analog delivery.}
   \label{fig:hda}
  \end{center}
\end{figure*}

\subsection{Hybrid Digital--Analog Delivery}
Although integration with digital operations, e.g., coset coding, was initially proposed in 2011~\cite{bib:dcast}, pioneering studies in~\cite{bib:hybrid_first_conf,bib:hybrid_first} proposed the superposition of analog-coded and digital-coded symbols, i.e., hybrid digital--analog~(HDA) delivery, to exploit the benefits of both conventional digital-based and soft delivery schemes.  
Fig.~\ref{fig:hda} shows an overview of the HDA delivery schemes. The HDA delivery schemes consist of the digital and analog coding parts. At the sender side, the video frames are first encoded by the digital video encoder and the digitally-coded bitstream is channel coded, modulated, and assigned transmission power by the sender. Meanwhile, the residuals are coded, power-assigned, and modulated by the soft delivery scheme. Both outputs from the digital and analog coding parts are superposed and transmitted over wireless channels. 
In this case, the transmitted signal $x_i$ is the sum of BPSK-modulated vector signal $x^{\langle\mathsf{d}\rangle}_{i}$ and output vector signal of the soft delivery scheme $x^{\langle\mathsf{a}\rangle}_{i}$ as follows:
\begin{equation}
    x_i = x^{\langle\mathsf{d}\rangle}_{i} + \jmath x^{\langle\mathsf{a}\rangle}_{i},
\end{equation}
The BPSK-modulated
symbol and the analog-modulated symbol are scaled by $P_\mathrm{d}$ and
$g_{i}$, respectively.
\begin{equation}
x^{\langle\mathsf{d}\rangle}_{i} = \sqrt{P_\mathsf{d}} \cdot b_{i},
 \qquad
x^{\langle\mathsf{a}\rangle}_{i} = g_{i} \cdot s_{i},
\label{eq:xdxa}
\end{equation}
where $b_{i}\in \mathbb{X} = \left\{\pm 1 \right\}$ is the BPSK-modulated symbol and $\jmath=\sqrt{-1}$ denotes the imaginary unit.
Here, the near-optimal solution of $g_i$ under the transmission power budget $P_\mathsf{a}$ is expressed as:
\begin{equation}
g_{i} = {\lambda}_{i}^{-1/4} \sqrt{\frac{P_\mathsf{a}}{\sum_j{\lambda}_{j}}}.
\end{equation}
We note that the budgets of the transmission power for the digital and analog parts need to satisfy the total power budget $P_\mathsf{t}$:
\begin{equation}
 P_\mathsf{t} = P_\mathsf{a} + P_\mathsf{d}.
\end{equation}

At the receiver side, it first decodes the digital-modulated symbols and then obtains the analog-modulated symbols by subtracting the digital-modulated symbols from the received symbols. Finally, the receiver reconstructs the baseline quality of the video frames from the output of the digital part and enhances the video quality by adding the output of the analog part. 

\begin{figure*}[t]
  \begin{center}
   \includegraphics[width=\hsize]{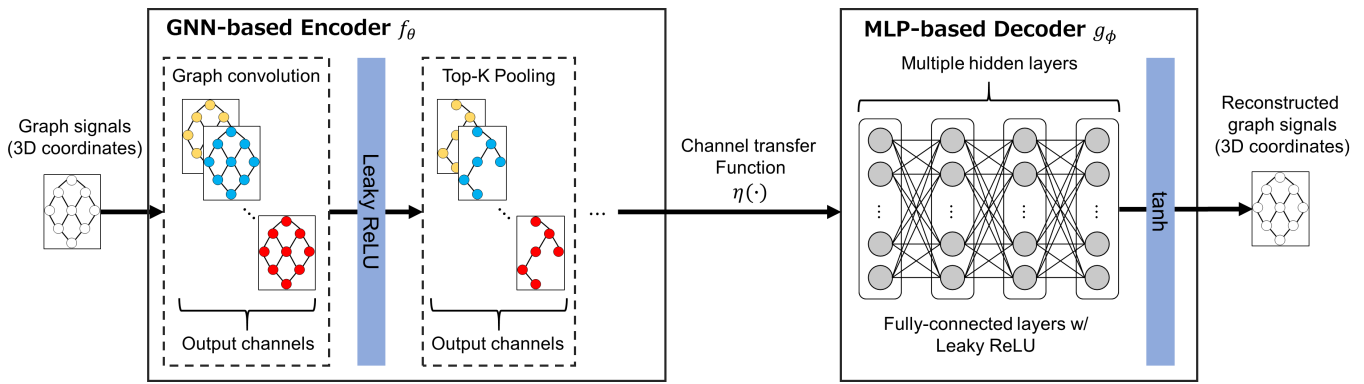}
   \caption{GNN-based end-to-end encoder and decoder for wireless 3D point cloud delivery~\cite{bib:GNNCast}.}
   \label{fig:gnn}
  \end{center}
\end{figure*}

A key issue in HDA delivery is the assignment of transmission power to the digital and analog parts~\cite{bib:hda_practical}. Specifically, the power assigned to the digital part must guarantee the correct decoding of the symbols. By contrast, the digital decoder treats the superimposed analog-modulated symbols $x^{\langle\mathsf{a}\rangle}_{i}$ as noise. To achieve better decoding performance, the I component of $x^{\langle\mathsf{a}\rangle}_{i}$ should be kept as small as possible.
In~\cite{bib:hybridcast}, they only select the high-frequency coefficients,
which are expected to be very small values for superposition. The remaining low-frequency coefficients are delivered using pseudo-analog modulation.
The HDA framework in~\cite{bib:hda_combined} regards the superposed symbols as three main parts: 1) orthogonal analog symbols, 2) digital symbols, and 3) nonorthogonal analog symbols superimposed onto digital symbols. They designed resource allocation among these three parts to achieve a better balance between lowering interference and improving reconstruction quality. 
Another study~\cite{bib:hda_model} designs a prediction model
to describe the relationship between the variance of residuals and the quantization parameter, and determines the optimal transmission power for the analog part, which maximizes the reconstruction quality with the correct decoding of the digital part.
The HDA delivery scheme in~\cite{bib:Zhang2019} treats the imperfect decoding of the digital part and finds the best assignment of the transmission power for the digital and analog parts. This prevents too much power assignment for the digital part to ensure a low bit error rate~(BER). 
In contrast to the aforementioned studies, \cite{bib:hda3} treats the bandwidth of other digital traffic as hidden resources for HDA video delivery. Specifically, they superimpose the analog-modulated symbols and digital symbols of the other digital traffic to utilize the hidden resource under the constraint that the BER requirement of the other digital traffic is not compromised.

Other studies have redesigned the power allocation in HDA delivery for practical wireless channel environments, including fading~\cite{bib:hybrid_fading,bib:hda_fading2}, OFDM~\cite{bib:hda_ofdm,bib:Yahampath2020}, MIMO~\cite{bib:hybrid_mimo}, and relay networks~\cite{bib:hybrid_second}. 
For example, the power allocation with perfect channel state information~(CSI) is designed in~\cite{bib:hybrid_fading, bib:hda_ofdm, bib:hybrid_mimo}, whereas the power allocation with imperfect CSI is designed in~\cite{bib:hda_fading2,bib:Yahampath2020}. 
In view of the packet loss resilience in HDA delivery, the study in~\cite{bib:fuji_cs} introduced compressive sensing for the residuals.  

Other studies~\cite{bib:swift,bib:Li2020,bib:holoplus} extend HDA video delivery for immersive contents. Swift in~\cite{bib:swift} considers stereo video delivery and designs a zigzag coding structure for the stereo video to utilize both intra- and inter-view correlations. In the zigzag coding structure, the odd frames in the left view and the even frames in the right view are encoded digitally, and the rest of the frames are encoded in analog. Here, the reconstructions of the digitally coded frames are used as side information to further remove redundant information from the analog-coded frames.
Another study ~\cite{bib:Li2020} extends HDA delivery for MVD videos and solves the view synthesis optimization to yield the best quality from an intermediate virtual viewpoint. HoloCast+ in~\cite{bib:holoplus} designs HDA delivery for point cloud delivery.

\subsection{Applied Deep Neural Network}
In recent studies, DNN-based nonlinear operations have been integrated with soft delivery. The multi-layer perceptron auto-encoder is first adopted to reduce the overhead of soft delivery~\cite{bib:fuji_DNN}. Specifically, the proposed encoder obtains a few latent variables from the pixel values, and the proposed decoder decodes the accurate power information from the received latent variables for proper power allocation. The reconstruction quality can be maintained even with only one metadata across one GoP.
Another study ~\cite{bib:DeepSoft1} integrates a DCNN-based auto-encoder into a soft delivery scheme. The proposed encoder directly compresses each image into a limited number of latent variables, and the proposed decoder reconstructs the image from the latent variables. Here, the latent variables are transmitted over wireless channels using pseudo-analog modulation. Even though the latent variables are obtained by nonlinear functions and delivered over wireless channels with a lower SNR, cliff and leveling effects can be prevented via pseudo-analog modulation. 
Other studies have introduced the DNN architecture for power allocation~\cite{bib:Tang2020_2} and decoding operations~\cite{bib:soft-dip}. 
The study in~\cite{bib:Tang2020_2} uses a you-only look-once~(YOLO) structure~\cite{bib:yolo} to extract the region of interest~(ROI) and non-ROI parts from each image and then assign unequal transmission power across ROI and non-ROI parts for perceptual quality enhancement.  
The proposed scheme in~\cite{bib:soft-dip} integrates DCNN-based image denoising, specifically deep image prior (DIP)~\cite{bib:DIP}, into soft delivery.
The DIP finds linear and nonlinear noise effects for reconstructing clean images from noisy images. The proposed scheme can remove fading and noise effects from the received images using DCNN-based image restoration.
Another study ~\cite{bib:GNNCast} introduces graph neural networks~(GNN)~\cite{bib:GNN} for wireless point cloud delivery. 
The GNN is a novel model for graph representation learning that allows the analysis of the irregular geometric structure of graph data. 
Fig.~\ref{fig:gnn} showed their GNN-based auto-encoder (GAE)~\cite{bib:ChenDYLFT:20,bib:PCT} to encode 3D point clouds into a limited number of latent variables. 
One of the benefits of the GAE is that it allows graph signal reconstruction from a limited number of latent variables without requiring additional metadata.

\section{Conclusion}
\label{sec:conclusion}
Herein, we present an exhaustive survey and research outlook
of the soft delivery schemes. We first review conventional digital-based video delivery schemes and the critical issues of the schemes, including cliff, leveling, and staircase effects. Next, we provide an overview of the soft delivery schemes and the taxonomy of the existing schemes from the perspectives of energy compaction, power allocation, bandwidth utilization, packet loss resilience, overhead reduction, receiver heterogeneity, and implementation. Finally, future research directions, including extension of immersive content, HDA delivery, and DNN-based soft delivery, are discussed.

\bibliographystyle{IEEEtran}
\bibliography{./main}

\begin{thebibliography}{100}
\providecommand{\url}[1]{#1}
\csname url@samestyle\endcsname
\providecommand{\newblock}{\relax}
\providecommand{\bibinfo}[2]{#2}
\providecommand{\BIBentrySTDinterwordspacing}{\spaceskip=0pt\relax}
\providecommand{\BIBentryALTinterwordstretchfactor}{4}
\providecommand{\BIBentryALTinterwordspacing}{\spaceskip=\fontdimen2\font plus
\BIBentryALTinterwordstretchfactor\fontdimen3\font minus
  \fontdimen4\font\relax}
\providecommand{\BIBforeignlanguage}[2]{{%
\expandafter\ifx\csname l@#1\endcsname\relax
\typeout{** WARNING: IEEEtran.bst: No hyphenation pattern has been}%
\typeout{** loaded for the language `#1'. Using the pattern for}%
\typeout{** the default language instead.}%
\else
\language=\csname l@#1\endcsname
\fi
#2}}
\providecommand{\BIBdecl}{\relax}
\BIBdecl

\bibitem{bib:cisco2017}
Cisco, ``{Cisco Visual Networking Index:} global mobile data traffic forecast
  update 2017--2022,'' Mar. 2019.

\bibitem{bib:pcs_hevc}
D.~Grois, D.~Marpe, A.~Mulayoff, B.~Itzhaky, and O.~Hadar, ``Performance
  comparison of {H.265/MPEG-HEVC}, {VP9}, and {H.264/MPEG-AVC} encoders,'' in
  \emph{IEEE Picture Coding Symposium}, 2013, pp. 394--397.

\bibitem{bib:LiDAR_journal}
S.~Schwarz, M.~Preda, V.~Baroncini, M.~Budagavi, P.~Cesar, P.~A. Chou, R.~A.
  Cohen, M.~Krivokuca, S.~Lassere, Z.~Li, J.~Llach, K.~Mammou, R.~Mekuria,
  O.~Nakagami, E.~Siahaan, A.~Tabatabai, A.~M. Tourapis, and V.~Zakharchenko,
  ``Emerging {MPEG} standards for point cloud compression,'' \emph{IEEE Journal
  of Emerging and Selected Topics in Circuits and Systems}, vol.~9, no.~1, pp.
  133--148, 2019.

\bibitem{bib:Danillo2020}
D.~B. Graziosi, O.~Nakagami, S.~Kuma, A.~Zaghetto, T.~Suzuki, and A.~Tabatabai,
  ``An overview of ongoing point cloud compression standardization activities:
  Video-based ({V-PCC}) and geometry-based ({G-PCC}),'' \emph{APSIPA
  Transactions on Signal and Information Processing}, vol.~9, pp. 1--17, 2020.

\bibitem{bib:cliff2011}
S.~Kokalj-Filipovi{\'{c}}, E.~Soljanin, and Y.~Gao, ``Cliff effect suppression
  through multiple-descriptions with split personality,'' in \emph{IEEE
  International Symposium on Information Theory}, 2011, pp. 948--952.

\bibitem{bib:softcast1}
S.~Jakubczak and D.~Katabi, ``A cross-layer design for scalable mobile video,''
  in \emph{ACM Annual International Conference on Mobile Computing and
  Networking}, 2011, pp. 289--300.

\bibitem{bib:softcast2}
S.~Jakubczak, J.~Z. Sun, D.~Katabi, and V.~K. Goyal, ``Performance regimes of
  uncoded linear communications over {AWGN} channels,'' in \emph{45th Annual
  Conference on Information Sciences and Systems}, 2011, pp. 1--6.

\bibitem{bib:softcast3}
S.~Jakubczak and D.~Katabi, ``{SoftCast}: One-size-fits-all wireless video,''
  \emph{Computer Communication Review}, vol.~40, no.~4, pp. 449--450, 2010.

\bibitem{bib:wireless_streaming1}
T.~Stockhammer, H.~Jenkac, and G.~Kuhn, ``Streaming video over variable
  bit-rate wireless channels,'' \emph{IEEE Transactions on Multimedia}, vol.~6,
  no.~2, pp. 268--277, 2004.

\bibitem{bib:wireless_streaming2}
Z.~Guo, Y.~Wang, E.~Erkip, and S.~Panwar, ``Wireless video multicast with
  cooperative and incremental transmission of parity packets,'' \emph{IEEE
  Transactions on Multimedia}, vol.~17, no.~8, pp. 1335--1346, 2015.

\bibitem{bib:wireless_streaming3}
S.~Almowuena, M.~M. Rahman, C.~H. Hsu, A.~A. Hassan, and M.~Hafeeda,
  ``Energy-aware and bandwidth-efficient hybrid video streaming over mobile
  networks,'' \emph{IEEE Transactions on Multimedia}, vol.~18, no.~1, pp.
  102--115, 2016.

\bibitem{bib:MPEG_Standard}
W.~Thomas, S.~G. J, B.~Gisle, and L.~Ajay, ``Overview of the {H.264/AVC} video
  coding standard,'' \emph{IEEE Transactions of Circuits And Systems for Video
  Technology}, vol.~13, no.~7, pp. 560--576, 2003.

\bibitem{bib:VLC}
Q.~Fan, D.~J. Lilja, and S.~S. Sapatnekar, ``Adaptive-length coding of image
  data for low-cost approximate storage,'' \emph{IEEE Transactions on
  Computers}, vol.~69, no.~2, pp. 239--252, 2020.

\bibitem{bib:rateless}
Y.~Li, J.~Wu, B.~Tan, M.~Wang, and W.~Zhang, ``Compressive spinal codes,''
  \emph{IEEE Transactions on Vehicular Technology}, vol.~68, no.~12, pp.
  11\,944--11\,954, 2019.

\bibitem{bib:rateless2}
L.~Liu, J.~Wu, and J.~Wu, ``{COQRC}: A rateless video transmission solution,''
  in \emph{International Conference on Computing, Networking and
  Communications}, 2018, pp. 463--467.

\bibitem{bib:iCast}
L.~Wang, H.~Yang, X.~Qi, J.~Xu, and K.~Wu, ``{ICast:} fine-grained wireless
  video streaming over internet of intelligent vehicles,'' \emph{IEEE Internet
  of Things Journal}, vol.~6, no.~1, pp. 111--123, 2019.

\bibitem{bib:simcast}
G.~Wang, K.~Wu, Q.~Zhang, and L.~M. Ni, ``{SimCast}: Efficient video delivery
  in {MU-MIMO WLANs},'' in \emph{IEEE Conference on Computer Communications},
  2014, pp. 2454--2462.

\bibitem{bib:flexcast}
S.~T. Aditya and S.~Katti, ``{FlexCast}: Graceful wireless video streaming,''
  in \emph{Annual International Conference on Mobile Computing and Networking},
  2011, pp. 277--288.

\bibitem{bib:svc1}
H.~Schwarz, D.~Marpe, and T.~Wiegand, ``Overview of the scalable video coding
  extension of the {H.264/AVC} standard,'' \emph{IEEE Transactions on Circuits
  and Systems for Video Technology}, vol.~17, no.~9, pp. 1103--1120, 2007.

\bibitem{bib:hm1}
T.~Kratochvil and R.~Stukavec, ``Hierarchical modulation in {DVB-T/H} mobile
  {TV} transmission over fading channels,'' in \emph{International Symposium on
  Information Theory and Its Applications}, 2008, pp. 1--6.

\bibitem{bib:hm2}
C.~Hellge, S.~Mirta, T.~Schierl, and T.~Wiegand, ``Mobile {TV} with {SVC} and
  hierarchical modulation for {DVB-H} broadcast services,'' in \emph{IEEE
  International Symposium on Broadband Multimedia Systems and Broadcasting}, 06
  2009, pp. 1--5.

\bibitem{bib:hm3}
M.~Ghandi and M.~Ghanbari, ``Layered {H.264} video transmission with
  hierarchical {QAM},'' \emph{Journal of Visual Communication and Image
  Representation}, vol.~17, no.~2, pp. 451--466, 2006.

\bibitem{bib:mos}
A.~Trioux, G.~Valenzise, M.~Cagnazzo, M.~Kieffer, F.-X. Coudoux, P.~Corlay, and
  M.~Gharbi, ``Subjective and objective quality assessment of the {SoftCast}
  video transmission scheme,'' in \emph{IEEE International Conference on Visual
  Communications and Image Processing}, 2020, pp. 96--99.

\bibitem{bib:preprocessing}
A.~Trioux, F.~X. Coudoux, P.~Corlay, and M.~Gharbi, ``A comparative
  preprocessing study for {SoftCast} video transmission,'' in
  \emph{International Symposium on Signal, Image, Video and Communications},
  2018, pp. 54--59.

\bibitem{bib:decorrelation1}
R.~Xiong, F.~Wu, J.~Xu, X.~Fan, C.~Luo, and W.~Gao, ``Analysis of decorrelation
  transform gain for uncoded wireless image and video communication,''
  \emph{IEEE Transactions on Image Processing}, vol.~25, no.~4, pp. 1820--1833,
  2016.

\bibitem{bib:3ddwt-soft}
H.~Cui, R.~Xiong, C.~Luo, Z.~Song, and F.~Wu, ``Denoising and resource
  allocation in uncoded video transmission,'' \emph{IEEE Journal on Selected
  Topics in Signal Processing}, vol.~9, no.~1, pp. 102--112, 2015.

\bibitem{bib:3ddwt-soft2}
Q.~Wang, X.~Lin, Y.~Liu, L.~Zhang, and X.~Wu, ``A scalable mobile video
  broadcast scheme using {3D} wavelet transform,'' in \emph{IEEE Vehicular
  Technology Conference}, 2014, pp. 3--7.

\bibitem{bib:3ddwt-soft3}
------, ``A scalable framework for mobile video broadcast using {MCTF} and
  {2D-DWT},'' in \emph{International Symposium on Wireless Personal Multimedia
  Communications}, 2015, pp. 118--123.

\bibitem{bib:wavecast}
X.~Fan, R.~Xiong, F.~Wu, and D.~Zhao, ``{WaveCast: Wavelet} based wireless
  video broadcast using lossy transmission,'' in \emph{IEEE Visual
  Communications and Image Processing}, 2012, pp. 1--6.

\bibitem{bib:cactus}
H.~Cui, Z.~Song, Z.~Yang, C.~Luo, R.~Xiong, and F.~Wu, ``Cactus: A hybrid
  digital-analog wireless video communication system,'' in \emph{ACM
  International Conference on Modeling, Analysis \& Simulation of Wireless and
  Mobile Systems}, 2013, pp. 273--278.

\bibitem{bib:adaptive_hybrid}
X.~Zhao, H.~Lu, C.~W. Chen, and J.~Wu, ``Adaptive hybrid digital-analog video
  transmission in wireless fading channel,'' \emph{IEEE Transactions on
  Circuits and Systems for Video Technology}, vol.~26, no.~6, pp. 1117--1130,
  2016.

\bibitem{bib:2ddwt-soft}
X.~Lin, N.~Fan, Y.~Liu, S.~Cai, and X.~Wang, ``Soft wireless image/video
  broadcast based on component protection,'' in \emph{IEEE International
  Conference on Network Infrastructure and Digital Content}, 2014, pp. 84--89.

\bibitem{bib:AGDCast}
C.~He, H.~Qin, Z.~He, and K.~Niu, ``Adaptive {GoP} dividing video coding for
  wireless broadcast based on power allocation optimization,'' in
  \emph{International Conference on Wireless Communications and Signal
  Processing}, 2016, pp. 1--5.

\bibitem{bib:layeredsoftcast}
Z.~Song, R.~Xiong, S.~Ma, X.~Fan, and W.~Gao, ``Layered image/video {SoftCast}
  with hybrid digital-analog transmission for robust wireless visual
  communication,'' in \emph{IEEE International Conference on Multimedia and
  Expo}, 2014, pp. 1--6.

\bibitem{bib:layeredcast1}
J.~Zhao, J.~Xie, and R.~Xiong, ``Residual signals modeling for layered
  image/video {Softcast} with hybrid digital-analog transmission,'' \emph{IEEE
  International Conference on Image Processing}, pp. 3284--3288, 2018.

\bibitem{bib:structure}
D.~He, C.~Luo, C.~Lan, F.~Wu, and W.~Zeng, ``Structure-preserving hybrid
  digital-analog video delivery in wireless networks,'' \emph{IEEE Transactions
  on Multimedia}, vol.~17, no.~9, pp. 1658--1670, 2015.

\bibitem{bib:visual_hybrid}
Y.~Li, Y.~Liu, Y.~Wang, and Z.~Li, ``Visual information exploited hybrid
  digital-analog scheme for wireless video multicast,'' in \emph{Visual
  Communications and Image Processing}, 2016, pp. 1--4.

\bibitem{bib:G-Cast}
R.~Xiong, H.~Liu, S.~Ma, X.~Fan, F.~Wu, and W.~Gao, ``{G-CAST:} gradient based
  image {SoftCast} for perception-friendly wireless visual communication,'' in
  \emph{Data Compression Conference}, 2014, pp. 133--142.

\bibitem{bib:CG-Cast}
H.~Liu, R.~Xiong, X.~Fan, D.~Zhao, Y.~Zhang, and W.~Gao, ``{CG-cast:} scalable
  wireless image softcast using compressive gradient,'' \emph{IEEE Transactions
  on Circuits and Systems for Video Technology}, vol.~29, no.~6, pp.
  1832--1843, 2019.

\bibitem{bib:dcast}
A.~Zhang, X.~Fan, R.~Xiong, and D.~Zhao, ``Distributed soft video broadcast
  with variable block size motion estimation,'' in \emph{IEEE International
  Conference on Visual Communications and Image Processing}, 2013, pp. 1--5.

\bibitem{bib:dcast2}
X.~Fan, F.~Wu, D.~Zhao, O.~C. Au, and W.~Gao, ``Distributed soft video
  broadcast ({DCAST}) with explicit motion,'' in \emph{Data Compression
  Conference Proceedings}, 2012, pp. 199--208.

\bibitem{bib:dcast3}
X.~Fan, F.~Wu, and D.~Zhao, ``D-cast: {DSC} based soft mobile video
  broadcast,'' in \emph{International Conference on Mobile and Ubiquitous
  Multimedia}, 2011, pp. 226--235.

\bibitem{bib:dcast4}
X.~Fan, F.~Wu, D.~Zhao, and O.~C. Au, ``Distributed wireless visual
  communication with power distortion optimization,'' \emph{IEEE Transactions
  on Circuits and Systems for Video Technology}, vol.~23, no.~6, pp.
  1040--1053, 2013.

\bibitem{bib:coset1}
M.~Lv, Y.~Liu, and Y.~Wang, ``Scalable wireless video broadcast based on
  unequal protection,'' in \emph{Visual Communication and Image Processing},
  2017, pp. 1--4.

\bibitem{bib:magnitude_shift}
X.~Lin, Y.~Liu, and L.~Zhang, ``Scalable video {SoftCast} using magnitude
  shift,'' in \emph{IEEE Wireless Communications and Networking Conference},
  2015, pp. 1996--2001.

\bibitem{bib:hyperspectrum}
A.~Hagag, X.~Fan, and F.~E. {Abd El-Samie}, ``Hyperspectral image coding and
  transmission scheme based on wavelet transform and distributed source
  coding,'' \emph{Multimedia Tools and Applications}, vol.~76, no.~22, pp.
  23\,757--23\,776, 2017.

\bibitem{bib:layered_coset}
X.~Fan, R.~Xiong, D.~Zhao, and F.~Wu, ``Layered soft video broadcast for
  heterogeneous receivers,'' \emph{IEEE Transactions on Circuits and Systems
  for Video Technology}, vol.~25, no.~11, pp. 1801--1814, 2015.

\bibitem{bib:layered_coset2}
M.~Lv, Y.~Liu, and Y.~Wang, ``Adaptive scalable wireless video coding based on
  unequal protection and quadtree partition,'' in \emph{International
  Conference on Network Infrastructure and Digital Content}, vol.~9, 2017, pp.
  214--218.

\bibitem{bib:relay1}
J.~Shen, F.~Liang, C.~Luo, H.~Li, and W.~Zeng, ``Cooperative hybrid
  digital-analog video transmission in {D2D} networks,'' \emph{IEEE
  International Conference on Image Processing}, pp. 3274--3278, 2018.

\bibitem{bib:relay2}
Y.~Wang, M.~Sun, and Y.~Liu, ``Distributed and adaptive analog coding for video
  broadcast in wireless cooperative system,'' \emph{Wireless Personal
  Communications}, vol. 102, no.~3, pp. 2287--2306, 2018.

\bibitem{bib:relay3}
M.~Sun, Y.~Wang, H.~Yu, and Y.~Liu, ``Distributed cooperative video coding for
  wireless video broadcast system,'' in \emph{IEEE International Conference on
  Multimedia and Expo}, 2015, pp. 1--6.

\bibitem{bib:hybrid_theory}
V.~Prabhakaran, R.~Puri, and K.~Ramchandran, ``Hybrid digital-analog codes for
  source-channel broadcast of {Gaussian} sources over {Gaussian} channels,''
  \emph{IEEE Transactions on Information Theory}, vol.~57, no.~7, pp.
  4573--4588, 2011.

\bibitem{bib:decorrelation2}
R.~Xiong, F.~Wu, J.~Xu, and W.~Gao, ``Performance analysis of transform in
  uncoded wireless visual communication,'' in \emph{IEEE International
  Symposium on Circuits and Systems}, no.~2, 2013, pp. 1159--1162.

\bibitem{bib:gop}
A.~Trioux, F.~X. Coudoux, P.~Corlay, and M.~Gharbi, ``{Temporal information
  based GoP adaptation for linear video delivery schemes},'' \emph{Signal
  Processing: Image Communication}, vol.~82, pp. 1--17, 2020.

\bibitem{bib:sfc1}
------, ``A reduced complexity/side information preprocessing method for high
  quality {Softcast}-based video delivery,'' in \emph{European Workshop on
  Visual Information Processing}, 2019, pp. 205--210.

\bibitem{bib:coset_theory1}
A.~Sehgal, A.~Jagmohan, and N.~Ahuja, ``{Wyner-Ziv} coding of video: an
  error-resilient compression framework,'' \emph{IEEE Transactions on
  Multimedia}, vol.~6, no.~2, pp. 249--258, 2004.

\bibitem{bib:coset_theory2}
S.~Pradhan and K.~Ramchandran, ``{Distributed source coding using syndromes
  (DISCUS): design and construction},'' \emph{IEEE Transactions on Information
  Theory}, vol.~49, no.~3, pp. 626--643, 2003.

\bibitem{bib:SIRcast}
W.~Huang, X.~Fan, and D.~Zhao, ``Soft mobile video broadcast based on side
  information refining,'' in \emph{IEEE International Conference on Visual
  Communications and Image Processing}, 2013.

\bibitem{bib:sir}
R.~Martins, C.~Brites, J.~a. Ascenso, and F.~Pereira, ``Refining side
  information for improved transform domain {Wyner-Ziv} video coding,''
  \emph{IEEE Transactions on Circuits and Systems for Video Technology},
  vol.~19, no.~9, p. 1327–1341, 2009.

\bibitem{bib:fading1}
H.~Cui, D.~Liu, Y.~Han, and J.~Wu, ``Robust uncoded video transmission under
  practical channel estimation,'' in \emph{IEEE Global Communications
  Conference}, 2016, pp. 1--6.

\bibitem{bib:soft-fading}
F.~Zhang, A.~Wang, H.~Wang, S.~Li, X.~Ma, Z.~Idglqj, F.~Sdshu, S.~D. Qhz, X.~S.
  Wr, G.~L. Frpsduhg, Z.~Wkh, L.~Q. Wkh, V.~Ri, and X.~Dqg, ``Channel-aware
  video {SoftCast} scheme,'' in \emph{IEEE China Summit and International
  Conference on Signal and Information Processing}, 2015, pp. 578--581.

\bibitem{bib:carrier}
Z.~Zhang, D.~Liu, and X.~Wang, ``Joint carrier matching and power allocation
  for wireless video with general distortion measure,'' \emph{IEEE Transactions
  on Mobile Computing}, vol.~17, no.~3, pp. 577--589, 2018.

\bibitem{bib:mimo1}
J.~Wu, B.~Tan, J.~Wu, and R.~Wang, ``Efficient soft video {MIMO} design to
  combine diversity and spatial multiplexing gain,'' \emph{IEEE Internet of
  Things Journal}, vol.~6, no.~3, pp. 5461--5472, 2019.

\bibitem{bib:waterfilling}
S.~Zheng, M.~Cagnazzo, and M.~Kieffer, ``Precoding matrix design in linear
  video coding,'' in \emph{IEEE International Conference on Acoustics, Speech
  and Signal Processing}, 2018, pp. 1198--1202.

\bibitem{bib:per-carrier}
S.~Zheng, M.~Antonini, M.~Cagnazzo, L.~Guerrieri, M.~Kieffer, I.~Nemoianu,
  R.~Samy, and B.~Zhang, ``Softcast with per-carrier power-constrained
  channels,'' in \emph{IEEE International Conference on Image Processing},
  2016, pp. 2122--2126.

\bibitem{bib:parcast}
X.~L. Liu, W.~Hu, Q.~Pu, F.~Wu, and Y.~Zhang, ``{ParCast:} soft video delivery
  in {MIMO-OFDM WLANs},'' in \emph{Proceedings of the 18th annual international
  conference on Mobile computing and networking}, 2012, pp. 233--244.

\bibitem{bib:ParCast+}
X.~L. Liu, W.~Hu, C.~Luo, Q.~Pu, F.~Wu, and Y.~Zhang, ``{ParCast+}: Parallel
  video unicast in {MIMO-OFDM WLANs},'' \emph{IEEE Transactions on Multimedia},
  vol.~16, no.~7, pp. 2038--2051, 2014.

\bibitem{bib:ecast}
Z.~Zhang, D.~Liu, X.~Ma, and X.~Wang, ``{ECast:} an enhanced video transmission
  design for wireless multicast systems over fading channels,'' \emph{IEEE
  Systems Journal}, vol.~11, no.~4, pp. 2566--2577, 2015.

\bibitem{bib:impulse}
S.~Zheng, M.~Cagnazzo, and M.~Kieffer, ``{Channel Impulsive Noise Mitigation
  for Linear Video Coding Schemes},'' \emph{IEEE Transactions on Circuits and
  Systems for Video Technology}, vol.~30, no.~9, pp. 3196--3209, 2020.

\bibitem{bib:noma2}
X.~Jiang, H.~Lu, C.~W. Chen, and F.~Wu, ``Receiver-driven video multicast over
  {NOMA} systems in heterogeneous environments,'' in \emph{IEEE International
  Conference on Computer Communications}.\hskip 1em plus 0.5em minus
  0.4em\relax IEEE, 2019, pp. 982--990.

\bibitem{bib:noma1}
J.~Wu, B.~Tan, J.~Wu, and M.~Wang, ``Video multicast: Integrating scalability
  of soft video delivery systems into {NOMA},'' \emph{IEEE Wireless
  Communications Letters}, vol.~8, no.~6, pp. 1722--1726, 2019.

\bibitem{bib:underwater}
R.~Zhang, Y.~Kong, X.~Ma, and D.~Wang, ``Adaptive video transmission designs
  over underwater acoustic channels,'' in \emph{International Conference on
  Computing, Networking and Communications}, 2018, pp. 295--299.

\bibitem{bib:UAV}
X.-w. Tang, X.-l. Huang, and F.~Hu, ``{QoE}-driven {UAV}-enabled pseudo-analog
  wireless video broadcast: A joint optimization of power and trajectory,''
  \emph{IEEE Transactions on Multimedia}, vol.~23, pp. 2398--2412, 2021.

\bibitem{bib:mmwave}
Y.~Gui, L.~Hancheng, F.~Wu, and C.~W. Chen, ``{LensCast}: Robust wireless video
  transmission over {mmWave MIMO} with lens antenna array,'' \emph{IEEE
  Transactions on Multimedia}, vol.~PP, no.~99, pp. 1--16, 2020.

\bibitem{bib:ssim-softcast}
J.~Zhao, R.~Xiong, C.~Luo, F.~Wu, and W.~Gao, ``Wireless image and video soft
  transmission via perception-inspired power distortion optimization,'' in
  \emph{IEEE Visual Communications and Image Processing}, 2018, pp. 1--4.

\bibitem{bib:foveacast}
J.~Shen, L.~Yu, L.~Li, and H.~Li, ``Foveation-based wireless soft image
  delivery,'' \emph{IEEE Transactions on Multimedia}, vol.~20, no.~10, pp.
  2788--2800, 2018.

\bibitem{bib:saliency}
H.~Hadizadeh, ``Saliency-guided wireless transmission of still images using
  {SoftCast},'' in \emph{International Symposium on Telecommunications}, 2017,
  pp. 506--509.

\bibitem{bib:scast}
Y.~Li, Z.~Li, Y.~Liu, and Y.~Wang, ``{SCAST:} wireless video multicast scheme
  based on segmentation and {Softcast},'' in \emph{IEEE Wireless Communications
  and Networking Conference}, 2017, pp. 1--6.

\bibitem{bib:foveation}
Z.~Wang and A.~C. Bovik, ``Embedded foveation image coding,'' \emph{IEEE
  Transactions on Image Processing}, vol.~10, no.~10, pp. 1397--1410, 2001.

\bibitem{bib:salience_model}
L.~Itti, C.~Koch, and E.~Niebur, ``A model of saliency-based visual attention
  for rapid scene analysis,'' \emph{IEEE Transactions on Pattern Analysis and
  Machine Intelligence}, vol.~20, no.~11, pp. 1254--1259, 1998.

\bibitem{bib:tradeoff}
D.~Liu, J.~Wu, H.~Cui, D.~Zhang, C.~Luo, and F.~Wu, ``Cost-distortion
  optimization and resource control in pseudo-analog visual communications,''
  \emph{IEEE Transactions on Multimedia}, vol.~20, no.~11, pp. 3097--3110,
  2018.

\bibitem{bib:MUCast}
C.~He, Y.~Hu, Y.~Chen, X.~Fan, H.~Li, and B.~Zeng, ``{MUcast}: Linear uncoded
  multiuser video streaming with channel assignment and power allocation
  optimization,'' \emph{IEEE Transactions on Circuits and Systems for Video
  Technology}, vol.~30, no.~4, pp. 1136--1146, 2020.

\bibitem{bib:MUCast2}
------, ``Exploiting channel assignment and power allocation for linear uncoded
  multiuser video streaming,'' in \emph{IEEE International Conference on
  Communications}, 2019.

\bibitem{bib:unequalblock}
Z.~Li, Y.~Liu, and Y.~Wang, ``Unequal block for low bandwidth adaption in
  wireless video broadcast,'' in \emph{International Conference on Network
  Infrastructure and Digital Content}, 2017, pp. 386--390.

\bibitem{bib:SKmapping}
M.~Cagnazzo and M.~Kieffer, ``{Shannon-Kotelnikov} mappings for softcast-based
  joint source-channel video coding,'' in \emph{IEEE International Conference
  on Image Processing}, 2015, pp. 1085--1089.

\bibitem{bib:cs2}
Y.~Gui, H.~Lu, X.~Jiang, F.~Wu, and C.~W. Chen, ``Compressed pseudo-analog
  transmission system for remote sensing images over bandwidth-constrained
  wireless channels,'' \emph{IEEE Transactions on Circuits and Systems for
  Video Technology}, vol.~30, no.~9, pp. 3181--3195, 2020.

\bibitem{bib:satelite-cs}
Y.~Wang, H.~Lu, Z.~Li, and J.~Li, ``Robust satellite image transmission over
  bandwidth-constrained wireless channels,'' in \emph{IEEE International
  Conference on Communications}, 2017, pp. 1--6.

\bibitem{bib:visual-cs}
A.~S. Yami and H.~Hadizadeh, ``Visual attention-driven wireless multicasting of
  images using adaptive compressed sensing,'' in \emph{Artificial Intelligence
  and Signal Processing}, oct 2018, pp. 37--42.

\bibitem{bib:cs1}
H.~Hadizadeh and I.~V. Bajic, ``Soft video multicasting using adaptive
  compressed sensing,'' \emph{IEEE Transactions on Multimedia}, vol.~23, pp.
  12--25, 2021.

\bibitem{bib:cs4}
W.~Yin, X.~Fan, Y.~Shi, R.~Xiong, and D.~Zhao, ``Compressive sensing based soft
  video broadcast using spatial and temporal sparsity,'' \emph{Mobile Networks
  and Applications}, vol.~21, no.~6, pp. 1002--1012, 2016.

\bibitem{bib:sparsecast}
T.~Y. Tung and D.~Gunduz, ``{SparseCast:} hybrid digital-analog wireless image
  transmission exploiting frequency-domain sparsity,'' \emph{IEEE
  Communications Letters}, vol.~22, no.~12, pp. 2451--2454, 2018.

\bibitem{bib:cs6}
S.~Liu, K.~Niu, and C.~Dong, ``Channel polarization based block compressive
  sensing {SoftCast} system,'' in \emph{IEEE International Conference on
  Computer and Communications}, 2019, pp. 778--783.

\bibitem{bib:adaptcast}
G.~Angelopoulos, M.~Medard, and A.~P. Chandrakasan, ``{AdaptCast:} an
  integrated source to transmission scheme for wireless sensor networks,'' in
  \emph{IEEE International Conference on Communications}, 2015, pp. 2894--2899.

\bibitem{bib:CS_theory_orig}
D.~L. Donoho, ``Compressed sensing,'' \emph{IEEE Transactions on Information
  Theory}, vol.~52, no.~4, pp. 1289--1306, 2006.

\bibitem{bib:CS_theory}
E.~J. Candes and M.~B. Wakin, ``An introduction to compressive sampling,''
  \emph{IEEE Signal Processing Magazine}, vol.~25, no.~2, pp. 21--30, 2008.

\bibitem{bib:kmvcast3}
X.~L. Huang, J.~Wu, and F.~Hu, ``Knowledge-enhanced mobile video broadcasting
  framework with cloud support,'' \emph{IEEE Transactions on Circuits and
  Systems for Video Technology}, vol.~27, no.~1, pp. 6--18, 2017.

\bibitem{bib:dacmobi}
J.~Wu, J.~Wu, H.~Cui, C.~Luo, X.~Sun, and F.~Wu, ``{DAC-Mobi}: Data-assisted
  communications of mobile images with cloud computing support,'' \emph{IEEE
  Transactions on Multimedia}, vol.~18, no.~5, pp. 893--904, 2016.

\bibitem{bib:dacran}
J.~Wu, D.~Liu, X.~L. Huang, C.~Luo, H.~Cui, and F.~Wu, ``{DaC-RAN}: A
  data-assisted cloud radio access network for visual communications,''
  \emph{IEEE Wireless Communications}, vol.~22, no.~3, pp. 130--136, 2015.

\bibitem{bib:kmvcast2}
X.~L. Huang, X.~Huan, J.~Wu, Q.~Sun, and Y.~Yuan, ``Performance analysis of
  {KMV}-cast with imperfect prior knowledge,'' in \emph{IEEE Global
  Communications Conference}, 2016, pp. 1--5.

\bibitem{bib:kmv-cast1}
X.~L. Huang, X.~Tang, X.~Huan, P.~Wang, and J.~Wu, ``Improved {KMV-Cast} with
  {BM3D} denoising,'' \emph{Mobile Networks and Applications}, vol.~23, no.~1,
  pp. 100--107, 2018.

\bibitem{bib:analog_code}
X.~Lin, Y.~Liu, and M.~Sun, ``Analog channel coding for wireless image/video
  {SoftCast} by data division,'' in \emph{International Conference on
  Telecommunications}, 2015, pp. 353--357.

\bibitem{bib:analog-coded}
B.~Tan, J.~Wu, Y.~Li, H.~Cui, W.~Yu, and C.~W. Chen, ``{Analog Coded SoftCast}:
  A network slice design for multimedia broadcast/multicast,'' \emph{IEEE
  Transactions on Multimedia}, vol.~19, no.~10, pp. 2293--2306, 2017.

\bibitem{bib:chaotic}
Y.~Liu, J.~li, X.~Lu, C.~Yuen, and J.~Wu, ``A family of chaotic pure analog
  coding schemes based on baker's map function,'' \emph{EURASIP Journal on
  Advances in Signal Processing}, vol. 2015, 2015.

\bibitem{bib:channel}
C.~He, H.~Wang, Y.~Hu, Y.~Chen, X.~Fan, H.~Li, and B.~Zeng, ``{MCast:}
  high-quality linear video transmission with time and frequency diversities,''
  \emph{IEEE Transactions on Image Processing}, vol.~27, no.~7, pp. 3599--3610,
  2018.

\bibitem{bib:progressive}
C.~Lan, D.~He, C.~Luo, F.~Wu, and W.~Zeng, ``Progressive pseudo-analog
  transmission for mobile video live streaming,'' in \emph{IEEE Visual
  Communications and Image Processing}, Singapore, dec 2015, pp. 1--4.

\bibitem{bib:analog_progressive}
D.~He, C.~Lan, C.~Luo, E.~Chen, F.~Wu, and W.~Zeng, ``Progressive pseudo-analog
  transmission for mobile video streaming,'' \emph{IEEE Transactions on
  Multimedia}, vol.~19, no.~8, pp. 1894--1907, 2017.

\bibitem{bib:CS_analog1}
A.~Wang, B.~Zeng, and H.~Chen, ``Wireless multicasting of video signals based
  on distributed compressed sensing,'' \emph{Signal Processing: Image
  Communication}, vol.~29, no.~5, pp. 599--606, 2014.

\bibitem{bib:cs3}
A.~Wang, Q.~Wu, X.~Ma, and B.~Zeng, ``A wireless video multicasting scheme
  based on multi-scale compressed sensing,'' \emph{EURASIP Journal on Advances
  in Signal Processing}, vol. 2015, no.~1, pp. 1--11, 2015.

\bibitem{bib:cs5}
S.~Liu, A.~Wang, H.~Wang, S.~Li, M.~Li, and J.~Liang, ``Adaptive residual-based
  distributed compressed sensing for soft video multicasting over wireless
  networks,'' \emph{Multimedia Tools and Applications}, vol.~76, no.~14, pp.
  15\,587--15\,606, 2017.

\bibitem{bib:BCS_SPL}
S.~Mun and J.~E. Fowler, ``Block compressed sensing of images using directional
  transforms,'' in \emph{IEEE International Conference on Image Processing},
  2009, pp. 3021--3024.

\bibitem{bib:overhead1}
Z.~Song, R.~Xiong, X.~Fan, S.~Ma, and W.~Gao, ``Transform domain energy
  modeling of natural images for wireless {SoftCast} optimization,'' in
  \emph{IEEE International Symposium on Circuits and Systems}, 2014, pp.
  1114--1117.

\bibitem{bib:overhead2}
R.~Xiong, J.~Zhang, F.~Wu, J.~Xu, and W.~Gao, ``Power distortion optimization
  for uncoded linear transformed transmission of images and videos,''
  \emph{IEEE Transactions on Image Processing}, vol.~26, no.~1, pp. 222--236,
  2017.

\bibitem{bib:l-shape}
R.~Xiong, F.~Wu, X.~Fan, C.~Luo, S.~Ma, and W.~Gao, ``Power-distortion
  optimization for wireless image/video {SoftCast} by transform coefficients
  energy modeling with adaptive chunk division,'' in \emph{IEEE International
  Conference on Visual Communications and Image Processing}, 2013, pp. 1--6.

\bibitem{bib:fuji_GMRF}
T.~Fujihashi, T.~Koike-Akino, T.~Watanabe, and P.~V. Orlik, ``High-quality soft
  video delivery with {GMRF}-based overhead reduction,'' \emph{IEEE
  Transactions on Multimedia}, vol.~20, no.~2, pp. 473--483, 2018.

\bibitem{bib:analogcast}
J.~Li, X.~E. Wen, H.~Jia, X.~Xie, and W.~Gao, ``{AnalogCast}: Full linear
  coding and pseudo analog transmission for satellite remote-sensing images,''
  in \emph{IEEE International Conference on Acoustics, Speech and Signal
  Processing}, 2016, pp. 1362--1366.

\bibitem{bib:blind}
T.~Zhang and S.~Mao, ``Metadata reduction for soft video delivery,'' \emph{IEEE
  Networking Letters}, vol.~1, no.~2, pp. 84--88, 2019.

\bibitem{bib:hetero_resolution}
Y.~Gui, H.~Lu, F.~Wu, and C.~W. Chen, ``Robust video broadcast for users with
  heterogeneous resolution in mobile networks,'' \emph{IEEE Transactions on
  Mobile Computing}, vol.~PP, no.~99, pp. 1--16, 2020.

\bibitem{bib:hetero1}
X.~Lin, W.~Hu, C.~Luo, and F.~Wu, ``Compressive image broadcasting in {MIMO}
  systems with receiver antenna heterogeneity,'' \emph{Signal Processing: Image
  Communication}, vol.~29, no.~3, pp. 361--374, 2014.

\bibitem{bib:airscale}
H.~Cui, C.~Luo, C.~W. Chen, and F.~Wu, ``Scalable video multicast for {MU-MIMO}
  systems with antenna heterogeneity,'' \emph{IEEE Transactions on Circuits and
  Systems for Video Technology}, vol.~26, no.~5, pp. 992--1003, 2016.

\bibitem{bib:free}
T.~Fujihashi, T.~Koike-Akino, T.~Watanabe, and P.~V. Orlik, ``Soft video
  delivery for free viewpoint video,'' in \emph{IEEE International Conference
  on Communications}, 2017, pp. 1--7.

\bibitem{bib:FreeCast}
------, ``{FreeCast}: Graceful free-viewpoint video delivery,'' \emph{IEEE
  Transactions on Multimedia}, vol.~21, no.~4, pp. 1000--1010, 2019.

\bibitem{bib:freecast2}
T.~Zhang and S.~Mao, ``{Joint Power and Channel Resource Optimization in Soft
  Multi-View Video Delivery},'' \emph{IEEE Access}, vol.~7, pp.
  148\,084--148\,097, 2019.

\bibitem{bib:Luo2019}
L.~Luo, T.~Yang, C.~Zhu, Z.~Jin, and S.~Tang, ``Joint texture/depth power
  allocation for {3-D} video {SoftCast},'' \emph{IEEE Transactions on
  Multimedia}, vol.~21, no.~12, pp. 2973--2984, 2019.

\bibitem{bib:depth}
T.~Yang, L.~Luo, C.~Zhu, and S.~Tang, ``Block {DCT} based optimization for
  wireless {SoftCast} of depth map,'' \emph{IEEE Access}, vol.~7, pp.
  29\,484--29\,494, 2019.

\bibitem{bib:Nu_MVV1}
T.~T. Nu, T.~Fujihashi, and T.~Watanabe, ``{Soft Video Uploading for Low-Power
  Crowdsourced Multi-view Video Streaming},'' \emph{IEICE Transactions on
  Communications}, no.~5, pp. 524--536, 2020.

\bibitem{bib:Nu_MVV2}
------, ``Power-efficient video uploading for crowdsourced multi-view video
  streaming,'' in \emph{IEEE Global Communications Conference}, 2018, pp. 1--7.

\bibitem{bib:360-fuji}
T.~Fujihashi, M.~Kobavashi, K.~Endo, S.~Saruwatari, S.~Kobayashi, and
  T.~Watanabe, ``Graceful quality improvement in wireless 360-degree video
  delivery,'' in \emph{IEEE Global Communications Conference}, 2018, pp. 1--7.

\bibitem{bib:omnicast}
J.~Zhao, R.~Xiong, and J.~Xu, ``{OmniCast}: Wireless pseudo-analog transmission
  for omnidirectional video,'' \emph{IEEE Journal on Emerging and Selected
  Topics in Circuits and Systems}, vol.~9, no.~1, pp. 58--70, mar 2019.

\bibitem{bib:360cast}
Y.~Lu, T.~Fujihashi, S.~Saruwatari, and T.~Watanabe, ``{360Cast}:
  Foveation-based wireless soft delivery for 360-degree video,'' in \emph{IEEE
  International Conference on Communications}, jun 2020, pp. 1--6.

\bibitem{bib:360cast+}
L.~Yujun, T.~Fujihashi, S.~Saruwatari, and T.~Watanabe, ``{360Cast+}: Viewport
  adaptive soft delivery for 360-degree videos,'' \emph{IEEE Access}, vol.~9,
  pp. 52\,684--52\,697, 2021.

\bibitem{bib:holocast}
T.~Fujihashi, T.~Koike-Akino, T.~Watanabe, and P.~Orlik, ``{HoloCast}: Graph
  signal processing for graceful point cloud delivery,'' in \emph{IEEE
  International Conference on Communications}, 2019, pp. 1--7.

\bibitem{bib:HoloCastGivens}
------, ``Overhead reduction in graph-based point cloud delivery,'' in
  \emph{IEEE International Conference on Communications}, 2020, pp. 1--7.

\bibitem{bib:Tang2020_1}
X.~W. Tang and X.~L. Huang, ``A design of {SDR}-based pseudo-analog wireless
  video transmission system,'' \emph{Mobile Networks and Applications},
  vol.~25, no.~6, pp. 2495--2505, 2020.

\bibitem{bib:implementation1}
S.~Chen, J.~Wu, H.~Ren, J.~Wu, B.~Zhang, and F.~Zhu, ``Hardware implementation
  of a pseudo-analog wireless video transmission system,'' in \emph{IEEE
  International Conference on Communication Technology Proceedings}, 2019, pp.
  519--524.

\bibitem{bib:implementation2}
F.~Gao, H.~Gao, and J.~Wu, ``A reconfigurable {SoC} for {SoftCast} wireless
  video transmission,'' in \emph{IEEE International Conference on Industrial
  Internet}.\hskip 1em plus 0.5em minus 0.4em\relax IEEE, 2018, pp. 169--170.

\bibitem{bib:implementation5}
Y.~Jiang, P.~Xia, J.~Wu, S.~Chen, and B.~Zhang, ``Pseudo-analog wireless stereo
  video transmission in hardware acceleration,'' in \emph{International
  Conference on Wireless Communications and Signal Processing}, 2017, pp. 1--6.

\bibitem{bib:implementation3}
Z.~Chen, X.~Zhang, S.~Wang, Y.~Xu, J.~Xiong, and X.~Wang, ``Enabling practical
  large-scale {MIMO} in {WLANs} with hybrid beamforming,'' \emph{IEEE/ACM
  Transactions on Networking}, vol.~29, no.~4, pp. 1605--1619, 2021.

\bibitem{bib:implementation4}
Z.~Ding, J.~Wu, W.~Yu, Y.~Han, and X.~Chen, ``Pseudo analog video transmission
  based on {LTE} physical layer,'' in \emph{IEEE/CIC International Conference
  on Communications in China}, 2016, pp. 1--6.

\bibitem{bib:ftv1}
Z.~Chen, X.~Zhang, Y.~Xu, J.~Xiong, Y.~Zhu, and X.~Wang, ``{MuVi}: Multiview
  video aware transmission over mimo wireless systems,'' \emph{IEEE
  Transactions on Multimedia}, vol.~19, no.~12, pp. 2788--2803, 2017.

\bibitem{bib:ftv3}
M.~Tanimoto, ``{FTV}:free-viewpoint television,'' \emph{Signal Processing:
  Image Communication}, vol.~27, no.~6, pp. 555--570, 2012.

\bibitem{bib:ftv4}
O.~Stankiewicz, M.~Domanski, A.~Dziembowski, A.~Grzelka, D.~Mieloch, and
  J.~Samelak, ``A free-viewpoint television system for horizontal virtual
  navigation,'' \emph{IEEE Transactions on Multimedia}, vol.~20, no.~9, pp.
  2182--2195, Aug. 2018.

\bibitem{bib:mpeg}
R.~Mekuria and L.~Bivolarsky, ``Overview of the {MPEG} activity on point cloud
  compression,'' in \emph{Data Compression Conference}, 2016, p. 620.

\bibitem{bib:dibr}
C.~Fehn, ``Depth-image-based rendering {(DIBR)}, compression, and transmission
  for a new approach on {3D-TV},'' in \emph{Stereoscopic Displays and Virtual
  Reality Systems}, vol. 5291, 2004, pp. 93--105.

\bibitem{bib:dibr1}
S.~Li, C.~Zhu, and M.-T. Sun, ``Hole filling with multiple reference views in
  {DIBR} view synthesis,'' \emph{IEEE Transactions on Multimedia}, vol.~20,
  no.~8, pp. 1948--1959, 2018.

\bibitem{bib:mvd}
Y.~Chen, M.~M. Hannuksela, T.~Suzuki, and S.~Hattori, ``Overview of the {MVC+D}
  {3D} video coding standard,'' \emph{Journal of Visual Communciation and Image
  Representation}, vol.~25, no.~4, pp. 679--688, may 2014.

\bibitem{bib:mvc_d}
A.~D. Abreu, P.~Frossard, and F.~Pereira, ``Optimizing multiview video plus
  depth prediction structures for interactive multiview video streaming,''
  \emph{IEEE Journal of Selected Topics in Signal Processing}, vol.~9, no.~3,
  pp. 487--500, 2015.

\bibitem{bib:3davc}
Y.~Chen and S.~Yen, ``{3D-AVC} draft text 6, document {JCT3V-D1002}.doc,''
  JCT-3V, Apr. 2013.

\bibitem{bib:viewport_only}
F.~Qian, B.~Han, L.~Ji, and V.~Gopalakrishnan, ``Optimizing 360 video delivery
  over cellular networks,'' in \emph{5th Workshop on All Things Cellular:
  Operations, Applications and Challenges}, 2016, pp. 1--6.

\bibitem{bib:sickness}
J.~Moss and E.~Muth, ``Characteristics of head-mounted displays and their
  effects on simulator sickness,'' \emph{The Journal of the Human Factors and
  Ergonomics Society}, vol.~53, no.~3, pp. 308--319, 2011.

\bibitem{bib:viewport}
D.~Ochi, Y.~Kunita, A.~Kameda, A.~Kojima, and S.~Iwaki, ``Live streaming system
  for omnidirectional video,'' in \emph{IEEE Virtual Reality}, 2015, pp.
  349--350.

\bibitem{bib:holodisplay}
P.~A. Blanche, A.~Bablumian, R.~Voorakaranam, C.~Christenson, W.~Lin, T.~Gu,
  D.~Flores, P.~Wang, W.~Y. Hsieh, M.~Kathaperumal, B.~Rachwal, O.~Siddiqui,
  J.~Thomas, R.~A. Norwood, M.~Yamamoto, and N.~Peyghambarian, ``Holographic
  three-dimensional telepresence using large-area photorefractive polymer,''
  \emph{Nature}, vol. 468, no. 7320, pp. 80--83, 2010.

\bibitem{bib:holodisplay2}
H.~Yu, K.~Lee, J.~Park, and Y.~Park, ``Ultrahigh-definition dynamic {3D}
  holographic display by active control of volume speckle fields,''
  \emph{Nature Photonics}, vol.~11, no.~3, pp. 186--192, 2017.

\bibitem{bib:draco}
\BIBentryALTinterwordspacing
``Draco {3D} data compression.'' [Online]. Available:
  \url{https://google.github.io/draco/}
\BIBentrySTDinterwordspacing

\bibitem{bib:kd}
O.~{Devillers} and P.~{Gandoin}, ``Geometric compression for interactive
  transmission,'' in \emph{Visualization}, 2000, pp. 319--326.

\bibitem{bib:PCL}
J.~Kammerl, N.~Blodow, R.~B. Rusu, S.~Gedikli, M.~Beetz, and E.~Steinbach,
  ``Real-time compression of point cloud streams,'' in \emph{IEEE International
  Conference on Robotics and Automation}, 2012, pp. 778--785.

\bibitem{bib:PCL2}
K.~Muller, H.~Schwarz, D.~Marpe, C.~Bartnik, S.~Bosse, H.~Brust, T.~Hinz,
  H.~Lakshman, P.~Merkle, F.~H. Rhee, G.~Tech, M.~Winken, and T.~Wiegand,
  ``{3D} is here: Point cloud library ({PCL}),'' in \emph{IEEE International
  Conference on Robotics and Automation}, 2011, pp. 1--4.

\bibitem{bib:octree}
R.~schnabel and R.~Klein, ``Octree-based point-cloud compression,'' in
  \emph{Eurographics Symposium on Point-Based Graphics}, 2006, pp. 111--121.

\bibitem{bib:DeQueiroz2016}
D.~Queiroz, R.~L., and P.~A. Chou, ``Compression of {3D} point clouds using a
  region-adaptive hierarchical transform,'' \emph{IEEE Transactions on Image
  Processing}, vol.~25, no.~8, pp. 3947--3956, aug 2016.

\bibitem{bib:DigitalGFT}
C.~{Zhang}, D.~{Flor{\^e}ncio}, and C.~{Loop}, ``Point cloud attribute
  compression with graph transform,'' in \emph{IEEE International Conference on
  Image Processing}, 2014, pp. 2066--2070.

\bibitem{bib:graph_PCC2}
P.~de~Oliveira~Rente, C.~Brites, J.~Ascenso, and F.~Pereira, ``Graph-based
  static {3D} point clouds geometry coding,'' \emph{IEEE Transactions on
  Multimedia}, vol.~21, no.~2, pp. 284--299, 2019.

\bibitem{bib:graph_PCC3}
C.~Zhang, D.~Florencio, and C.~Loop, ``Point cloud attribute compression with
  graph transform,'' \emph{IEEE International Conference on Image Processing},
  pp. 2066--2070, 2014.

\bibitem{bib:givens}
M.~A. Sadrabadi, A.~Khandani, and F.~Lahouti, ``Channel feedback quantization
  for high data rate {MIMO} systems,'' \emph{IEEE Transactions on Wireless
  Communications}, vol.~5, no.~12, pp. 3335--3338, 2006.

\bibitem{bib:givens2}
J.~C. Roh and B.~D. Rao, ``Efficient feedback methods for {MIMO} channels based
  on parameterization,'' \emph{IEEE Transactions on Wireless Communications},
  vol.~6, no.~1, pp. 282--292, 2007.

\bibitem{bib:hybrid_first_conf}
L.~Yu, H.~Li, and W.~Li, ``Hybrid digital-analog scheme for video transmission
  over wireless,'' in \emph{IEEE International Symposium on Circuits and
  Systems}, 2013, pp. 1163--1166.

\bibitem{bib:hybrid_first}
------, ``Wireless scalable video coding using a hybrid digital-analog
  scheme,'' \emph{IEEE Transactions on Circuits and Systems for Video
  Technology}, vol.~24, no.~2, pp. 331--345, 2014.

\bibitem{bib:DCNN2}
Y.~Chen and T.~Pock, ``{Trainable Nonlinear Reaction Diffusion:} a flexible
  framework for fast and effective image restoration,'' \emph{IEEE Transactions
  on Pattern Analysis and Machine Intelligence}, vol.~39, no.~6, pp.
  1256--1272, 2017.

\bibitem{bib:DCNN3}
Y.~Tai, J.~Yang, X.~Liu, and C.~Xu, ``{MemNet:} a persistent memory network for
  image restoration,'' in \emph{IEEE International Conference on Computer
  Vision}, 2018, pp. 4549--4557.

\bibitem{bib:GNNCast}
T.~Fujihashi, T.~K. Akino, S.~Chen, and T.~Watanabe, ``Wireless {3D} point
  cloud delivery using deep graph neural networks,'' in \emph{IEEE
  International Conference on Communications}, 2021, pp. 1--6.

\bibitem{bib:hda_practical}
C.~Lan, C.~Luo, W.~Zeng, and F.~Wu, ``A practical hybrid digital-analog scheme
  for wireless video transmission,'' \emph{IEEE Transactions on Circuits and
  Systems for Video Technology}, vol.~28, no.~7, pp. 1634--1647, 2018.

\bibitem{bib:hybridcast}
Z.~Song, R.~Xiong, S.~Ma, and W.~Gao, ``Hybridcast: A wireless image/video
  {Softcast} scheme using layered representation and hybrid digital-analog
  modulation,'' in \emph{IEEE International Conference on Image Processing},
  2014, pp. 6001--6005.

\bibitem{bib:hda_combined}
B.~Tan, J.~Wu, R.~Wang, W.~Luo, and J.~Liu, ``An optimal resource allocation
  for hybrid digital-analog with combined multiplexing,'' \emph{IEEE Internet
  of Things Journal}, vol.~6, no.~1, pp. 1125--1135, 2019.

\bibitem{bib:hda_model}
B.~Tan, H.~Cui, J.~Wu, and C.~W. Chen, ``An optimal resource allocation for
  superposition coding-based hybrid digital–analog system,'' \emph{IEEE
  Internet of Things Journal}, vol.~4, no.~4, pp. 945--956, 2017.

\bibitem{bib:Zhang2019}
J.~Zhang, A.~Wang, J.~Liang, H.~Wang, S.~Li, and X.~Zhang, ``Distortion
  estimation-based adaptive power allocation for hybrid digital-analog video
  transmission,'' \emph{IEEE Transactions on Circuits and Systems for Video
  Technology}, vol.~29, no.~6, pp. 1806--1818, 2019.

\bibitem{bib:hda3}
F.~Liang, C.~Luo, R.~Xiong, W.~Zeng, and F.~Wu, ``Superimposed modulation for
  soft video delivery with hidden resources,'' \emph{IEEE Transactions on
  Circuits and Systems for Video Technology}, vol.~28, no.~9, pp. 2345--2358,
  2018.

\bibitem{bib:hybrid_fading}
J.~Shen, L.~Yu, and H.~Li, ``Hybrid digital-analog scheme for video
  transmission over fading channel,'' in \emph{IEEE International Symposium on
  Circuits and Systems}, 2016, pp. 1582--1585.

\bibitem{bib:hda_fading2}
P.~Yahampath, ``Hybrid digital-analog coding with bandwidth expansion for
  correlated {Gaussian} sources under {Rayleigh} fading,'' \emph{EURASIP
  Journal on Advances in Signal Processing}, vol. 2017, no.~1, pp. 1--16, 2017.

\bibitem{bib:hda_ofdm}
------, ``Digital-analog superposition coding for {OFDM} channels with
  application to video transmission,'' in \emph{IEEE International Conference
  on Acoustics, Speech and Signal Processing}, 2018, pp. 1802--1806.

\bibitem{bib:Yahampath2020}
------, ``Video coding for {OFDM} systems with imperfect {CSI}: A hybrid
  digital–analog approach,'' \emph{Signal Processing: Image Communication},
  vol.~87, pp. 1--22, 2020.

\bibitem{bib:hybrid_mimo}
Y.~Liu, X.~Lin, N.~Fan, and L.~Zhang, ``Hybrid digital-analog video
  transmission in wireless multicast and multiple-input multiple-output
  system,'' \emph{Journal of Electronic Imaging}, vol.~25, no.~1, pp. 1--14,
  2016.

\bibitem{bib:hybrid_second}
L.~Yu, H.~Li, and W.~Li, ``Wireless cooperative video coding using a hybrid
  digital-analog scheme,'' \emph{IEEE Transactions on Circuits and Systems for
  Video Technology}, vol.~25, no.~3, pp. 436--450, 2015.

\bibitem{bib:fuji_cs}
T.~Fujihashi, T.~Koike-Akino, T.~Watanabe, and P.~V. Orlik, ``Compressive
  sensing for loss-resilient hybrid wireless video transmission,'' in
  \emph{IEEE Global Communications Conference}, 2015, pp. 1--5.

\bibitem{bib:swift}
D.~He, C.~Luo, F.~Wu, and W.~Zeng, ``Swift: A hybrid digital-analog scheme for
  low-delay transmission of mobile stereo video,'' in \emph{ACM International
  Conference on Modeling, Analysis, and Simulation of Wireless and Mobile
  Systems}, 2015, pp. 327--336.

\bibitem{bib:Li2020}
P.~Li, F.~Yang, J.~Zhang, Y.~Guan, A.~Wang, and J.~Liang,
  ``Synthesis-distortion-aware hybrid digital analog transmission for {3D}
  videos,'' \emph{IEEE Access}, vol.~8, pp. 85\,128--85\,139, 2020.

\bibitem{bib:holoplus}
T.~Fujihashi, T.~Koike-Akino, T.~Watanabe, and P.~V. Orlik, ``{HoloCast+:}
  hybrid digital-analog transmission for graceful point cloud delivery with
  graph fourier transform,'' \emph{IEEE Transactions on Multimedia}, vol.~PP,
  no.~99, pp. 1--13, 2021.

\bibitem{bib:fuji_DNN}
T.~Fujihashi, T.~Koike-Akino, P.~V. Orlik, and T.~Watanabe, ``{DNN}-based
  overhead reduction for high-quality soft delivery,'' in \emph{IEEE Global
  Communications Conference}, 2019, pp. 1--6.

\bibitem{bib:DeepSoft1}
E.~Bourtsoulatze, D.~B. Kurka, and D.~Gunduz, ``Deep joint source-channel
  coding for wireless image transmission,'' \emph{IEEE Transactions on
  Cognitive Communications and Networking}, vol.~5, no.~3, pp. 567--579, 2019.

\bibitem{bib:Tang2020_2}
X.-W. Tang, X.-L. Huang, F.~Hu, and Q.~Shi, ``Human-perception-oriented pseudo
  analog video transmissions with deep learning,'' \emph{IEEE Transactions on
  Vehicular Technology}, vol.~69, no.~9, pp. 9896--9909, 2020.

\bibitem{bib:soft-dip}
T.~Fujihashi, T.~Koike-Akino, T.~Watanabe, and P.~V. Orlik, ``High-quality soft
  image delivery with deep image denoising,'' in \emph{IEEE International
  Conference on Communications}, 2020, pp. 1--6.

\bibitem{bib:yolo}
J.~Redmon and A.~Farhadi, ``{YOLO9000}: better, faster, stronger,'' in
  \emph{IEEE Conference on Computer Vision and Pattern Recognition}, 2017, pp.
  7263--7271.

\bibitem{bib:DIP}
D.~Ulyanov, A.~Vedaldi, and V.~Lempitsky, ``Deep image prior,'' in \emph{IEEE
  Conference on Computer Vision and Pattern Recognition}, 2018.

\bibitem{bib:GNN}
C.~T. Duong, T.~D. Hoang, H.~H. Dang, Q.~V.~H. Nguyen, and K.~Aberer, ``On node
  features for graph neural networks,'' \emph{arXiv e-prints}, pp. 1--6, Nov.
  2019.

\bibitem{bib:ChenDYLFT:20}
S.~Chen, C.~Duan, Y.~Yang, D.~Li, C.~Feng, and D.~Tian, ``Deep unsupervised
  learning of {3D} point clouds via graph topology inference and filtering,''
  \emph{{IEEE} Transactions on Image Processing}, vol.~29, pp. 3183--3198,
  2020.

\bibitem{bib:PCT}
S.~{Chen}, S.~{Niu}, T.~{Lan}, and B.~{Liu}, ``{PCT}: Large-scale {3D} point
  cloud representations via graph inception networks with applications to
  autonomous driving,'' in \emph{IEEE International Conference on Image
  Processing}, 2019, pp. 4395--4399.

\end{thebibliography}

\vfill


\end{document}